\documentclass[fleqn,usenatbib]{mnras}

\usepackage{newtxtext,newtxmath}

\usepackage[T1]{fontenc}

\DeclareRobustCommand{\VAN}[3]{#2}
\let\VANthebibliography\thebibliography
\def\thebibliography{\DeclareRobustCommand{\VAN}[3]{##3}\VANthebibliography}

\usepackage{graphicx}	
\usepackage{amsmath}	
\usepackage{amssymb}	
\usepackage{subcaption}
\usepackage{xcolor}
\usepackage{multirow}
\usepackage{orcidlink}
\definecolor{ao}{rgb}{0.0, 0.5, 0.0}
\definecolor{bv}{rgb}{0.54, 0.17, 0.89}
\definecolor{r}{rgb}{0.8, 0.0, 0.0}
\definecolor{notegreen}{rgb}{0.235,0.651,0.282}

\newcommand{\oiii}{[\mbox{O\,\sc{iii}}]}

\newcommand{\EWoiiihb}{$W_{\lambda}$([\mbox{O\,\sc{iii}}]+$\mathrm{H\,\beta}$)}

\newcommand{\xiion}{$\xi_\mathrm{ion}$}

\newcommand{\wha}{$W_{\lambda}$(H\,$\alpha$)}

\title[The ionizing properties of SFGs]{The JWST EXCELS Survey: A spectroscopic investigation of the ionizing properties of star-forming galaxies at 1<z<8}
\author[R. Begley et al.]{R. Begley\orcidlink{0000-0003-0629-8074}$^{1}$\thanks{E-mail:rbeg@roe.ac.uk}, R. J. McLure$^{1}$\orcidlink{0009-0005-9742-2318}, F. Cullen\orcidlink{0000-0002-3736-476X}$^{1}$, A. C. Carnall\orcidlink{0000-0002-1482-5818}$^{1}$, T. M. Stanton\orcidlink{0000-0002-0827-9769}$^{1}$, D. Scholte\orcidlink{0000-0002-6867-1244}$^{1}$, \and D. J. McLeod\orcidlink{0000-0003-4368-3326}$^{1}$, J. S. Dunlop$^{1}$, K. Z. Arellano-Córdova\orcidlink{0000-0002-2644-3518}$^{1}$, C. Bondestam$^{1}$, C. T. Donnan\orcidlink{0000-0002-7622-0208}$^{2}$, \and M. L. Hamadouche$^{3}$, H.-H. Leung\orcidlink{0000-0003-0486-5178}$^{1}$, A. E. Shapley\orcidlink{0000-0003-3509-4855}$^{4}$, S. Stevenson$^{1}$\\\\
$^{1}$Institute for Astronomy, University of Edinburgh, Royal Observatory, Edinburgh EH9 3HJ, UK\\
$^{2}$NSF's National Optical-Infrared Astronomy Research Laboratory, 950 N. Cherry Ave., Tucson, AZ 85719, USA\\
$^{3}$Department of Astronomy, University of Massachusetts, Amherst, MA 01003, USA\\
$^{4}$Department of Physics \& Astronomy, University of California, 430 Portola Plaza, Los Angeles, CA 90095, USA}

\date{Accepted XXX. Received YYY; in original form ZZZ}
\pubyear{2025}

\begin{document}
\label{firstpage}
\pagerange{\pageref{firstpage}--\pageref{lastpage}}
\maketitle

\begin{abstract}
Charting the Epoch of Reionization demands robust assessments of what drives the production of ionizing photons in high-redshift star-forming galaxies, and requires better predictive capabilities from current observations. 
Using a sample of $N=159$ star-forming galaxies (SFGs) at $1\lesssim z \lesssim 8$, observed with ultra-deep medium-resolution spectroscopy from the \emph{JWST}/NIRSpec EXCELS survey, we perform a statistical analysis of their ionizing photon production efficiencies ($\xi_\mathrm{ion}$). 
We consider $\xi_\mathrm{ion}$, accurately measured with spectroscopic Balmer line measurements, in relation to a number of key galaxy observables including; nebular emission line equivalent widths ($W_\lambda(\mathrm{H\alpha})$ and $W_\lambda$(\mbox{[O\,\textsc{iii}]})), UV luminosity ($M_\mathrm{UV}$) and spectral slope ($\beta_\mathrm{UV}$), as well as dust attenuation ($E(B-V)_\mathrm{neb}$) and redshift. 
Implementing a Bayesian linear regression methodology, we fit $\xi_\mathrm{ion}$ against the principal observables while fully marginalising over all measurement uncertainties, mitigating against the impact of outliers and determining the intrinsic scatter. Significant relations between $\xi_\mathrm{ion}$ and $ W_\lambda(\mathrm{H\alpha})$, $W_\lambda$(\mbox{[O\,\textsc{iii}]}) and $\beta_\mathrm{UV}$ are recovered. Moreover, the weak trends with $M_\mathrm{UV}$ and redshift can be fully explained by the remaining property dependencies. Expanding our analysis to multivariate regression, we determine that $W_\lambda(\mathrm{H\alpha})$ or $W_\lambda$(\mbox{[O\,\textsc{iii}]}), along with $\beta_\mathrm{UV}$ and $E(B-V)_\mathrm{neb}$, are the most important observables for accurately predicting $\xi_\mathrm{ion}$. The latter identifies the most common outliers as SFGs with relatively high $E(B-V)_\mathrm{neb}\gtrsim0.5$, possibly indicative of intense obscured star-formation or strong differential attenuation. Combining these properties enable $\xi_\mathrm{ion,0}$ to be inferred with an accuracy of $\simeq0.15\,$dex, with a population intrinsic scatter of $\sigma_\mathrm{int}\simeq0.035\,$dex.
\end{abstract}




\section{Introduction}\label{sec:intro}

The rate of ionizing photon production (Lyman continuum; LyC, $\mathrm{h}\nu\geq13.6\,\mathrm{eV}$) and the subsequent escape fraction are critical quantities in determining the total number of ionizing photons available \citep{robertson+15,robertson+23} to drive and then maintain the ionized state of the Universe \citep[][]{goto+21,bosman+21}. 
Sufficient evidence has now been collated to confidently suggest that it was star-forming galaxies \citep[SFGs;][]{chary+16,robertson+23}, rather than quasars \citep[][]{aird+15,kulkarni+19,matsuoka+23,trebitsch+23}, that forced this transition in the ionization state of the Universe, during the so-called Epoch of Reionization (EOR).

The population demographics of galaxies at redshifts $5\leq z \leq 10$, during which the bulk of reionization takes place \citep[][]{mason+18,planck20,robertson+23,simmonds+24b,roberts-borsani+24,tang+25}, is now increasingly well constrained.  Determinations of the UV luminosity function at these epochs provide these key constraints \citep[e.g., see][]{mclure+13,bouwens+15,donnan+22,mcleod+24,adams+25,whitler+25}, revealing large populations of low-luminosity sources ($M_\mathrm{UV}\gtrsim -18$) consequential of a relatively steep faint-end slope \citep{finkelstein+15,atek+23} as well as a surprising abundance of $M_\mathrm{UV}-$bright sources \citep[i.e., $M_\mathrm{UV}\lesssim-20$][]{bowler+15,whitler+25}.

Connecting the SFG demographics to models tracking the timeline and topology of reionization, then becomes a challenge of piecing together the relative ionizing photon budget contributions of different sub-populations, which in turn begs the question - which reionization-era SFGs are the most conducive to strong ionizing outputs?

The escape of ionizing photons (parameterised as the LyC escape fraction; $f\mathrm{_{esc}^{LyC}}$) can only be directly constrained from the local Universe \citep[e.g., see][]{izotov+18,flury+22b,maulick+25} out to intermediate redshifts ($z\lesssim4$) as a result of the increasingly opaque IGM at higher redshifts \citep[][]{inoue+08}. Photometric and spectroscopic studies at $z\simeq3-4$, provide estimates of $f\mathrm{_{esc}^{LyC}}\simeq3-10$ per cent \citep[][]{steidel+18,pahl+21,pahl+23,begley+22}, in good agreement with predictions from simulations \citep[e.g., see][]{ma+20,trebitsch+23}. Consequentially for current models of the reionization timeline utilising UV luminosity functions, these works \citep[see also;][]{marchi+18,fletcher+19} also find that higher  $f\mathrm{_{esc}^{LyC}}$ is more likely in $M_\mathrm{UV}-$faint sources \citep[however, see][for opposing arguments]{naidu+20,jung+23}.

Significant effort has also been invested in constructing accurate and robust indirect methods of measuring LyC leakage, primarily from low-to-intermediate redshift galaxy observations \citep[e.g., see][]{jaskot+19,flury+22b,mascia+23,saldana-lopez+23}. In principle, such indirect probes then allow key insights into the LyC escape and ionizing capabilities of galaxies beyond $z\sim4$ and into the EOR \citep[][]{chisholm+22,begley+23,mascia+24}. However, it is worth highlighting there are still significant open questions regarding Lyman continuum escape. Firstly, the scatter in many of the derived relations between $f\mathrm{_{esc}^{LyC}}$ and other observed properties is significant \citep[e.g., see][]{flury+22b,jaskot+24,jaskot+25}. Moreover, there are still significant gaps in the parameter spaces (e.g., $M_{\mathrm{UV}}$) accessible from current simulations and observations, in addition to more complex temporal and small-scale geometrical factors thought to significantly impact LyC escape \citep{kimm+22,maji+22,katz+23,choustikov+24} that are extremely challenging to observationally explore.

Nonetheless, taken together, the numerous population of faint galaxies, along with their apparently favourable $f\mathrm{_{esc}^{LyC}}$ conditions, would suggest that they are likely the dominant contributors to the ionizing photon budget. However, such conclusions can only be maintained if these faint populations are also significant \emph{producers} of ionizing photons. The ionizing photon production efficiency ($\xi_\mathrm{ion}$), defined as $\xi_\mathrm{ion}=N(\mathrm{H^0})/L_\mathrm{UV}$ where $N(\mathrm{H^0})$ is the rate of ionizing photon production and $L_\mathrm{UV}$ is the intrinsic monochromatic UV luminosity ($\lambda_\mathrm{rest}=1500\,$\AA), is the most common parameterisation of ionizing photon production, and is typically required to be $\mathrm{log_{10}}(\xi_\mathrm{ion,0}\,/\,\mathrm{erg\, s^{-1}\, Hz})\gtrsim25.2$ on average to satisfy reionization modelling requirements \citep[][]{robertson+15}.

The first years of imaging and spectroscopy data from JWST enabled significant progress to be made characterising the ionizing photon production efficiencies of galaxies from cosmic noon ($z\sim2$) through to the EOR \citep[][]{simmonds+24b,begley+25,pahl+25a,llerena+25,pahl+25b,papovich+25}. Importantly, this collation of the latest literature suggests that the SFG population has sufficiently high $\xi_\mathrm{ion}$ on average to drive reionization to a timely completion, but with significant population scatter \citep[e.g., $
\langle \mathrm{log_{10}}(\xi_\mathrm{ion,0}\,/\,\mathrm{erg\, s^{-1}\, Hz})\rangle\simeq25.3$ and $\sigma\simeq0.3\,$dex, see][where $\xi_\mathrm{ion,0}$ is computed implicitly assuming $f_\mathrm{esc}^{\mathrm{LyC}}=0$, as common in the literature]{begley+25}. This helps avoid a photon budget crisis \citep[][]{munoz+24} in which there are too many ionizing photons based on pre-JWST expectations \citep[][]{finkelstein+19,stefanon+22,chen+24}.

There remain however, critical impasses to fully understanding what the quintessential `reionization-driving galaxy’ is. Recent work by \citet{endsley+22,begley+25} and \citet{pahl+25a} \citep[see also;][]{simmonds+24b} find evidence for a $\xi_\mathrm{ion}-M_\mathrm{UV}$ anti-correlation during the EOR, in which $M_\mathrm{UV}-$bright sources display larger $\xi_{\mathrm{ion}}$, consistent with being observed in a recent burst phase (in addition to increasingly bursty star-formation towards higher redshifts). In contrast, other works suggest the fainter population has elevated $\xi_\mathrm{ion}$ \citep[][]{prieto-lyon+23,llerena+23,simmonds+24,papovich+25}. Differences in these inferred relations could be due to a multitude of factors, likely including sample bias effects from spectroscopically versus photometrically selected SFG populations and SED-modelling versus directly derived ionizing photon production rates.

These conflicting trends likely arise from a combinations of both the diversity in the true underlying physical conditions \citep[e.g., stellar populations, metallicities, burstiness, initial mass function among others][]{conroy+13,eldridge+17,robertson+21,roberts-borsani+24,hayes+25} as well as a range of possible observational and measurement effects (e.g., selection biases, spectroscopic versus photometric samples, treatment of dust correction, to name a few). 

It is clear that a more detailed understanding of what dictates the production (and escape) of ionisation photons across the galaxy population, and how we can better calibrate this with the observed and derived physical properties of the SFG population is required. To this end, in this work we explore the physical drivers of $\xi_\mathrm{ion}$ using medium resolution spectroscopy of $N= 159$ SFGs observed in the JWST/NIRSpec Early eXtragalactic Continuum and Emission Line Science (EXCELS; GO 3543; PIs: Carnall, Cullen; \citet{carnall+24}) program, spanning a wide swathe of cosmic time ($z\simeq 1-8$). By statistically investigating the factors associated with $\xi_\mathrm{ion}$, we aim to decipher what facilitates increased ionizing photon production in star-forming galaxies.

The structure of the paper is as follows. In Section \ref{ses:data_and_sample} we describe the EXCELS dataset and galaxy measurements, including the emission line fluxes, UV continuum slopes, and ionizing photon production efficiencies.
In Section \ref{sec:analysis}, we introduce our Bayesian Linear regression methodology, while Section \ref{sec:results} presents the results of fitting $\xi_\mathrm{ion,0}$ against the key observed properties (i.e., see Section \ref{subsec:analysis:sv_fitting}). Section \ref{subsec:modelling:mv} expands this to a multivariate analysis, including discussions on the best predictors for $\xi_\mathrm{ion,0}$ and the impact of outliers.
Lastly, in Section \ref{sec:conclusions}, we summarise the main conclusions of our analysis.

Throughout the paper we adopt the following cosmological parameters: $H_0=70\,\mathrm{km\,s^{-1}\,Mpc^{-3}}$, $\Omega_\mathrm{m}=0.3$, $\Omega_\Lambda=0.7$ and all magnitudes are quoted in the AB system \citep[][]{oke_gunn+83}.

\section{Data and sample properties}\label{ses:data_and_sample}

\subsection{The EXCELS survey}
The spectroscopic sample that forms the basis of this work was observed in the JWST NIRSpec EXCELS survey. The primary goal of the EXCELS survey was to target massive $3\leq z\leq5$ quiescent galaxies with high signal-to-noise, medium resolution ($R=\lambda/\Delta\lambda\simeq1000$) spectroscopy, in addition to hundreds of star-forming galaxies across $z\simeq2-8$ observed in the remaining MSA slits.
For complete details on the selection of EXCELS survey targets, observing configurations and data processing, readers are directed to \citet{carnall+24} \citep[see also][]{scholte+25,stanton+25}. Below we provide a brief overview. 
\begin{figure*}
    \centering
    \includegraphics[width=2.\columnwidth]{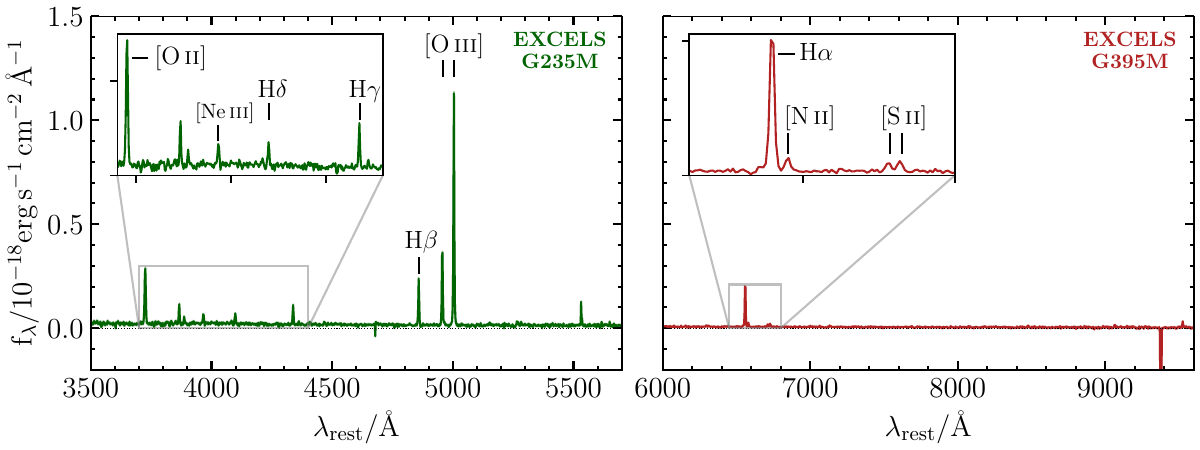}
    \caption{Example spectroscopy available for a galaxy in the EXCELS sample (ID: EXCELS-59720). The G235M/F170LP and G395M/F290LP spectra from EXCELS for this SFG are shown (\textit{left} and \textit{right}, respectively), highlighting the plethora of rest-frame optical emission lines (including but not limited to; {\oiii}, $\mathrm{H\alpha}$, $\mathrm{H\beta}$, $\mathrm{H\gamma}$, $\mathrm{H\delta}$; see insets) potentially observable with the deep NIRSpec spectroscopy.}
    \label{fig:example_excels_spectra}
\end{figure*}

\subsubsection{Observing configuration and data reduction}\label{subsubsec:excelstechnical}

The EXCELS survey consists of four NIRSpec/MSA pointings within the UDS (Ultra-Deep Survey) footprint of the Cycle 1 JWST/NIRCam survey PRIMER (Public Release IMaging for Extragalactic Research; GO 1837; PI: Dunlop). Each pointing is observed with the G140M/F100LP, G235M/F170LP and G395M/F290LP gratings (with unique masks in each) providing continuous wavelength coverage over $\lambda_\mathrm{obs}\simeq1-5\,\mu\mathrm{m}$, and employing a 3-shutter slitlet configuration and a 3-point dither pattern. The sources are observed for $\simeq4\,$hours in the G140M and G395M gratings, and $\simeq5.5$ hours in the G235M grating, reaching $3\sigma$ line flux limits of $\simeq5\times10^{-19}\,\mathrm{erg\,s^{-1}\,cm^{-2}}$ \citep[see][for full details]{carnall+24}.

Unique NIRSpec/MSA masks were designed for each grating configuration and pointing, optimised to maximise the number of targets observed with wavelength coverage spanning key spectral diagnostics, specifically at $\lambda_\mathrm{rest}\simeq3500-7000\,\mathrm{\AA}$ (i.e., the rest-frame near-UV to optical regime).
The EXCELS target selection was primarily based on the VANDELS survey catalogues \citep{mclure+18,garilli+21}, in addition to galaxies selected from PRIMER/UDS photometric catalogues \citep[see][and additionally \citealt{mcleod+24,begley+25}; McLeod et al. in preparation for further details]{carnall+24}.

The level 1 NIRSpec data products were reduced using the JWST reduction pipeline, including the advanced snowball rejection (v1.15.1 with the $\mathrm{CRDS{\_}CTX=}\,$jwst\_1258.pmap \emph{JWST} Calibration Reference Data System files). Following this, the level 2 and level 3 pipeline procedures are implemented in their default configuration. A custom 1D optimal extraction of the 2D spectra is implemented, whereby the extraction centroid is set as the flux-weighted mean position \citep{horne+86} of the source within the NIRSpec/MSA slitlet.

Lastly, a two-stage flux calibration procedure is implemented to ensure accurate absolute flux calibrations for the EXCELS spectra accounting for wavelength-dependent slit losses arising from the broadening PSF at longer wavelengths, as well as small source centroid offsets between gratings. In the first stage, a relative flux calibration between the gratings is performed by scaling the G140M and/or G395M spectra to match the G235M spectra using the median flux of the overlapping spectra regions. Subsequently, an absolute spectrophotometric calibration is applied by scaling the spectra to align with the available broadband photometry described in Section \ref{subsec:physicalproperties} \citep[see also][following an analogous method used in the CECILIA survey]{rogers+24}.
Each EXCELS spectrum is integrated through the available set of filters overlapping in wavelength, with a linear interpolation applied to derive a wavelength-dependent absolute flux calibration. The calibration is verified using a subsample of EXCELS sources with high signal-to-noise continuum detections and/or emission lines in overlapping wavelength ranges between the gratings (e.g., see Fig. \ref{fig:example_excels_spectra}), indicating that the flux calibrations are accurate to $\lesssim5-8$ per cent.

\subsection{Supplementary photometric catalogues}\label{subsec:photometrycatalogues}

The EXCELS spectroscopic sample in this work benefits from having deep multi-wavelength imaging (as a result of its original selection). Specifically, all EXCELS sources are covered by the PRIMER/UDS footprint and therefore have imaging in eight JWST/NIRCam filters (F090W, F115W, F150W, F200W, F277W, F356W, F410M, F444W), in addition to deep optical imaging from HST/ACS in three filters (F435W, F606W, F814W) from CANDELS \citep[Cosmic Assembly Near-IR Deep Extragalactic Legacy Survey;][]{grogin+11,koekemoer+11}. Sources are detected in the PRIMER NIRCam/F356W image using \textsc{Source Extractor} \citep{bertin+96}, with photometry measured using $0.5-\mathrm{arcsec}$ diameter apertures (enclosing $\simeq85\,$per cent of the total flux assuming a point source) in dual-image mode on the full set of available images after they have been PSF homogenised to the F444W image \citep[see][]{mcleod+24}.

The associated photometric uncertainties are deduced from local depth maps, where each object flux error in a given band is calculated as the scaled median absolute deviation ($\sigma_\mathrm{local}=1.483\times\sigma_\mathrm{MAD}$) of the fluxes measured in the nearest $150-200$ empty-sky apertures \citep[e.g., see][]{mcleod+21}. A more comprehensive description of the PRIMER photometric catalogues and their creation can be found in \citet{begley+25} \citep[see also][McLeod et al. 2025, in preparation, for further methodology details]{mcleod+24}.

\subsection{EXCELS emission line measurements}\label{subsec:linemeasurements}

To measure robust line fluxes and equivalent widths for each source, a two-step fitting procedure is implemented - first estimating the continuum flux, and then fitting the continuum-subtracted (line only) spectrum with a linear combination of Gaussian line profiles. The methodology used to obtain the emission line and continuum models for the calibrated 1D EXCELS spectra, as shown in Fig. \ref{fig:example_excels_spectra}, are described comprehensively in \citet[][]{scholte+25} \citep[see also][]{stanton+25}, with a summary of the key elements included here.

The continuum flux is estimated from the running mean of the flux values within a $\Delta_{\rm{rest}}=350\,${\AA} top hat function, after clipping values outside the $16^\mathrm{th}-84^\mathrm{th}$ percentile range to mitigate the impact of strong spectral features or noise spikes on the continuum measurement \citep{stoughton+02}. Given the underestimated uncertainties in the pixel flux measurements produced from the JWST data reduction pipeline, an empirical correction based on the residuals of the continuum subtracted spectra was applied. This is implemented as a multiplicative factor to the pipeline flux uncertainties, given as $c_F= \frac{1}{2}(R_{84}-R_{16})/\tilde\sigma_{F}$ where $R_{X}$ is the $X^{\rm th}$ percentile of the residuals spectrum ($R$) and $\tilde{\sigma}_{F}$ is the median of the pipeline flux uncertainty. Typical values of the empirical correction factor are $c_F\simeq1.5-2.0$ (with median $\bar{ c}_F\simeq1.6$), in agreement with similar independent analyses \citep{maseda+23,carnall+24}

The (continuum-subtracted) emission line spectrum is fit using a linear combination of Gaussian line profiles, where each individual line amplitude is a free parameter. The total width of the line profile ($\sigma_{\rm{tot}}$) is given as the convolution of the intrinsic line width ($\sigma_{\rm{intr}}$) and the instrumental broadening ($\sigma_{\rm{g}}(\lambda)$) of the NIRSpec grating used: $\sigma_{\rm{tot}}^2 = \sigma_{\rm{intr}}^2 + \sigma_{\rm{g}}^2(\lambda) $. A common line velocity (relative to the input $z_{\rm{spec}}$) and intrinsic line width is used for each galaxy spectral fit resulting in $2+n_{\rm{line}}$ free parameters (where $n_\mathrm{line}$ is the number of emission line profiles included in the model).
The fitting is performed using a weighted least-squares minimisation via \textsc{SciPy} \citep{branch+99,virtanen+20}. The line fluxes are calculated from the line profile weighted integral over the individual wavelength pixels in each fitted spectrum, with the uncertainties propagated from the same weighted pixel fluxes \citep[e.g., see][]{scholte+25}.

To account for any additional uncertainties arising from the absolute flux calibration employed, an additional $\simeq8\,$per cent error is included in quadrature when absolute flux values, or line flux ratios between lines spanning multiple gratings, are used \citep[][see also Section \ref{subsubsec:excelstechnical}]{scholte+25}.

\subsection{Sample Selection}\label{subsec:sampleselection}

Observations taken in the EXCELS survey enabled spectroscopic redshifts to be derived for $N=349$ sources, of which $N=341$ were assigned high quality redshift flags of 3, 4 or 9 according to the framework outlined in \citet[][]{lefevre+05} \citep[see also;][]{pentericci+18}. The most secure quality flag, 4 indicating $\gtrsim99\,$per cent confidence in the derived redshift, was assigned to $N=309$ sources representing $\sim90\,$per cent of the full sample. Sources without a high quality redshift flag are not considered in this study.

Accurately inferring the ionizing photon production efficiency spectroscopically from nebular emission lines (see Section \ref{subsec:physicalproperties}) requires robust determination of the \emph{intrinsic} line fluxes corrected for the effects of dust attenuation. To calculate the nebular dust attenuation we use the Balmer decrement (see Section \ref{subsubsec:physicalproperties:dustcorrection} and Fig. \ref{fig:balmerdecrement_distributions}), and thus select galaxies with at least one Balmer line pair from $\mathrm{H\alpha}$, $\mathrm{H\beta}$, $\mathrm{H\gamma}$ and $\mathrm{H\delta}$. We additionally impose that the emission lines in a given pair are significantly detected, requiring $\mathrm{S/N}\geq3$ for each.
From this selection, we obtain a final sample of $N=159$ star-forming galaxies (noting that none of the targets classified as quiescent according to the priority classes defined in \citealt{carnall+24} are selected in the defined criteria). The sample contains galaxies across a wide redshift range ($1\lesssim z_{\mathrm{spec}}\lesssim8$), with a median redshift $\langle z_\mathrm{spec} \rangle=3.24$ and $\simeq73\,$per cent ($N=116$) of the sample between $z_{\rm{spec}}\simeq2-6$. The spectroscopic redshift distribution of the EXCELS sample is shown in Fig. \ref{fig:covariates_corner} alongside physical properties calculated in Section \ref{subsec:physicalproperties} below.
We note here that given the target selection from the VANDELS and PRIMER parent photometric catalogues \citep[][]{mclure+18,carnall+24} used in EXCELS, the sample assembled here is consistent with being drawn from the star-forming main sequence (i.e., as opposed to a narrow-band selected catalogue preferentially targetting emission line galaxies).

\subsection{Physical properties}\label{subsec:physicalproperties}

\subsubsection{Nebular dust attenuation}\label{subsubsec:physicalproperties:dustcorrection}

\begin{figure}
    \centering
    \includegraphics[width=1\columnwidth]{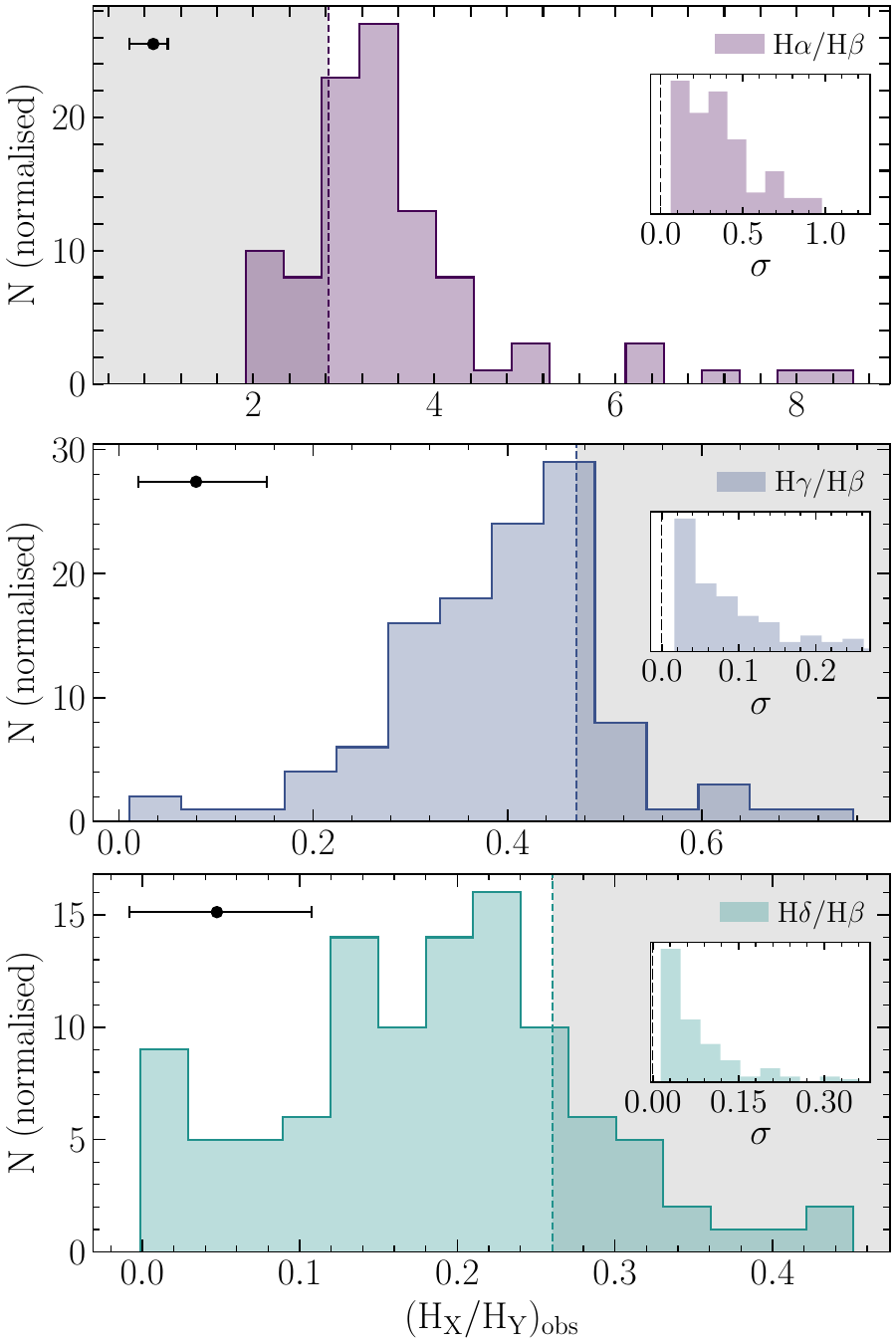}
    \caption{The observed Balmer decrement (see Section \ref{subsubsec:physicalproperties:dustcorrection} for details) distributions for the EXCELS galaxy sample ($N=159$) subsets with $\mathrm{H\alpha/H\beta}$ ($N=105$; \textbf{top panel}), $\mathrm{H\gamma/H\beta}$ ($N=121$; \textbf{centre panel}) and $\mathrm{H\delta/H\beta}$ ($N=112$; \textbf{bottom panel}). The inset panels show the distributions of measurement errors ($\mathrm{\sigma}$) for each Balmer decrement pair $\mathrm{(H_X/H_Y)}_\mathrm{obs}$, with typical errors of $\simeq^{+0.24}_{-0.15}$ for $\mathrm{H\alpha/H\beta}$, and of $\simeq^{+0.06}_{-0.07}$ for both $\mathrm{H\gamma/H\beta}$ and $\mathrm{H\delta/H\beta}$. The dashed lines and preceding grey shaded regions illustrate the intrinsic Balmer line ratios under the Case B recombination assumptions employed here.} 
    \label{fig:balmerdecrement_distributions}
\end{figure}

To correct the nebular emission line fluxes for the effects of dust attenuation, we compare the observed Balmer decrements in each source to the expected intrinsic line ratios. In each case, we use all the available Balmer line pairs from $H_X=\,$\{$\mathrm{H\alpha}$, $\mathrm{H\beta}$, $\mathrm{H\gamma}$, $\mathrm{H\delta}$\} where the lines in each pair are robustly detected ($\mathrm{S/N}\geq3$). The theoretical intrinsic line ratios are calculated with \textsc{PyNeb} \citep[v1.1.27][]{luridiana+15,morisset+20}, assuming case B recombination \citep{storey+95} with electron temperatures and densities of $T_e=12,000\,\mathrm{K}$ and $n_e=300\,\mathrm{cm^{-3}}$. These temperature and density assumptions are based on the average values found in EXCELS galaxies \citep[e.g., see][]{scholte+25,stanton+25,arellano-cordova+25}, and produce only marginally different intrinsic line ratios compared to the default case B assumptions commonly used (i.e., $T_e=10,000\,\mathrm{K}$ and $n_e=100\,\mathrm{cm^{-3}}$; with $\Delta (\mathrm{H\alpha/H\beta})_{\mathrm{intr}}\simeq0.03$ and $\Delta (\mathrm{H\gamma/H\beta})_{\mathrm{intr}}\simeq0.002$). In Fig. \ref{fig:balmerdecrement_distributions}, we show distributions of the observed Balmer decrements for $\mathrm{H\alpha/H\beta}$, $\mathrm{H\gamma/H\beta}$ and $\mathrm{H\delta/H\beta}$ for our EXCELS sample, with the theoretical intrinsic limits under our case B recombination assumptions also marked. The vast majority of the sample ($\simeq92\,$per cent) have observed Balmer decrements consistent with Case B recombination assumptions. Only 13 sources show discrepancies at the $>2\sigma$ level - for these sources we assume $E(B-V)_\mathrm{neb}=0.0$ (as commonly adopted in the literature; e.g., \citealt{pahl+25a}), with a more detailed treatment of different recombination cases left to future work \citep[but see][for recent relevant discussions]{scarlata+24,yanagisawa+24,arellano-cordova+25,mcclymont+25}.

The reddening from each galaxy, $E(B-V)_{\mathrm{neb}}$, is calculated as the variance weighted mean of values from each available Balmer line pair $E(B-V)_{\mathrm{neb},\,i}$, where:
\begin{equation}
E(B-V)_{\mathrm{neb},\,i}=2.5\frac{\mathrm{log}_{10} \left( \:\frac{(f_{\mathrm{H_X}}/f_{\mathrm{H_Y}})_{\mathrm{obs}}}{(f_{\mathrm{H_X}}/f_{\mathrm{H_Y}})_{\mathrm{intr}}}\: \right) }{k_\lambda(\lambda_{\mathrm{H_Y}})-k_\lambda(\lambda_{\mathrm{H_X}})}
\end{equation}
where $(f_{\mathrm{H_X}}/f_{\mathrm{H_Y}})_{\mathrm{obs}}$ and $(f_{\mathrm{H_X}}/f_{\mathrm{H_Y}})_{\mathrm{intr}}$ are the observed and intrinsic Balmer flux ratio, respectively, and $k_\lambda$ parameterises the dust attenuation law. We assume the nebular dust attenuation follows a \citet{cardelli+89} attenuation curve, having been shown to be a good approximation for high-redshift star-forming galaxies \citep{reddy+20}. In reality, the dust attenuation law likely varies across the galaxy population \citep[e.g., ][]{salim+20,sanders+24,reddy+25}, however, inferring the dust attenuation law on an individual galaxy-by-galaxy basis would require larger dynamic range in wavelength for each source than is currently available. Nevertheless, detailed analysis of the individual dust attenuation laws for a subset of EXCELS galaxies provides supporting evidence in favour of the \citet{cardelli+89}$\,-\,$like prescription on average (Stanton et al. in prep.). 

We take a Monte Carlo approach to estimate the uncertainties on $E(B-V)_\mathrm{neb}$ by repeating the calculation $\sim10^3$ times and perturbing the line flux measurements by their errors in each iteration, to construct a $E(B-V)_\mathrm{neb}$ distribution for each EXCELS galaxy. The fiducial $E(B-V)_\mathrm{neb}$ values and errors are given as the median and $\pm1\sigma$ ($16^{\rm{th}}-84^{\rm{th}}$ percentiles) of the distributions\footnote{Computing the median $E(B-V)_\mathrm{neb}$ \textit{after} removing MCMC draws that are inconsistent with case B recombination would increase $E(B-V)_\mathrm{neb}$ in the low attenuation tail ($E(B-V)_\mathrm{neb}<0.1$) by $\simeq0.05$. This corresponds to a $\simeq0.02\,$dex shift in $\xi_\mathrm{ion,0}$, which is well within our individual galaxy uncertainties.}, noting however, that a Monte Carlo approach is taken for all subsequent calculations based on the intrinsic nebular line fluxes using the measured $E(B-V)_\mathrm{neb}$ probability distribution functions. Lastly, we note that the probability distribution functions are truncated with a floor of $E(B-V)_\mathrm{neb}\geq0.0$, setting unphysical reddening values to zero.

\subsubsection{UV continuum slope}\label{subsubsec:physicalproperties:uvslope}

The  UV continuum slope, $\beta_{\mathrm{UV}}$, where $f_\lambda\propto\lambda^{\beta_{\mathrm{UV}}}$, is measured following the method described in \citet[][]{begley+25} \citep[see also][]{cullen+24}. Briefly here, only photometry covering rest-frame wavelengths $\lambda_{\mathrm{rest}}\leq3000\,${\AA} is considered (specifically, selecting filters with $\lambda_{95^\mathrm{th}}\leq3000\,${\AA}, where $\lambda_{95^\mathrm{th}}$ is the $95^\mathrm{th}$ percentile wavelength of the filters cumulative transmission). We then adopt a power-law functional form to model the flux redward of $\mathrm{Ly\alpha}$ ($f_\lambda=a\cdot\lambda^{\beta_\mathrm{UV}}$: $\lambda>1216\,$\AA), with the addition of the \citet{inoue+14} IGM prescription at shorter wavelengths. We note that the \citet{inoue+14} IGM attenuation prescription is that of the \emph{average} transmission function at a given redshift, and so to account for the increasing IGM stochasticity with redshift we include a scaling parameter $\phi$, which we marginalise over in our fitting. The final fitted model is then $f_\lambda=a\cdot T{_\lambda}(z,\phi) \cdot\lambda^{\beta_\mathrm{UV}}$, where $T{_\lambda}=1$ at $\lambda>1216\,${\AA}, and $T_\lambda=\phi\cdot g_\mathrm{Inoue14}(\lambda,z)$, otherwise.

The best-fitting model parameters are estimated from their posterior distributions, sampled using the MCMC ensemble sampler \textsc{emcee} \citep{foreman-mackey+2013}. During the sampling, we fix the model redshift to the EXCELS spectroscopic redshift ($z=z_\mathrm{spec}$), and adopt uniform priors across the remaining parameters with the UV continuum slope allowed to vary in the range $-10\leq \beta_\mathrm{UV}\leq 10$. We note that the $\phi$ parameter, which captures the effect of the stochasticity of the IGM transmission, is treated as a dummy parameter and does not appreciably affect our measurements.

The $\beta_{\mathrm{UV}}$ measurements are based on the available photometric catalogues (see Section \ref{subsec:photometrycatalogues}), which provide robust HST$+$JWST photometry over the observed wavelength range $0.36\,\mu\mathrm{m}\lesssim \lambda_{\mathrm{obs}}\lesssim5.0\,\mu\mathrm{m}$. As such, galaxies in our sample at $z\gtrsim2.95$ will have robust $\beta_\mathrm{UV}$ constraints owing to sufficient photometric coverage redward and blueward of the Lyman break (typically with $3-4$ bands redward of $\lambda_\mathrm{rest}=1216\,${\AA}, and up to 7 bands when also considering those blueward.). On the other hand, lower redshift sources in our sample lack the photometry fully redward of $\lambda_{\mathrm{Ly\alpha}}=1216\,${\AA} and have more uncertain UV continuum slopes as a consequence.

Lastly, we measure the absolute rest-frame UV magnitude ($M_\mathrm{UV}\equiv M_{1500}$) by first sampling 1000 realisations from the posterior UV continuum models of each source (described above). Each model is then integrated through a $\Delta\lambda=100\,${\AA} top-hat filter centered on $\lambda_\mathrm{rest}=1500\,${\AA} \citep{begley+23} to produce a distribution of $M_\mathrm{UV}$ for each galaxy. The $\beta_\mathrm{UV}$ and $M_\mathrm{UV}$ for the EXCELS galaxies are then given as the median of their respective distributions, with the $\pm1\sigma$ errors denoted as the $16^\mathrm{th}-84^\mathrm{th}$ percentiles. The median absolute UV magnitude for our final sample of $N=159$ EXCELS galaxies is $\langle M_\mathrm{UV} \rangle=-19.5$, with a $16^\mathrm{th}-84^\mathrm{th}$ percentile range of $M_\mathrm{UV}\in [-20.3,-18.4]$ and reaching a faint (bright) extreme of $M_\mathrm{UV}\simeq-16.5$ ($-21.2$). Similarly, across our EXCELS sample we find a median UV continuum slope of $\langle \beta_\mathrm{UV} \rangle=-1.50$, and a $16^\mathrm{th}-84^\mathrm{th}$ percentile range of $\beta_\mathrm{UV}\in [-2.21,-0.88]$. The EXCELS sample spans a large dynamic range in UV continuum slopes, with sources as blue (red) as $\beta_\mathrm{UV}\simeq-3.0$ ($+1.0$).

\begin{figure*}
    \centering
    \includegraphics[trim={4cm 4cm 4cm 6cm},clip,width=1.\textwidth]{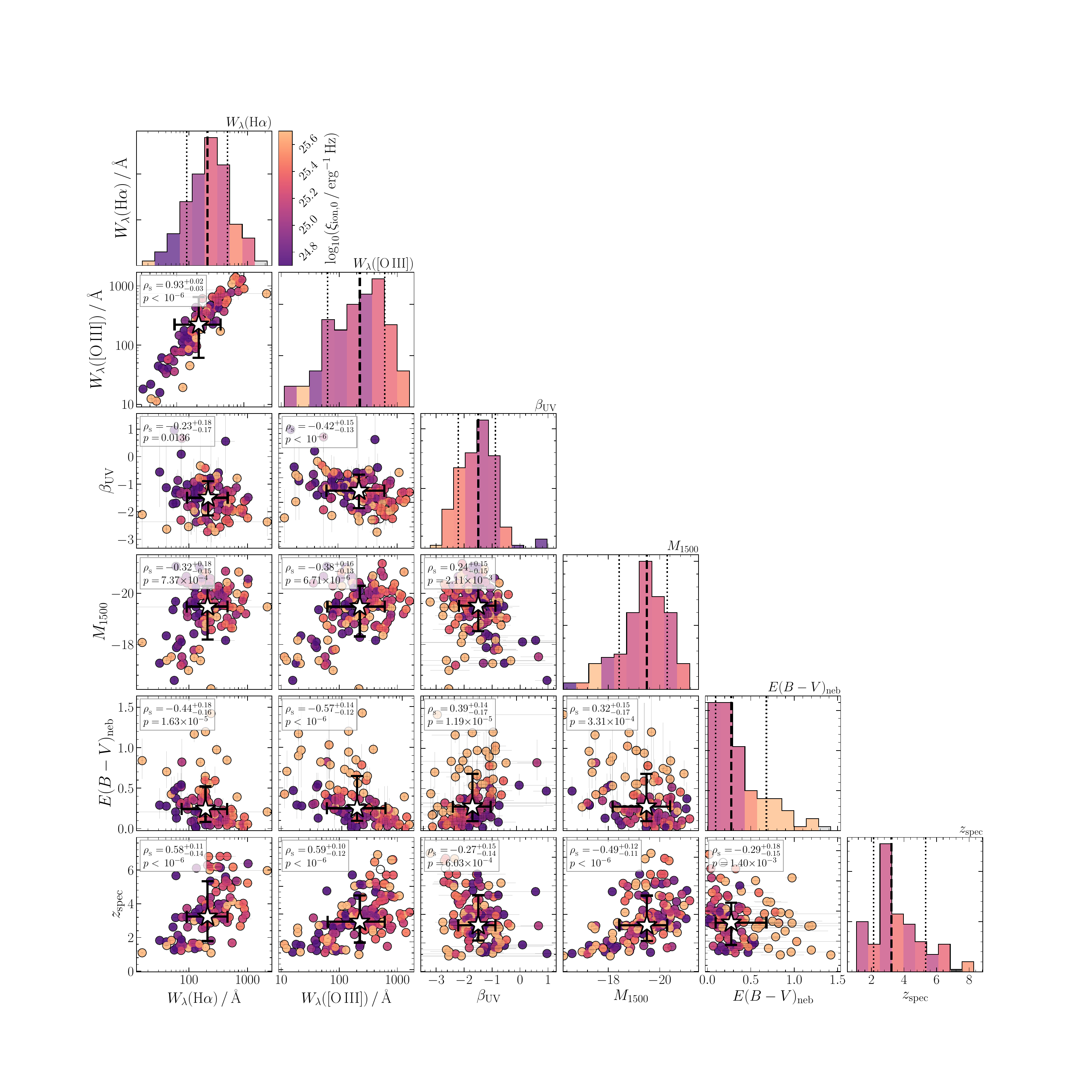}
    \caption{A corner plot of the six key galaxy observables $\mathbf{X}$ (where $X_i=\{ W_\lambda(\mathrm{H\alpha}),\, W_\lambda([\mathrm{O\,III}]),\, \beta_{\mathrm{UV}},\, M_{1500},\, E(B-V),\, z_\mathrm{spec} \}$, shown in order as columns left to right, and as rows top to bottom), from which we derive relationships with $\xi_\mathrm{ion,0}$ in Section \ref{sec:analysis}. Along the diagonal axes, we show the 1D marginal histograms for each covariate, highlighting the large dynamic range of physical properties spanned by our spectroscopic sample. Markers throughout are coloured by the inferred ionizing photon production efficiency ($\xi_\mathrm{ion,0}$; here specifically derived from $\mathrm{H\alpha}$, where available), with the histogram bars along the diagonal axes indicating the median $\xi_\mathrm{ion,0}$ in the given bin. In each panel,the star symbol represents the property medians and $16^\mathrm{th}-84^{\rm th}$ percentile range. Additionally, we present the Spearman rank correlation coefficient ($\rho_\mathrm{S}$) and associated $p-$value between parameter pairs, alongside the $95$ per cent confidence intervals on $\rho_\mathrm{S}$, computed with \textsc{Pingouin} \citep[e.g., see][for details]{vallat+18}. As shown in Section \ref{sec:analysis}, the apparent strength of the predictive power for a single parameter $X_i$ on $\xi_\mathrm{ion,0}$ can significantly weaken as a result of the collinearity between the observables in the multivariate case (see also Table \ref{tab:correlations_table}).}
    \label{fig:covariates_corner}
\end{figure*}

\begin{figure}
    \centering
    \includegraphics[width=1\columnwidth]{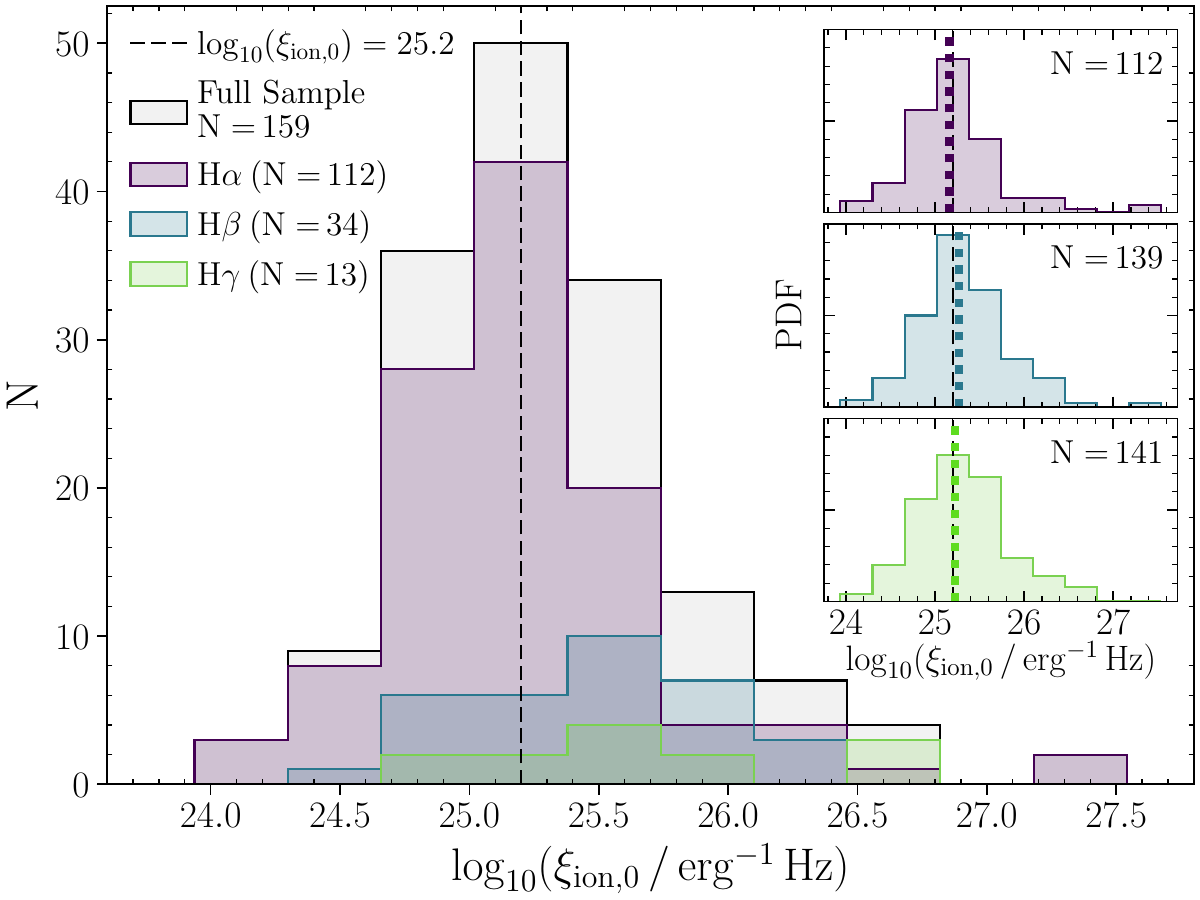}
    \caption{The distribution of ionizing photon production efficiencies ($\mathrm{\xi_\mathrm{ion,0}}$) for the sample of $N=159$ EXCELS galaxies (light grey), calculated from the intrinsically brightest available Balmer line from $\mathrm{H\,\alpha}$, $\mathrm{H\,\beta}$, $\mathrm{H\,\gamma}$ (see Section \ref{subsubsec:physicalproperties:ionizingphotonproduction} for details). From the total sample, $N=112$ are computed from $\mathrm{H\,\alpha}$, with a remaining $N=34$ from $\mathrm{H\,\beta}$ and $N=13$ from $\mathrm{H\,\gamma}$ (with each plotted individually in purple, blue and green, respectively). 
    The insets show the associated full $\xi_\mathrm{ion,0}$ distributions from each Balmer line (in galaxies where they are measurable), highlighting the excellent consistency across each. The coloured thick dotted lines indicate the distribution medians, and the black dashed vertical lines mark the canonical $\mathrm{log_{10}(\xi_\mathrm{ion,0}\,/\,erg^{-1}Hz})\simeq25.2$ value often cited as the benchmark "typical" limit for galaxies to have driven reionization completely \citep[][]{robertson+13}}.
    \label{fig:xiion_histogram}
\end{figure}

\subsubsection{Ionizing photon production efficiency}\label{subsubsec:physicalproperties:ionizingphotonproduction}

In this analysis, the ionizing photon production efficiency ($\xi_\mathrm{ion,0}\, /\, \mathrm{erg\cdot Hz^{-1}}$) complies with the common definition of: $\xi_\mathrm{ion,0}=N(\mathrm{H^0})/L_\mathrm{UV}^\mathrm{intr}$, where $N(\mathrm{H^0})\,/\, \mathrm{s^{-1}}$ is the production rate of hydrogen-ionizing photons ($h\nu\geq13.6\,\mathrm{eV}$) and $L_\mathrm{UV}^\mathrm{intr}\,/\,\mathrm{erg\cdot s^{-1}\cdot Hz^{-1}}$ is the UV continuum luminosity at $\lambda_\mathrm{obs}=1500${\AA} \textit{after} correcting for the effects of reddening from attenuating dust grains in the diffuse ISM.\footnote{Note, this is equivalent to the definition of $\xi\mathrm{_{ion}^{H\,II}}$ used in \citet{tang+19}.}

Firstly, we calculate the intrinsic $\mathrm{H\alpha}$ luminosity ($L\mathrm{^{intr}_{H\alpha}}$) by dust-correcting the observed line fluxes assuming a \citet{cardelli+89} dust curve and adopting the $E(B-V)_\mathrm{neb}$ values calculated from Balmer decrements as outlined in Section \ref{subsubsec:physicalproperties:dustcorrection}. The ionizing photon production rate can then be deduced from the intrinsic $\mathrm{H\alpha}$ luminosity under the assumption of Case B recombination \citep{osterbrock+06} using: $N(\mathrm{H^0})=7.35\times10^{-13} \cdot L\mathrm{^{intr}_{H\alpha}}$ (noting that we have implicitly dropped a factor of $1/(1-f_\mathrm{esc}^\mathrm{LyC})$ as our definition of $\xi_\mathrm{ion,0}$ assumes $f_\mathrm{esc}^\mathrm{LyC}=0$).

To compute the intrinsic UV luminosity ($L\mathrm{^{intr}_{UV}}$) we take advantage of the robust UV continuum slope measurements $\beta_\mathrm{UV}$ (calculated in Section \ref{subsubsec:physicalproperties:uvslope}). As detailed in \citet{begley+22} \citep[see also][]{mclure2018_irxbeta}, the UV dust attenuation can be derived from the observed UV continuum slope as: $A_{1500}= \kappa\cdot(\beta_\mathrm{UV}^\mathrm{obs}-\beta_\mathrm{UV}^\mathrm{intr})$, where $\beta_\mathrm{UV}^\mathrm{intr}$ is intrinsic UV spectral slope and $\kappa= d A_{1500}\,/\,d\beta$ as specified by the underlying assumed dust attenuation law. Here, we adopt a fiducial value of $\beta_\mathrm{UV}^\mathrm{intr}=-2.4$ which was found to be representative of SFGs at $z\simeq3-4$ in \citet[][]{begley+22}\footnote{Different assumptions for $\beta\mathrm{^{intr}_{UV}}$ will systematically shift our $\xi_\mathrm{ion,0}$ measurements. Changes of $\Delta(\beta\mathrm{^{intr}_{UV}})=\pm0.1$ will impact $L\mathrm{_{UV}^{intr}}$ by $\pm15\,$per cent, corresponding to $d\xi_\mathrm{ion,0}/d\beta\mathrm{_{UV}^{intr}}\simeq+0.5$. Specifically, adopting an alternative $\simeq0.1$ bluer intrinsic slope of $\beta\mathrm{^{intr}_{UV}}=-2.5$ will \textit{decrease} $\xi_\mathrm{ion,0}$ by $\simeq0.05\,$dex}. Additionally, we assume a dust law intermediate in greyness between a relatively flat \citet{calzetti+00}$-$like dust attenuation law and a steeper SMC dust extinction curve \citep{gordon+03}, using the \citet[][]{salim+18} dust curve parameterisation with $\delta=-0.25$.

From the full sample of $N=159$ galaxies studied in this work, we compute $\mathrm{H\alpha}-$based $\xi_\mathrm{ion,0}$ for $N=112$. For the remaining $N=47$ sources without $\mathrm{H\alpha}$ measurements (mainly due to EXCELS observing different sources with different combinations of NIRSpec gratings and a small number of sources where $\mathrm{H\alpha}$ is redshifted beyond the observed wavelength coverage), we calculate $\xi_\mathrm{ion,0}$ from the next available higher-order Balmer line ($N=34$ from $\mathrm{H\beta}$ and $N=13$ from $\mathrm{H\gamma}$). The full sample $\xi_\mathrm{ion,0}$ distribution, alongside the distributions calculated from $\mathrm{H\alpha}$, $\mathrm{H\beta}$, and $\mathrm{H\gamma}$ individually, are shown in Fig. \ref{fig:xiion_histogram}. 
The median ionizing photon production efficiency for the full EXCELS sample is $\mathrm{log_{10}(\xi_\mathrm{ion,0}\,/\,\mathrm{Hz\,erg^{-1}})=}25.22\pm0.01$, with a population scatter (quoted as the $\sigma_\mathrm{MAD}$) of $\Delta\xi_\mathrm{ion,0}\simeq0.42\:\mathrm{dex}$ (and $16^\mathrm{th}-84^\mathrm{th}$ percentile range of $\mathrm{log_{10}(\xi_\mathrm{ion,0}\,/\,\mathrm{Hz\,erg^{-1}})}\in  [24.82,\,25.74]$). We also highlight that the medians of the $\xi_\mathrm{ion,0}$ distributions calculated from each of the Balmer series lines individually (shown in the inset panels of Fig. \ref{fig:xiion_histogram}) agree within $\Delta\xi_\mathrm{ion,0}\lesssim0.1\,\mathrm{dex}$. Comparisons between the individual $\xi_\mathrm{ion,0}$ distributions reveal a marginal skew to increased $\xi_\mathrm{ion,0}$ values when inferred from higher order Balmer lines. However, this trend is minimal ($\lesssim0.1\:\mathrm{dex}$) and is consistent with the selection criteria requiring at least two Balmer series lines with $\mathrm{S/N}\geq3$ in conjunction with the fainter intrinsic luminosities of higher-order transitions.

\section{Analysis}\label{sec:analysis}

In the following section, we aim to establish relations between the ionizing photon production efficiency and a number of key galaxy observables $\mathbf{X}$; the equivalent widths of the strong optical nebular emission lines $\mathrm{H\,\alpha}$ and \oiii$\lambda5007$, the UV continuum slope and UV absolute magnitude, the nebular dust attenuation and redshift (that is $X_i=\{ W_\lambda(\mathrm{H\alpha}),\, W_\lambda([\mathrm{O\,III}]),\, \beta_{\mathrm{UV}},\, M_{1500},\, E(B-V),\, z_\mathrm{spec} \}$). These properties are chosen for a number of reasons, including they are: (i) routinely measurable in currently available JWST imaging and spectroscopic datasets, (ii) directly related to the physical drivers of ionizing photon production, and (iii) commonly considered in discussions related to the properties of the star-forming galaxies driving reionization and their evolution.

\subsection{Correlations between predictor variables}

Strong covariance between data in multivariate regression can degrade the quality of fits and inflate the uncertainties. Therefore, before introducing our fitting analysis in Section \ref{subsec:analysis:model} below and to assist in interpretation of the regression models, here we first examine the pairwise and conditional correlations between the six predictors in this analysis. 

\begin{figure}
    \centering
    \includegraphics[trim={0cm 0cm 0cm 0cm},clip,width=1.\columnwidth]{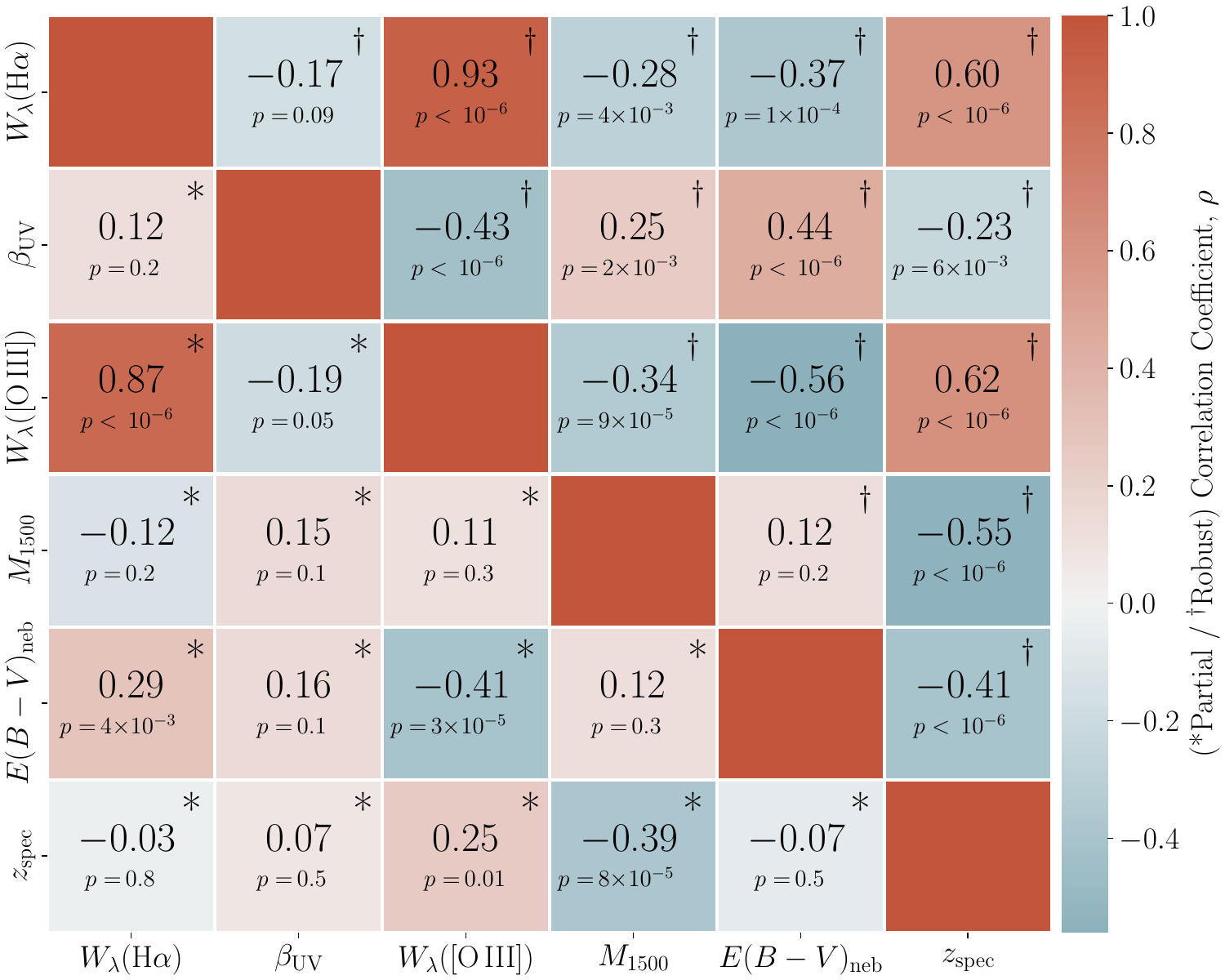}
    \caption{
    Heat map of the Spearman rank correlation ($\rho$) matrix for the six key predictors used in this work: $W_\lambda(\mathrm{H\alpha}),\: W_\lambda(\mathrm{[O\,III]}),\: \beta_\mathrm{UV},\: M_\mathrm{UV},\: E(B-V),\: z_\mathrm{spec}$. The lower panels (below the diagonal; denoted with a $*$) show the \textit{partial}$-\rho$ which removes the affect of all other covariates, while the upper panels (above the diagonal; denoted with a $^\dagger$) display the (`Skipped Spearman') robust$-\rho$, which mitigates against the impact of outliers. All quantities, including the $p-$values displayed below the coefficient at the centre of each panel, are computed using \textsc{Pingouin} \citep[][]{vallat+18}. In particular, the heat map brings attention to the mild-to-moderate (and in some cases very strong, e.g., between $\mathrm{H\,\alpha}$ and {\oiii} equivalent width) collinearity between the predictors, which may impact the regression parameters.
    }
    \label{fig:covariates_heatmap}
\end{figure}

Fig. \ref{fig:covariates_corner} shows a corner plot of the six galaxy observables used in this work, with the Spearman rank coefficients listed in each panel\footnote{We note here that a Pearson correlation coefficient more directly measures a \textit{linear} monotonic relationship, and as such the correlation coefficients tend to be weaker (and marginally less significant) when compared to the Spearman correlation coefficient.}. An important first point to highlight, as evident from the figure, is that the majority of observed properties show at least mild-to-moderate correlations $\mid\rho_{S}\mid\simeq0.2-0.6$ with robust significance ($p\lesssim10^{-3}$). We further verify these correlations from computing the Skipped correlation coefficient (otherwise known as the \textit{robust} Spearman correlation), which has reduced sensitivity to outliers. These are illustrated in the upper triangle panels of Fig. \ref{fig:covariates_heatmap}, and generally suggest the standard correlation coefficients are not being driven by outliers. 

These indicate that at some level, any single-observable fit (e.g., see Section \ref{subsec:analysis:sv_fitting} below) will implicitly fold in evolution in other covariates - a fact that should be held in consideration when being used for inferring ionizing photon production efficiencies. This would also imply that the covariances significantly influence the standard (Spearman) correlation coefficients. To test this, we also compute the partial correlation coefficient (controlling for all other variables), shown in the lower triangle of the heat map in Fig. \ref{fig:covariates_heatmap}. A number of key conclusions can be drawn from comparisons between the partial and standard correlation coefficients. Firstly, most of the observed correlations are considerably weaker as well as less significant based on the partial correlations. This trend is generally anticipated given all the variables show at least mild correlation with all other variables. A number of covariates show flipped correlations when controlling for those remaining (e.g., $W_\lambda(\mathrm{H\,\alpha})-\beta_\mathrm{UV}$, or $W_\lambda$([O\,\textsc{iii}])$-M_\mathrm{1500}$), and others become practically uncorrelated (e.g., $W_\lambda(\mathrm{H\,\alpha})-M_\mathrm{1500}$), which is indicative of the predictor bringing unique information with which to constrain $\xi_\mathrm{ion,0}$.

We note here, that some of the correlations will be influenced by the data limitations and sample selection adopted in the analysis. For example, an apparent correlation between redshift and absolute UV magnitude is expected and seen, given our faintest population at $M_{1500}\gtrsim-18$ is predominantly at $z\lesssim2$.

The most strongly correlated observed properties, including when accounting for other controlled variables with the partial correlation coefficient, are the equivalent widths of $\mathrm{H\,\alpha}$ and {\oiii}. Both these nebular emission lines trace the presence of intense, recent star-formation, leading to a strong correlation being expected \citep[e.g., see ][]{tang+19}. Practically for the regression modelling, the strong relationship motivates careful consideration when fitting together, including which is a more robust predictor, which we discuss at a later point in Section \ref{subsec:modelling:mv}.

\begin{figure*}
    \centering
    \includegraphics[width=2.1\columnwidth]{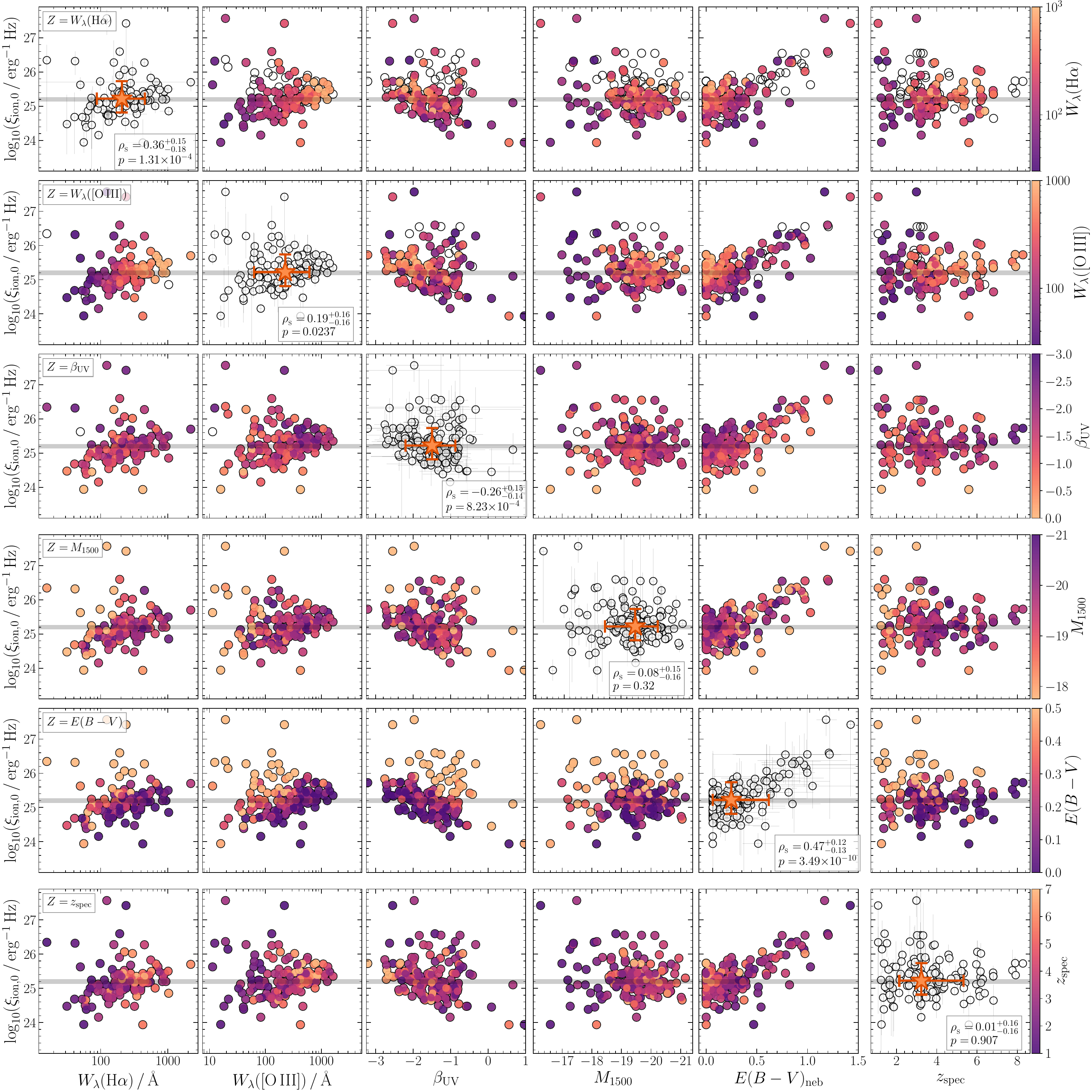}
    \caption{
    Figure demonstrating the relationships between the ionizing photon production efficiency, $Y=\xi_\mathrm{ion,0}$, and selected key observable covariates, $X_n=\{ W_\lambda(\mathrm{H\alpha}),\, W_\lambda([\mathrm{O\,III}]),\, \beta_{\mathrm{UV}},\, M_{1500},\, E(B-V),\, z_\mathrm{spec} \}$. Each column shows $\xi_\mathrm{ion,0}$ against a fixed covariate $X_i$, with each row having a designated covariate $X_j$ as the $z-\mathrm{axis}$ variable colouring the markers (with the associated colour bar at the end of each row). On-diagonal panels display the Spearman rank correlation coefficient, $\rho_{\mathrm{_{S}}}$, and associated $p-\mathrm{value}$ between the $X_i$ covariate in a given column with $\xi_{\mathrm{ion,0}}$. The upper and lower values of $\rho_{\mathrm{_{S}}}$ represent the 95 per cent parametric confidence intervals, calculated using \textsc{Pingouin} \citep{vallat+18}. We denote $\mathrm{log_{10}}(\xi_\mathrm{ion,0}\,/\,\mathrm{erg^{-1}Hz})=25.2$ as a horizontal grey line.
    }
    \label{fig:xiion_vs_covariates}
\end{figure*}

\subsection{Bayesian linear regression}\label{subsec:analysis:model}

To model the relation between $\xi_\mathrm{ion,0}$ and one or more of the predictor variables ($X_n=\{ W_\lambda(\mathrm{H\alpha}),\, W_\lambda([\mathrm{O\,III}]),\, \beta_{\mathrm{UV}},\, M_{1500},\, E(B-V),\, z_\mathrm{spec} \}$), we adopt a Bayesian linear regression methodology. Specifically, we aim to determine the regression parameters, including robust inferences of the intrinsic scatter, while marginalising over the per-galaxy uncertainties. 

Firstly, for each galaxy $i$, the regression model can be written:
\begin{equation}
    \mu_i(\theta) = \alpha+\sum_{k=1}^K \beta_k \cdot g_k(x_{i,k}^{\mathrm{true}})
\end{equation}
where $K$ is the number of predictor variables in the model each with a characteristic gradient $\beta_i$, resulting in $K+1$ model parameters $\theta$, including the intercept (i.e., $\theta=\{\alpha,\beta_1,...,\beta_K\}$). Here, $g_k(...)$ represents any transformation of the parameter $x_i$ (e.g., taking $\mathrm{log}_{10}$ in the case of $W_\lambda$, or centering the variable). 
To robustly account for errors in the observed data, we introduce latent ``true" $x_i$ variables which are related through a measurement model as: $x_i^{\mathrm{obs}}\sim \mathcal{N}(x_i^{\mathrm{true}},\sigma_i)$.

To better account for the possibility of significant outliers in the linear relationships (which are not captured by a Gaussian distribution; e.g., see Fig. \ref{fig:xiion_vs_covariates}), we generalise the typical Normal distribution likelihood to a Student$-t$ likelihood in our Bayesian formalism, given as:
\begin{equation}
\begin{split}
    p(y~|~\mu,\sigma_\mathrm{int},\nu) & = t_\nu(y~|~\mu,\sigma_\mathrm{int}^2) \\
    & = \frac{\Gamma(\tfrac{\nu+1}{2})}{\Gamma(\tfrac{\nu}{2})\cdot\sqrt{\nu\pi}\cdot\sigma_\mathrm{int}}\left[ {1+ \frac{1}{\nu}} \bigg( \frac{y-\mu}{\sigma_\mathrm{int}} \bigg) ^{2}  \right]^{-\tfrac{\nu+1}{2}}
\end{split}
\label{eq:student-t_kernel}
\end{equation}
where $\sigma_\mathrm{int}$ is the intrinsic scatter in the model (that is, explicitly stating our model is therefore $\mu_i(\theta) \Rightarrow \mu_i(\theta)+\epsilon_i$, and $\nu$ is the degrees of freedom in the Student$-t$ distribution. We fit for both $\sigma_\mathrm{int}$ and $\nu$ as free parameters, and note that the standard Normal likelihood is recovered when $\nu\rightarrow\infty$ (in practice $\nu\gg10$). However, the flexibility offered by a Student$-t$ model reduces the fitting sensitivity to outliers (unlike a Gaussian configuration) and yields more robust and accurate estimates of the parameter posterior distributions.

As discussed in Section \ref{subsubsec:physicalproperties:ionizingphotonproduction}, for each object we obtain a
\( \xi_{\mathrm{ion},0} \) distribution from our Monte Carlo error propagation (see examples shown in Fig.~\ref{fig:xiion_per_galaxy_distributions}).
Let \(D_i\) denote the upstream data for object \(i\) (photometry / line fluxes used to derive \(\xi_{\rm ion}\)) and write
\(y\equiv\log_{10}\xi_{\mathrm{ion},0}\). To accurately fold in the constraints provided by the per-object \(\xi_{\mathrm{ion},0}\)
distributions we begin from the exact marginal likelihood of the upstream data under the regression model:
\begin{equation}
    p(D_i\mid\theta) \;=\; \int p(D_i\mid y_i)\,p(y_i\mid\theta)\,dy_i,
\label{eq:true_marginalised_likelihood}
\end{equation}
where \(p(y\mid\theta)\) is the Student–\(t\) predictive density introduced in Eq.~\ref{eq:student-t_kernel} evaluated at \(\mu_i(\theta)\)
(i.e. \(p(y\mid\theta)\equiv t_\nu(y\mid\mu,\sigma_{\mathrm{int}}^2)\)). In practice we do not evaluate \(p(D_i\mid y_i)\) directly,
instead treating the Monte Carlo draws \(y_{i,1},\ldots,y_{i,J}\) from the per-object posterior \(p(y\mid D_i)\) as an empirical measurement-
uncertainty cloud and score the regression predictive density against it:
\begin{equation}
    \mathcal{L}_i(\theta) \;=\; \int p(y_i\mid\theta)\,p(y_i\mid D_i)\,dy_i
    \;\approx\; \frac{1}{J}\sum_{j=1}^{J} p\big(y_{i,j}\mid\theta\big).
\label{eq:mc_estimator}
\end{equation}
We evaluate this in log space using the numerically stable log-sum-exp identity:
\begin{equation}
\begin{split}
    \log p(D_i\mid \theta) &\approx \log\!\bigg(\frac{1}{J}\sum_{j=1}^{J}p(y_{i,j}\mid\theta)\bigg)\\
    &= \mathrm{logsumexp}\big(\{\log p(y_{i,j}\mid\theta)\}_{j=1}^J\big) - \log J .
\end{split}
\label{eq:loglike}
\end{equation}
Finally, the joint marginalised log-likelihood is the sum over galaxies,
\(\log p(\{D_i\}\mid\theta)=\sum_{i=1}^N\log p(D_i\mid\theta)\).

We note that Eq.~\ref{eq:mc_estimator} is formally equivalent (up to a \(\theta\)-independent constant) to the exact marginal likelihood in Eq.~\ref{eq:true_marginalised_likelihood} when the per-object posterior is data-dominated. In our analysis the \(y_{i,j}\) are produced by direct Monte Carlo propagation of the measured fluxes and Balmer-decrement dust corrections (i.e. they are not sampled assuming some strongly informative prior); we therefore adopt the estimator in Eq.~\ref{eq:mc_estimator} as our pragmatic choice for the likelihood.

\subsection{Implementation using \textsc{PyMC}}

We implement our Bayesian linear regression formalism using the \textsc{PyMC} framework \citep{salvatier+15}, and sample the posterior distribution $p(\theta\mid D_i)\propto p(\theta)\cdot p(D_i\mid\theta)$, where $p(D_i\mid\theta)$ written in Eq. \ref{eq:loglike}, using the No-U-Turn Sampler (NUTS). The NUTS sampler is a gradient-based (Hamiltonian Monte Carlo) algorithm for sampling Monte Carlo Markov Chains, whereby the step-size and number of steps per sample are automatically tuned \citep[e.g., see][for further details]{hoffman+11}. For the sampling, we use four independent sampling chains, with $N_\mathrm{tune}=1000$ tuning steps and $N_\mathrm{sample}=2000$ draws at a minimum.

We adopt a compound initialisation strategy for the NUTS sampler. First we perform a `warm-start' by running a variational inference (VI) optimization with \textsc{PyMC} ADVI (automatic differentiation variation inference); ADVI produces computationally cheap and sensible approximations of the posteriors. Samples drawn from these variational approximations are then used as the initial values for NUTS. Secondly, we use the \textsc{PyMC} `jitter$+$adapt\_diag' initialisation method, which enables appropriate tuning for the NUTS sampler. Together, these two steps reduce the possibility of early divergences in more complex parameter spaces and enable NUTS to sample more efficiently.

For the default fitting configuration used throughout this work, we adopt weakly informative priors for $p(\theta)$ on all regression parameters, setting the hyperparameters after performing prior predictive checks. This allows us to regularise the inference by selecting physically plausible ranges (aided by visually inspecting our data, e.g., Fig. \ref{fig:xiion_vs_covariates}), while enabling the data to drive the posterior sampling. Specifically, we set the intercept prior and gradient prior(s) as uniform:
\begin{align*}
    \alpha\sim\mathrm{Uniform(23,27)},\quad\quad \underbrace{\beta_k}_{k=1,...,K}\sim\mathrm{Uniform}(-2,+2) \quad
\end{align*}
The intrinsic scatter prior is stated as a half-normal, with the scale set empirically from the data:
\begin{align*}
    \sigma_\mathrm{int}\sim\mathrm{HalfNormal}(\sigma_\mathrm{hyp}), \quad 
    \sigma_\mathrm{hyp}=0.5\times\sigma_{\mathrm{MAD}}(\:\langle \mathrm{log_{10}}(\xi_{\mathrm{ion,0}})\rangle\:)
\end{align*}
which anchors the scatter to the typical per-galaxy spread (and disfavours extremely large values). The level of scatter in the heavy-tailed region of the Student$-t$ likelihood is parametrised with $\nu$ degrees of freedom and a prior as:
\begin{align*}
    \nu+2\sim \mathrm{Exponential(\lambda=\tfrac{1}{30})}
\end{align*}
which ensures finite variance ($\nu>2$) with the possibility of heavy tails if driven for by the data.

Lastly, we note that we performed a number of prior sensitivity tests in which we replaced the uniform regression priors with Gaussian priors (e.g., $\sim\mathcal{N}(0,1)$), and the intrinsic scatter prior with a number of alternatives (e.g., Uniform, Half-Cauchy, Half-Student$-t$). The posterior estimates of any of the key parameters described by $\theta$ remained stable to these changes, with the parameter medians shifting within their $\pm1\sigma$ uncertainties.

\begin{table}
    \caption{Correlation analysis for the EXCELS sample between the ionizing photon production efficiency, $Y=\xi_{\mathrm{ion,0}}$, and the six key physical parameters, $\mathbf{X}_n=\{W_\lambda(\mathrm{H\alpha}),\: W_\lambda(\mathrm{[O\,III]}),\: \beta_\mathrm{UV},\: M_\mathrm{UV},\: E(B-V),\: z_\mathrm{spec} \}$, displayed in Fig. \ref{fig:xiion_vs_covariates}. The \textbf{top} table displays the pairwise Pearson $r$ correlation coefficients $\rho(Y,X_i)$, whilst the \textbf{centre} table lists the Skipped-Spearman rank correlation coefficient $\rho_{\mathrm{robust}}(Y,X_i)$, which is considered a "robust" measurement accounting for outlier removal. Lastly, the \textbf{bottom} table presents the \emph{partial} (Spearman rank) correlation coefficient, $\rho(Y,X_i\mid \{ \mathbf{X}_j \}_{j\neq i})$, which controls for the multicollinearity between the physical variables. The table reports the individual correlation coefficients ($\rho$), the 95\% confidence intervals (95\% CI) and significance levels ($p-\mathrm{value}$) for each covariate, calculated using \textsc{Pingouin} Python package \citep[][]{vallat+18}.}
    \begin{tabular}{l | c c c}
    \hline\hline
    Covariate  & $\rho$ & 95\% CI & $p-\mathrm{value}$ \\
    \hline
    \multicolumn{4}{|c|}{Pearson Correlation: $\rho(Y,X_i)$} \\
    \hline
    $W_\lambda(\mathrm{H\alpha})$ & $+\,$0.172 & [$-\,$0.02, $+\,$0.35] & $0.072$\\
    $W_\lambda$([\mbox{O\,\sc{iii}}]) & $+\,$0.102 & [$-\,$0.09, $+\,$0.29] & $0.304$\\
    $\beta_\mathrm{UV}$ & $-\,0.397$ & [$-\,0.54$, $-\,0.23$] & $<10^{-4}$\\
    $M_{\mathrm{UV}}$ & $+\,$0.218 & [$-\,0.03$, $+\,$0.39] & 0.021 \\
    $E(B-V)$ & $+\,$0.657 & [$+\,0.54$, $+\,$0.75] & $<10^{-14}$\\
    $z_\mathrm{spec}$ & $-\,$0.015 & [$-\,0.20$, $+\,$0.17] & 0.879\\
    \hline
    \multicolumn{4}{|c|}{Skipped Spearman Correlation: $\rho_{\mathrm{robust}}(Y,X_i)$} \\
    \hline
    $W_\lambda(\mathrm{H\alpha})$ & $+\,$0.538 & [$+\,$0.38, $+\,$0.66] & $<10^{-8}$\\
    $W_\lambda$([\mbox{O\,\sc{iii}}]) & $+\,$0.535 & [$+\,$0.37, $+\,$0.67] & $<10^{-7}$\\
    $\beta_\mathrm{UV}$ & $-\,0.383$ & [$-\,0.54$, $-\,0.20$] & $<10^{-4}$\\
    $M_{\mathrm{UV}}$ & $+\,$0.110 & [$-\,0.09$, $+\,$0.30] & 0.278 \\
    $E(B-V)$ & $+\,$0.266 & [$+\,$0.07, $+\,$0.44] & $\simeq10^{-2}$\\
    $z_\mathrm{spec}$ & $+\,$0.268 & [$+\,$0.08, $+\,$0.44] & $\simeq10^{-2}$\\
    \hline
    \multicolumn{4}{|c|}{Partial Spearman Correlation: $\rho(Y,X_i \mid \{\mathbf{X}_j\}_{j\neq i})$} \\
    \hline
    $W_\lambda(\mathrm{H\alpha})$ & $+\,$0.315 & [$+\,$0.12, $+\,$0.48] & $\simeq10^{-3}$\\
    $W_\lambda$([\mbox{O\,\sc{iii}}]) & $+\,$0.089 & [$-\,$0.11, $+\,$0.28] & $0.386$\\
    $\beta_\mathrm{UV}$ & $-\,0.698$ & [$-\,0.79$, $-\,0.58$] & $<10^{-14}$\\
    $M_{\mathrm{UV}}$ & $+\,$0.327 & [$+\,$0.14, $+\,$0.49] & $\simeq10^{-3}$ \\
    $E(B-V)$ & $+\,$0.779 & [$+\,$0.69, $+\,$0.85] & $<10^{-20}$\\
    $z_\mathrm{spec}$ & $+\,$0.149 & [$-$0.05, $+\,$0.34] & 0.146\\
    \hline
    \end{tabular}
    \label{tab:correlations_table}
\end{table}

\section{Results}\label{sec:results}

\subsection{Single property predictions of $\xi_\mathrm{ion,0}$}\label{subsec:analysis:sv_fitting}

Recent imaging and spectroscopic surveys, such as those carried out with JWST, have enabled a rich set of physical properties to be measured or inferred on an individual galaxy basis across diverse populations; including the predictors used in this work (e.g., $\beta_{\mathrm{UV}}$, $M_\mathrm{UV}$). In Section \ref{subsec:modelling:mv} below, we explore combining various observed properties to obtain the most robust predictions of $\xi_\mathrm{ion,0}$. Here however, we present results from linear regressions between $\xi_\mathrm{ion,0}$ and single observables, including the intrinsic scatter, to facilitate more direct comparisons to the literature.

\subsubsection{Relation between $\xi_\mathrm{ion,0}$ and {\wha}}\label{subsubsec:analysis:sv_fitting:wha}

\begin{figure*}
    \centering
    \includegraphics[width=1\textwidth]{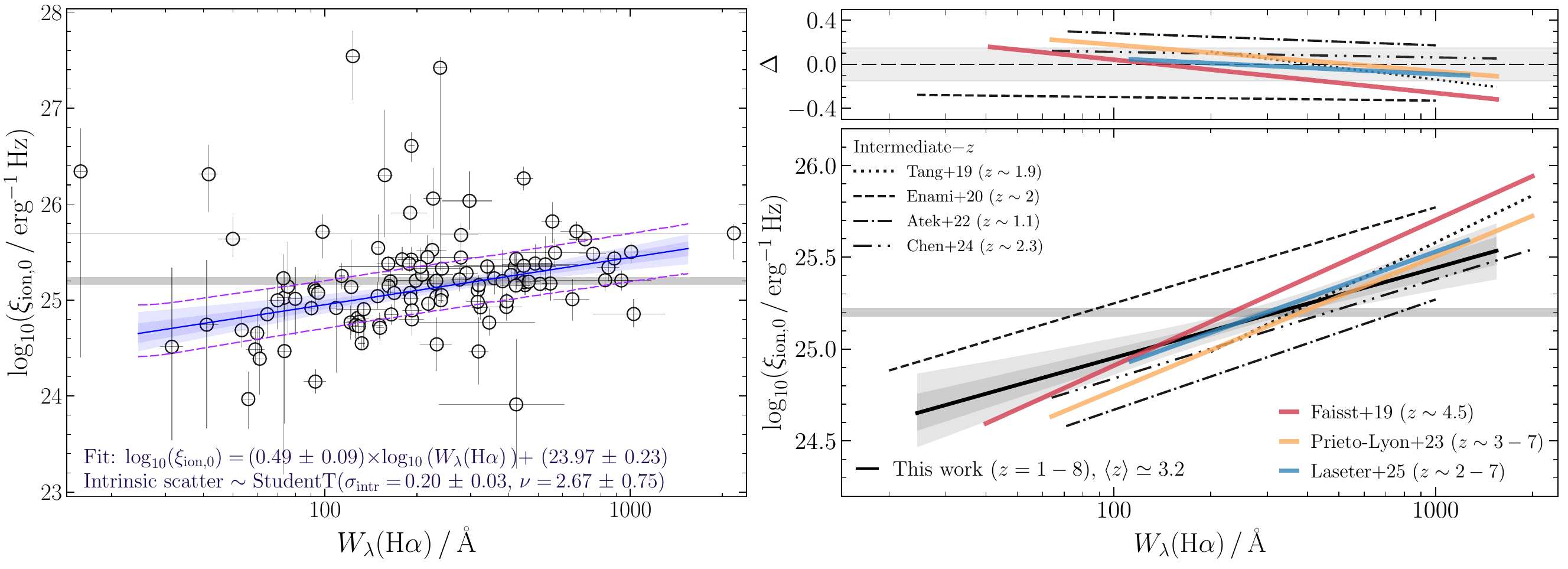}
    \caption{Relation between the the ionizing photon production efficiency $\xi_\mathrm{ion,0}$ and $\mathrm{H\,\alpha}$ equivalent width ($W_\lambda(\mathrm{H\,\alpha})$). In the \textbf{left panel} we show the median posterior relation (blue solid line, with blue shaded bands representing $\pm1\sigma$ and $±2\sigma$) from our Bayesian linear regression to the EXCELS data (black circles) described in Section \ref{subsec:analysis:model}, given by: $\log_{10}(\xi_\mathrm{ion,0}\,/\,\mathrm{Hz\,erg^{-1}})=(0.49\pm0.09)\times \mathrm{log}_{10}(W_\lambda(\mathrm{H\alpha})\,/\, \text{\AA})+(23.97\pm0.23)$. The intrinsic scatter in this relation is best described as $\epsilon\sim \mathrm{StudentT}(\sigma_\mathrm{int}=0.20\pm0.03,\, \nu=2.67\pm0.75)$, highlighting that the heavy-tailed residuals from outliers to the relation (see also Fig. \ref{fig:EWHa_regression_residuals}). The combined uncertainty from the intrinsic scatter and regression parameter posteriors are indicated by the magenta dashed lines. The \textbf{lower right panel} compares our relation (now in the background as a black line and grey shaded areas) with recent literature measurements \citep[][]{tang+19,faisst+19,emami+20,atek+22,prieto-lyon+23, chen+24,laseter+25}. In the \textbf{upper right panel}, we show the difference between the predicted $\xi_\mathrm{ion,0}$ using our inferred relation and those from the literature ($\Delta=\mathrm{log_{10}}(\xi_{\mathrm{ion,0}})_\mathrm{This\,work}-\mathrm{log_{10}}(\xi_{\mathrm{ion,0}})_\mathrm{Literature}$), as a function of $W_\lambda(\mathrm{H\alpha})$.
    } 
    \label{fig:EWHa_regression}
\end{figure*}

\begin{figure}
    \centering
    \includegraphics[width=1.05\linewidth]{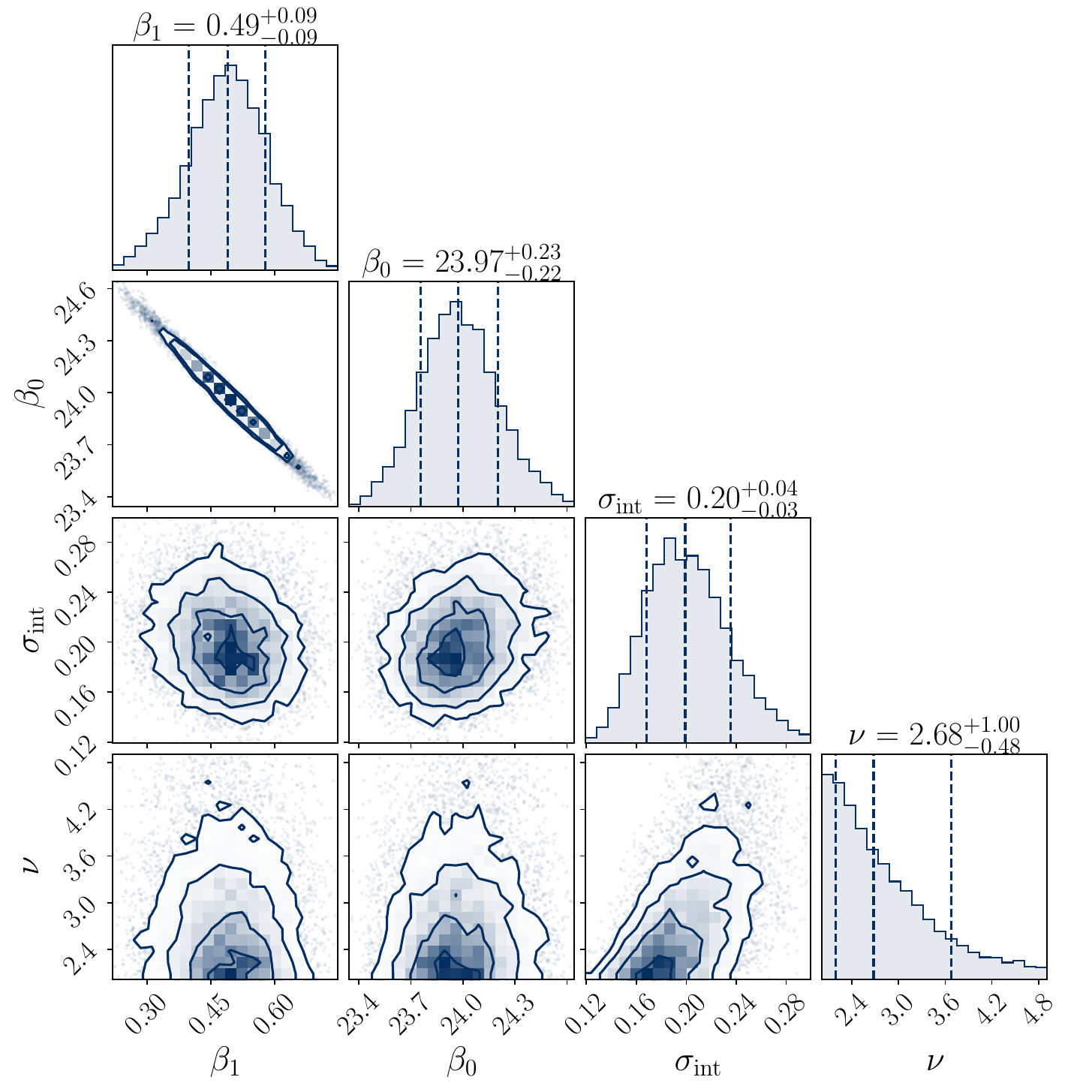}
    \includegraphics[width=1\linewidth]{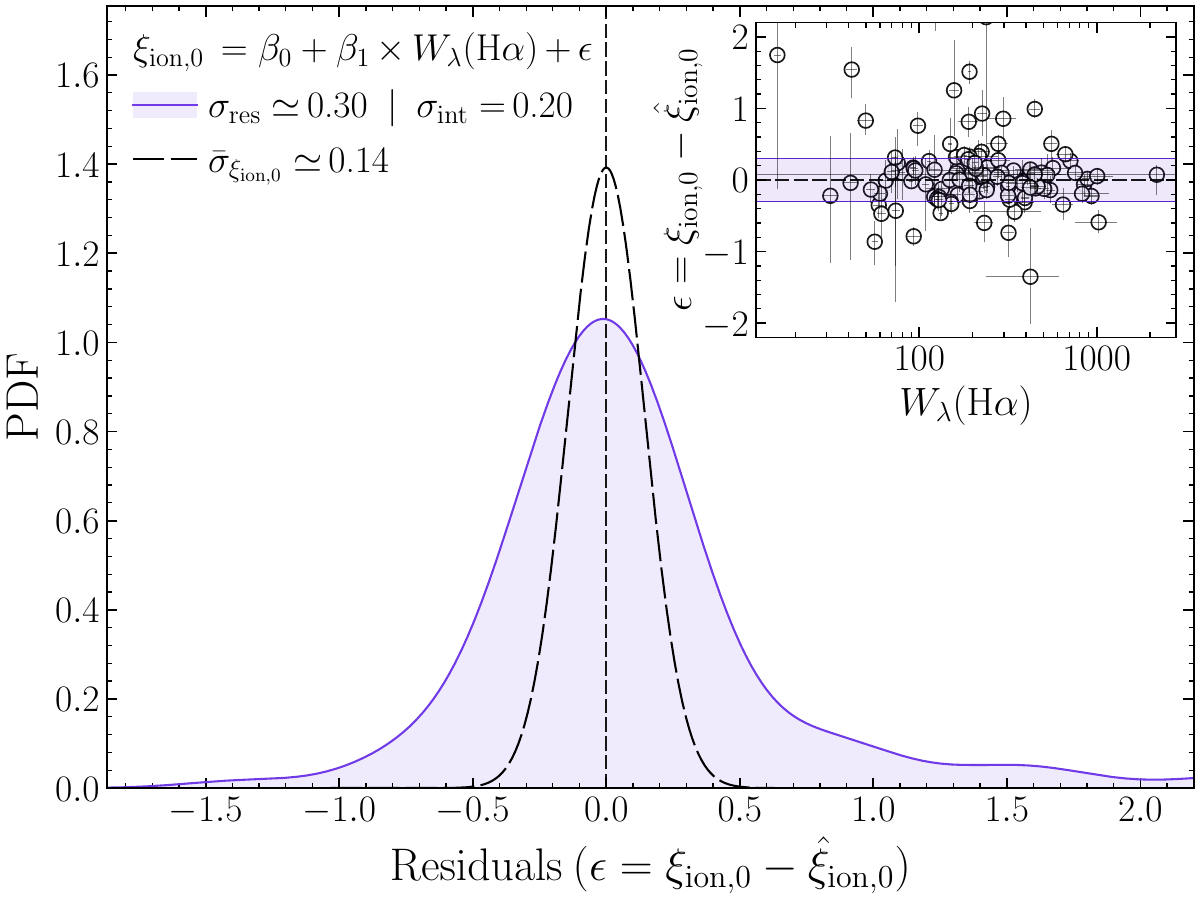}
    \caption{Supporting figures for the Bayesian linear regression of $\xi_\mathrm{ion,0}$ as a function of $W_\lambda(\mathrm{H\alpha})$ (e.g., see Section \ref{subsubsec:analysis:sv_fitting:wha} and Fig. \ref{fig:EWHa_regression_residuals}). \textbf{Top:} Corner plot showing the joint and marginal posterior probability distributions from our Bayesian fitting methodology (see Section \ref{subsec:analysis:model}) including the linear regression parameters (slope $\beta_1$, intercept $\beta_0$) and those describing the $\mathrm{StudentT}$ intrinsic scatter (and likelihood parameterisation; $\sigma_\mathrm{int}$ and $\nu$). \textbf{Bottom}: Distribution of the residuals ($\epsilon=\mathrm{log_{10}}(\xi_\mathrm{ion,0})-\mathrm{log_{10}}(\hat{\xi}_\mathrm{ion,0})$) from the $\xi_\mathrm{ion,0}-W_\lambda(\mathrm{H\,\alpha})$ linear regression. A clear non-normality is observed with the residuals having heavier tails, thus justifying the use of a $\mathrm{StudentT}$ likelihood. The residuals as a function of the independent variable (i.e., $W_\lambda(\mathrm{H\alpha})$, revealing no systematic offsets, is shown in the inset panel. 
    }
    \label{fig:EWHa_regression_residuals}
\end{figure}

Of the six predictors at the focus of this work, {\wha} has perhaps the clearest physical connection to the ionizing photon production, tracing the ratio of recent star-formation and hence current production of ionizing photons to the underlying continuum. The link is illustrated plainly in Fig. \ref{fig:xiion_vs_covariates} showing the joint distributions between {$\xi_\mathrm{ion,0}$} and one of the predictors in each column (coloured by another covariate in each row), with $\mathrm{H\,\alpha}$ equivalent width on the left side. In agreement with the previous literature \citep[e.g., see][]{faisst+19,atek+22,chen+24}, we see a clear positive correlation between $\xi_\mathrm{ion,0}$ and {\wha}, reporting a Spearman rank correlation coefficient of $\rho_\mathrm{S}=0.36$ ($p\sim10^{-4}$). This becomes more evident when mitigating the impact of outliers, as revealed by the increased ``Skipped" correlation coefficient in Table \ref{tab:correlations_table} ($\rho_\mathrm{robust}=0.54$, $p\lesssim10^{-8}$), and a similar coefficient when controlling for the remaining predictors ($\rho_\mathrm{partial}=0.32$, $p\sim10^{-3}$).\\

Applying our Bayesian fitting methodology and fitting the the logarithm of $W_\lambda(\mathrm{H\,\alpha})$ (as standard in the literature), we find the relation: $\log_{10}(\xi_\mathrm{ion,0}\,/\,\mathrm{Hz\,erg^{-1}})=(0.49\pm0.09)\times \mathrm{log}_{10}(W_\lambda(\mathrm{H\alpha})\,/\, \text{\AA})+(23.97\pm0.23)$, as shown in the left panel of Fig. \ref{fig:EWHa_regression}. The values and associated uncertainties are quoted as the median and $68$ per cent credible intervals ($16^\mathrm{th}-84^\mathrm{th}$ percentiles) from the parameter marginal posterior distributions - shown in the accompanying Fig. \ref{fig:EWHa_regression_residuals}\footnote{We note that we achieve satisfactory MCMC sampling performance in all the Bayesian linear regression results presented throughout, unless otherwise stated.}. Assessing the convergence diagnostics for the sampling, we confirm $\hat{R}\approx1.00$ (indicative of stable convergence and satisfactory mixing of the chains) for each of the parameters (in addition to large effective sample sizes, and small posterior mean Monte Carlo errors (MCSE); see also Fig. \ref{fig:appendix:bayes_diagnostics}).

As expected given the moderate positive correlation between the ionizing photon production efficiency and {\wha}, the median posterior gradient of the relation implies a $\sim0.5\,$dex increase in $\xi_\mathrm{ion,0}$ from $W_\lambda(\mathrm{H\alpha})=100-1000$\AA. The intrinsic scatter, $\sigma_\mathrm{int}=0.20\pm0.03$ (assuming a $\mathrm{StudentT}(\nu\sim2.67)$ model), is comparable to the typical measurement error of $\bar{\sigma}_{\mathrm{\xi_{ion,0}}}\sim0.15\,$dex which suggests that there is systematic source-to-source variation in $\xi_\mathrm{ion,0}$ not accounted for by {\wha}. A number of separate pieces of evidence substantiate this conclusion including; the heavier-than-normal residual tails (particularly at $\epsilon\gtrsim1.0\,$dex) seen in Fig. \ref{fig:EWHa_regression_residuals} (lower panel) from large outliers, the relatively moderate correlation coefficients shown in Table \ref{tab:correlations_table}, and the auxiliary predictor dependence seen in Fig. \ref{fig:xiion_vs_covariates}.

The right panel of Fig. \ref{fig:EWHa_regression} compares our work to other literature relations \citep[][]{tang+19,faisst+19,emami+20,atek+22,prieto-lyon+23,chen+24}. Broadly speaking, there is agreement that galaxies typically cross the canonical `ionising galaxy' threshold of $\mathrm{log_{10}}(\xi_\mathrm{ion,0}\,/\,\mathrm{erg^{-1}\,Hz})\simeq25.2$ when they display $W_\lambda(\mathrm{H\alpha})\simeq250-350\,${\AA}  (corresponding to a specific star-formation rate of $\mathrm{sSFR\simeq4-7}\,\mathrm{Gyr^{-1}}$; \citealt{marmol-queralto+16}). The relation presented here more closely aligns with the most recent literature of \citet{prieto-lyon+23} and \citet[][]{chen+24}, deviating by no more than $\sim0.1-0.15\,$dex at $W_\lambda(\mathrm{H\,\alpha})\gtrsim100\,${\AA} as shown in the top-right panel. Specifically comparing with the relation in \citet{chen+24}, we find a very close slope ($\beta_1=0.54\pm0.03$) but a slightly higher normalisation ($\Delta=+0.21\,${dex} but within $\sim1\sigma$). This difference in part may be due to different assumed dust attenuation laws \citep[e.g., see][]{reddy+16} - the relation we compare to used a \citet{calzetti+00} curve, however steeper laws (like the `intermediate' curve adopted here) would systematically increase $\xi_\mathrm{ion,0}$ by up to $0.1-0.2\,$dex. \citet{chen+24} find a relation almost identical to the one here ($\beta_1=0.51\pm0.04$, $\beta_0=23.96\pm0.11$) when using an SMC \citep[][]{gordon+03} curve (although we note that curve is steeper than ours). Other mild deviations could arise from the differences in sample selection. The sample studied in \citet{chen+24} comprises significantly of low-mass (photometrically measured) $\mathrm{H\alpha}$ emitters (HAEs) at $z\sim2.3$. A systematically lower-mass sample of HAEs may be younger or burstier, which would shift the relation to higher $W_\lambda(\mathrm{H\alpha})$ at a given $\xi_\mathrm{ion,0}$ (as seen in Fig. \ref{fig:EWHa_regression}). This effect could be exacerbated by their assumed fixed stellar-to-nebular dust reddening ratio ($\approx0.8$), which we find (calculating on a individual galaxy-by-galaxy basis) to have significant scatter (and a likely mass-dependence; see Fig. \ref{fig:appendix:stellar_to_nebular_dust}). Specifically, we find a median $\langle E(B-V)_\mathrm{neb}/E(B-V)_\mathrm{stellar}\rangle\simeq1.6$, but with a large galaxy-to-galaxy scatter (quantified by a $16^\mathrm{th}-84^\mathrm{th}$ percentile range of $\simeq0.0$ to $\gtrsim5.0$) and a heavily non-Gaussian distribution (with $\simeq25\,$per cent of the sample ($N=40$) consistent with zero nebular attenuation; i.e., $E(B-V)_\mathrm{neb}\leq0.02$).

In contrast, \citet{prieto-lyon+23} report a steeper slope ($\beta_1=0.73\pm0.04$) from a similar redshift range ($z\simeq3-7$). Several methodological differences could be responsible; they adopt an SMC-like dust curve for both nebular and stellar corrections (where a greyer law would increase the discrepancy), and their sample is substantially fainter ($\langle M_\mathrm{UV}\rangle\sim-18$, $\Delta\sim1.5\,$dex fainter). In this regime they also find higher median $\mathrm{log_{10}}(\xi_\mathrm{ion,0}\,/\,\mathrm{erg^{-1}\,Hz})\simeq25.33$ (which is $\simeq0.11\,$dex above EXCELS). Selection effects may also play a role, with their sample comprising LAEs and a high-purity photometric subsample, both of which preferentially select young, strongly star-forming systems with intense line emission and elevated $\xi_\mathrm{ion}$ \citep{erb+16,cullen+20,tang+21}, which may therefore bias the faint end toward objects with stronger lines, inflating inferred relations \citep[see][]{pahl+25a,simmonds+24b,begley+25}).

Possible variations in sample properties, as well as different methods and/or assumptions may also explain the contrast between this work and the other relations shown \citep[e.g.,][]{tang+19,faisst+19,emami+20,atek+22}. For example, \citet{atek+22} is the lowest-redshift study we compare to here. A relatively mild evolution in $\xi_\mathrm{ion,0}$ with redshift \citep[e.g., as seen in][]{pahl+25a,simmonds+24b,begley+25,llerena+25}, combined with a comparatively stronger redshift dependence with $W_\lambda(\mathrm{H\alpha})$ \citep{fumagalli+12,marmol-queralto+16,reddy+18b,khostovan+24}, could shift the relation left (i.e., higher $\xi_\mathrm{ion,0}$ at fixed $W_\lambda(\mathrm{H\alpha})$). Looking at the relation in \citet{tang+19}, the authors find a steeper trend ($\beta_1=0.86\pm0.06$). However, a direct comparison is difficult as their sample is very low mass, low-redshift, and generally focuses on emission line galaxies (specifically, {\EWoiiihb$\gtrsim200\,$\AA}).\\

Given the shallower relation obtained here relative to the other literature shown in Fig. \ref{fig:EWHa_regression}, we explore some variations in our fiducial fitting process.
Firstly, we test the sensitivity of our regression to the functional form of the likelihood model. As discussed above, the heavy-tailed residuals statistically prefer a $\mathrm{StudentT}$ scatter (with $\nu\sim2.67$), nonetheless repeating the fitting with a more typical Normal likelihood we find an even flatter slope ($\beta_1\simeq0.41$). This fit also implies a factor $\sim2\times$ larger intrinsic scatter, $\sigma_\mathrm{int}\simeq0.36$, but fails to accurately represent the observed heavy-tailed scatter. 
We also test an alternative fitting methodology using the \textsc{SciPy} ODR package, which is commonly adopted in the literature. This returns a slightly steeper (increasing by $\sim+1\sigma$) slope of $\beta_1\simeq0.6$, which is closer to some of the recent literature ($\beta_1\simeq0.6-0.85$). However this fitting method requires collapsing the posterior distributions of $\xi_\mathrm{ion,0}$ for each object to $y\pm\sigma_y$ assuming Gaussian uncertainties, which for some objects is an improper assumption and disregards the full information available (see Fig. \ref{fig:xiion_per_galaxy_distributions}).

Complementary to the aforementioned tests of simpler fitting methods, we also explore further modelling components. One such extension includes $\gamma$ parameterising the predictor dependence of the intrinsic scatter (known as model heteroscedasticity, i.e.,  given as $\sigma_\mathrm{int}(x)=\sigma_\mathrm{int,0}\cdot(x/x_0)^\gamma$). Here we find $\gamma\simeq-0.3\pm0.5$ (at $\lesssim1\sigma$ significance, with the remaining parameters consistent at the $\leq1\sigma$ level), implying more population scatter at lower $\mathrm{H\alpha}$ equivalent widths. Robust inferences of any possible evolution in $\sigma_\mathrm{int}$ with $W_\lambda(\mathrm{H\alpha})$ (or any other predictor) will require much larger sample sizes however.

As an alternative to the $\mathrm{StudentT}$ model, we also explore a Gaussian mixture model as another credible model configuration. Specifically, we parameterise the likelihood as two Normal distributions describing two distinct populations: $p(y\mid\theta,...)\sim(1-a)\cdot\mathcal{N}(\mu(\theta),\sigma^2_\mathrm{int,0})+a\cdot\mathcal{N}(\mu(\theta),\sigma^2_\mathrm{int,1})$, where where $a$ is the mixing fraction and $\sigma_\mathrm{int,0}>\sigma_\mathrm{int,1}$. Median values (from the sampled posterior distributions) indicate $\sigma_\mathrm{int,0}=0.06\pm0.05$, $\sigma_\mathrm{int,1}=0.31\pm0.05$, and $a=0.05\pm0.03$. We compare the performance of the mixture and $\mathrm{StudentT}$ models using leave-one-out cross validation (LOO), finding neither are statistically preferred\footnote{Here, we compare models using the \textit{expected log pointwise predictive density} ($\mathrm{elpd}$) calculated with the PSIS-LOO algorithm \citep[Pareto-smoothed importance sampling;][implemented within ArViz; \citealt{arviz}]{vehtari+15}. In this scheme, one model is considered moderately \textit{better} than another if $\Delta\mathrm{elpd}\gtrsim2\cdot\mathrm{SE_{\Delta}}$, where $\mathrm{SE}_\Delta$ is the standard error on the $\mathrm{elpd}$ estimate.}. As a result, the $\mathrm{StudentT}$ model is our fiducial setup given it is more interpretable and has lower complexity (with less free parameters).

\subsubsection{Relation between $\xi_\mathrm{ion,0}$ and $W_{\lambda}$([\mbox{O\,\sc{iii}}])}\label{subsubsec:analysis:sv_fitting:woiii}

\begin{figure}
    \centering
    \includegraphics[trim={0cm 0.35cm 0cm 2.4cm},clip,width=1.\columnwidth]{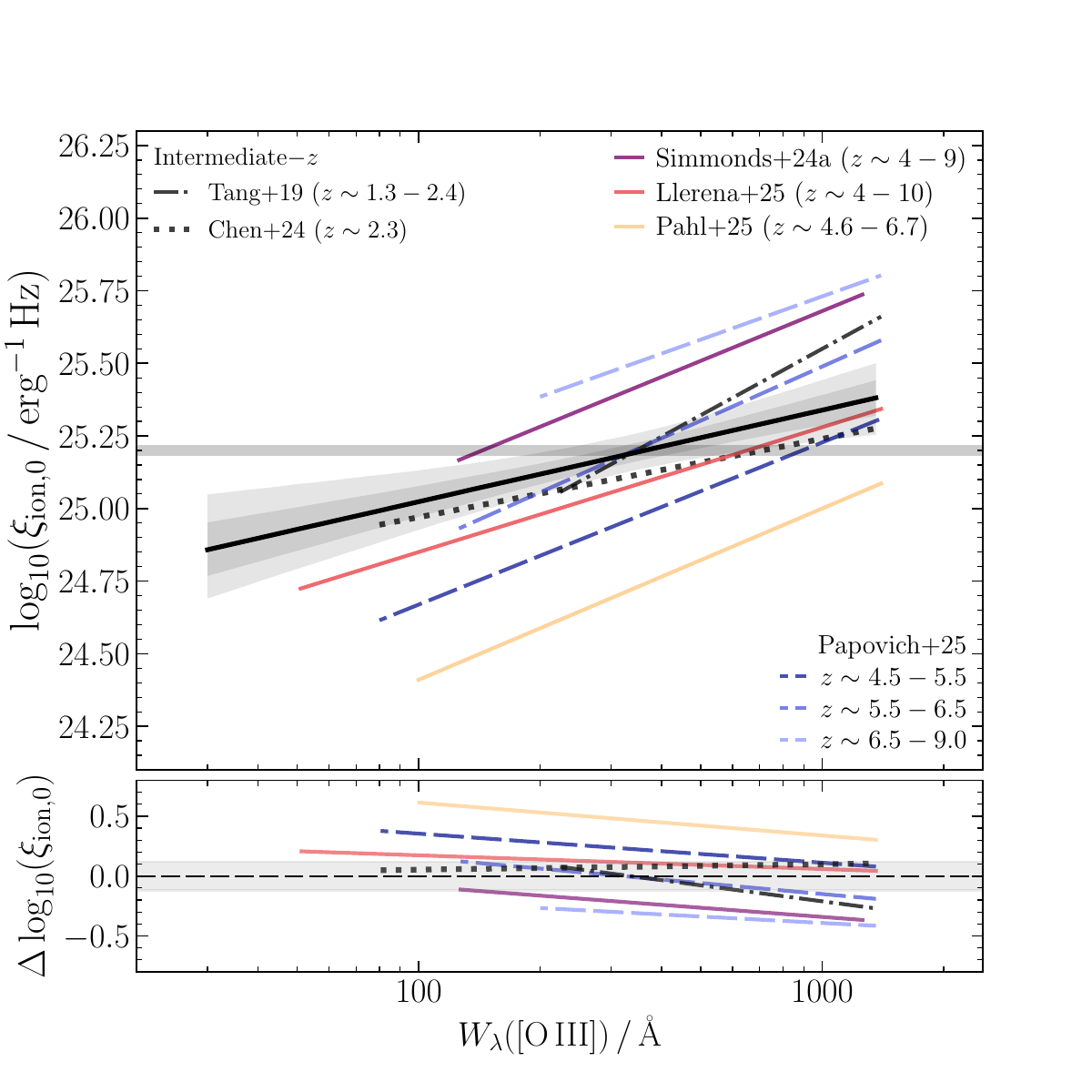}
    \caption{Relation between $\xi_\mathrm{ion,0}$ and $W_{\lambda}(\mathrm{[OIII]})$ (in the same format as the right panel of Fig. \ref{fig:betaUV_regression}), given as: $\log_{10}(\xi_\mathrm{ion,0}\,/\,\mathrm{Hz\,erg^{-1}})=(0.30\pm0.08)\times W_{\lambda}([\mathrm{O\,III}])+(24.44\pm0.21)$, with intrinsic scatter $\mathrm{StudentT}(\sigma_\mathrm{int}=0.20\pm0.03,\: \nu=2.56\pm0.63)$ (shown in the \textbf{top panel} as black line with the $\pm1\sigma,\,\pm2\sigma$ regions as grey shaded and the $\sigma_\mathrm{int}$ denoted as thin dashed lines). We compare to literature results at cosmic noon \citep[e.g., with samples of EELGS and HAEs / non-HAEs][as black dash-dotted and dotted lines, respectively]{tang+19,chen+24}, as well as those spanning a wider high-redshift regime \citep[$z\sim4-10$][; see legend]{simmonds+24b,pahl+25a,llerena+25,papovich+25}. In the \textbf{lower} panel, we show the difference in the predicted ionizing photon production efficiency as function of {$W_\lambda(\mathrm{[O\,III}])$} ($\Delta\mathrm{log_{10}}(\xi_\mathrm{ion,0})$).
    }
    \label{fig:EWOIII_regression}
\end{figure}

\begin{figure*}
    \centering
    \includegraphics[width=1.025\columnwidth]{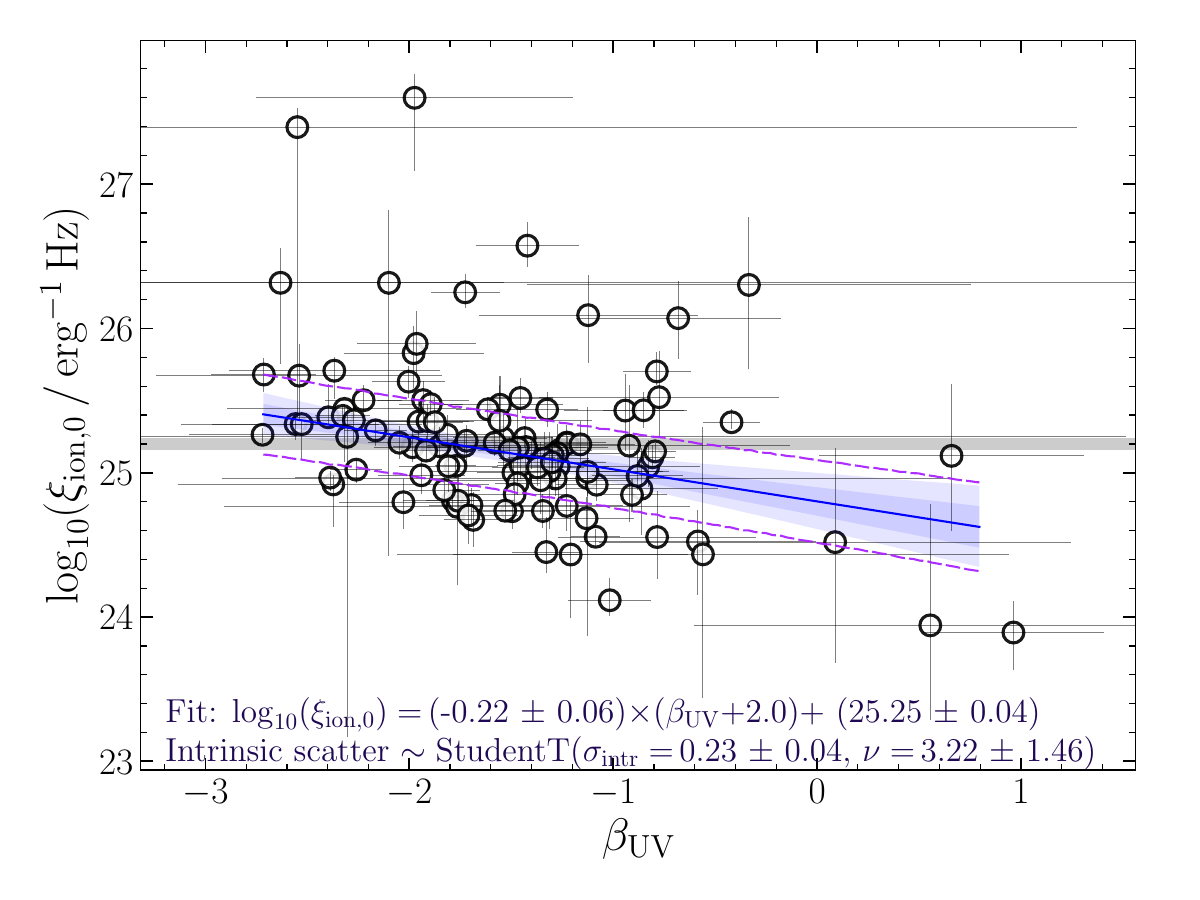}
    \includegraphics[width=1.025\columnwidth]{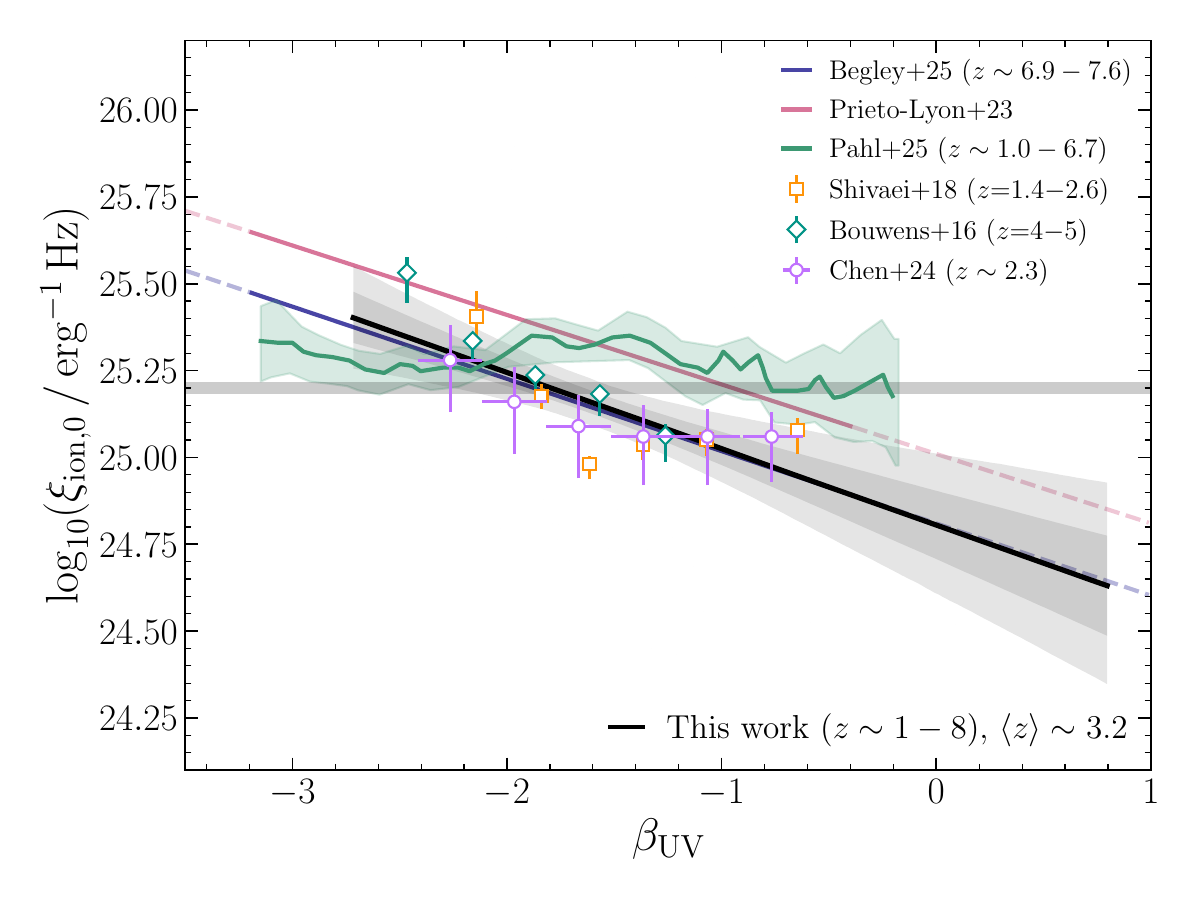}
    \caption{Inferred $\xi_\mathrm{ion,0}-\beta_\mathrm{UV}$ relation, shown in the same two-panel layout (and formatting) as Fig. \ref{fig:EWHa_regression}. On the \textbf{left panel}, we show the median posterior relation constrained by the data: $\log_{10}(\xi_\mathrm{ion,0}\,/\,\mathrm{Hz\,erg^{-1}})=(-0.22\pm0.06)\times \beta_\mathrm{UV}+(23.97\pm0.23)$, with intrinsic scatter $\mathrm{StudentT}(\sigma_\mathrm{int}=0.23\pm0.04,\: \nu=3.22\pm1.46)$. We compare to a number of different literature constraints in the \textbf{right panel}, including relations from \citet{prieto-lyon+23} and \citet{begley+25}; results from stacking \citep[][]{bouwens+16,shivaei+18,chen+24}; and the running mean across the spectroscopic sample analysed in \citet{pahl+25a}.
    }
    \label{fig:betaUV_regression}
\end{figure*}

Similar to $\mathrm{H\alpha}$, the \mbox{[O\,\textsc{iii}]}$\lambda5007$ emission line is directly produced as a result of young, massive stars powering their surrounding ionized nebulae. This is clearly visualised in Fig. \ref{fig:xiion_vs_covariates}, and backed-up quantitatively by the correlation coefficient with $\xi_\mathrm{ion,0}$ ($\rho_\mathrm{robust}=0.54$, $p<10^{-7}$). However, {$W_\lambda$(\mbox{[O\,\textsc{iii}]})} is also secondarily sensitive to the ionization parameter, spectral hardness and metallicity, all of which may enhance or suppress {$W_\lambda$(\mbox{[O\,\textsc{iii}]})} at fixed $\xi_\mathrm{ion}$ \citep[e.g., see][]{reddy+18b,laseter+25}. These confounding factors may explain why the partial correlation is much lower ($\rho_\mathrm{partial}=0.09$, $p=0.39$) than for $W_\lambda(\mathrm{H\alpha})$. Even when $\mathrm{H\alpha}$ and \mbox{[O\,\textsc{iii}]} strongly correlate ($\rho_\mathrm{S}=0.93$, $p<10^{-6}$) and thus provide overlapping information on $\xi_\mathrm{ion,0}$, $\mathrm{H\alpha}$ is less impacted by secondary factors and therefore is the more dominant predictor. 

Nonetheless, {\mbox{[O\,\textsc{iii}]}} remains an important avenue from which to empirically infer {\xiion} \citep[e.g., see][]{tang+19,begley+25}, and applying the Bayesian regression methodology we determine: $\mathrm{log_{10}}(\xi_\mathrm{ion,0}\,/\,\mathrm{erg^{-1}\,Hz})=(0.30\pm0.08)\times\mathrm{log_{10}(W_\lambda([\mathrm{O\,III}]\lambda5007))} + (24.44\pm0.21)$. We show this relation in comparison to recent literature in Fig. \ref{fig:EWOIII_regression}.  
Our relation suggests only a $\Delta\sim0.4\,$dex evolution from $W_\lambda([\mathrm{O\,\tiny{III}}])=100\,${\AA} to $W_\lambda([\mathrm{O\,\tiny{III}}])=1000\,${\AA}, with scatter at the $\sigma_\mathrm{int}\simeq0.2\,$dex level (and $\nu\sim2.56$ in $t_\nu$). We find the closest agreement with \citep[][]{chen+24}, differing by $\Delta\lesssim0.05-0.1\,$dex across the full dynamic range in our sample. This dependence is well matched by simulations such as FLARES, as shown in \citet{seeyave+23}. The relation presented here and in \citet{chen+24} \citep[see also][not shown, with similar slopes, but systematically offset as in Section \ref{subsubsec:analysis:sv_fitting:wha}]{emami+20,atek+22} is shallower than other results shown in Fig. \ref{fig:EWOIII_regression} ($\beta_1=0.30\pm0.08$ compared to $\beta_1\sim0.4-0.8$). This difference can likely be ascribed to dissimilarities in the samples used. Firstly, we highlight that the majority of the literature does not probe the low$-W_\lambda([\mathrm{O\,\tiny{III}}])$ ($\lesssim100$\AA) regime, whereas here this constitutes $\sim27\,$per cent of the sample subset with {\mbox{[O\,\textsc{iii}]}} measurements. Secondly, an underlying redshift dependence could be altering the inferred relations. In the higher redshift samples of \citet{simmonds+24,pahl+25a,llerena+25}, lower metallicities or harder ionizing spectra may produce a steeper $\xi_\mathrm{ion,0}-W_\lambda([\mathrm{O\,\tiny{III}}])$ dependence \citep[][and see Section \ref{subsubsec:redshift_dependence_evolution}]{nakajima+16,hayes+25}.

It is noteworthy that sample \textit{selection} is unlikely to be a significant contributing factor here. The EXCELS parent sample is selected mostly from large multi-band photometric catalogues (i.e., not emission line selected, and tracing the star-forming main sequence) while the spectroscopy has sufficient depth to provide sensitivity down to very faint flux limits ($f_{\mathrm{line}}\sim10^{-18}\,/\,\mathrm{erg\,s^{-1}\,cm^{-2}}$). Nonetheless, on the basis of our requirement to detect a minimum of two Balmer emission lines at $\geq3\sigma$, beyond a certain UV luminosity limit only the strongest emission lines will satisfy this criteria.

\subsubsection{Relation between $\xi_\mathrm{ion,0}$ and $\beta_{\mathrm{UV}}$}\label{subsubsec:analysis:sv_fitting:betauv}

Based on Fig. \ref{fig:xiion_vs_covariates} and the correlation coefficients in Table \ref{tab:correlations_table} (i.e., $\rho_\mathrm{partial}\simeq-0.7$, $p\ll10^{-6}$), the UV continuum slope has one of the strongest connections with $\xi_\mathrm{ion,0}$. Although strongly sensitive to the degree of dust attenuation, $\beta_\mathrm{UV}$ is likewise indicative of the age and metallicity of the stellar populations present; with bluer galaxies expected to be younger, have lower-metallicities and present harder ionizing spectra - all properties conducive to elevated $\xi_\mathrm{ion,0}$ \citep[][]{bouwens+14,bouwens+16,topping+24}.

From our regression, we infer the relation: $\log_{10}(\xi_\mathrm{ion,0}\,/\,\mathrm{Hz\,erg^{-1}})=(-0.22\pm0.06)\times \beta_\mathrm{UV}+(23.97\pm0.23)$, shown in Fig. \ref{fig:betaUV_regression}. Broadly speaking, galaxies in the general population with UV slopes bluer than $\beta_\mathrm{UV}\lesssim-1.6$ have ionizing photon production efficiencies around or above $\mathrm{log_{10}}(\xi_\mathrm{ion,0}\,/\,\mathrm{erg^{-1}\,Hz})\simeq25.2$, in approximate agreement with stacking-based analyses in the literature \citep[e.g.,][as shown in the right panel of Fig. \ref{fig:betaUV_regression}]{bouwens+16,chen+24}. 

This relation agrees remarkably well with the photometric analysis of star-forming galaxies at $z\simeq6.9-7.6$, presented in our previous work \citep[][]{begley+25}, supporting a redshift-invariant $\xi_\mathrm{ion,0}-\beta_\mathrm{UV}$ relation. This is at odds with the lack of any correlation recovered by \citet{pahl+25a}. Interestingly, we also find excellent agreement between the gradient ($\beta_1=\partial\xi_\mathrm{ion,0}/\partial\beta_\mathrm{UV}$) of the relation here and that presented in \citet{prieto-lyon+23}; but with a $\simeq0.2\,$dex offset in $\xi_\mathrm{ion,0}$ normalisation. We remind the reader of the $\simeq0.1\,$dex higher median $\xi_\mathrm{ion,0}$ (and $\simeq1.5\,$ mag fainter) sample in \citet{prieto-lyon+23}, and suggest the arguments made in Section \ref{subsubsec:analysis:sv_fitting:wha} could also explain this offset seen in $\xi_\mathrm{ion,0}-\beta_\mathrm{UV}$.

As with the previous predictors, we find a moderate and heavy tailed scatter given as $\sigma_\mathrm{int}=0.23\pm0.04$ and $\mathrm{StudentT}(\nu\simeq3.22)$. A key observation that we highlight here is that the outliers are discernible from the typical star-forming galaxy population that makes up the majority of the EXCELS sample - namely they show high $E(B-V)_\mathrm{neb}$ (e.g., see Fig. \ref{fig:xiion_vs_covariates}).
High $\xi_\mathrm{ion,0}$ in combination with high $E(B-V)_\mathrm{neb}$, which describes a large fraction of the heavy-tailed outliers, may be indicative of sources that contain young obscured star-forming regions. However, it is also probable that these types of galaxies do not yet contribute significantly to the ionizing photon budget, as the high $E(B-V)_\mathrm{neb}$ would indicate very little of the ionizing photons produced could escape due to the dust attenuation. Another important consideration is that we observe a moderate correlation between $E(B-V)_\mathrm{neb}$ and $\beta_\mathrm{UV}$ ($\rho_\mathrm{S}\simeq0.39$, $p\lesssim10^{-4}$), but again these sources are typically outliers in the $E(B-V)_\mathrm{neb}-\beta_\mathrm{UV}$ as well. This aligns with the large scatter seen between stellar and nebular attenuation, particularly at high redshifts \citep[][see also Fig. \ref{fig:appendix:stellar_to_nebular_dust}]{reddy+18b}, and that these galaxies lie in the extreme high ${E(B-V)_\mathrm{neb}}/{E(B-V)_\mathrm{stellar}}$ tail. Complicating the interpretation of these objects, is the underlying assumption of a foreground dust geometry when using the Balmer decrement (and Case B) to obtain the intrinsic Balmer line luminosity. It may be the case that a more mixed, clumpy dust geometry is more appropriate \citep[producing greyer attenuation curves; e.g., see][for a more in-depth discussion]{vijayan+24}, in which case the Balmer decrement method leads to over-estimated $\xi_\mathrm{ion,0}$.

\subsubsection{Relation between $\xi_\mathrm{ion,0}$ and $M_\mathrm{UV}$}\label{subsubsec:analysis:sv_fitting:muv}

\begin{figure}
    \centering
    \includegraphics[trim={0cm 0.7cm 0cm 0cm},clip,width=1.\columnwidth]{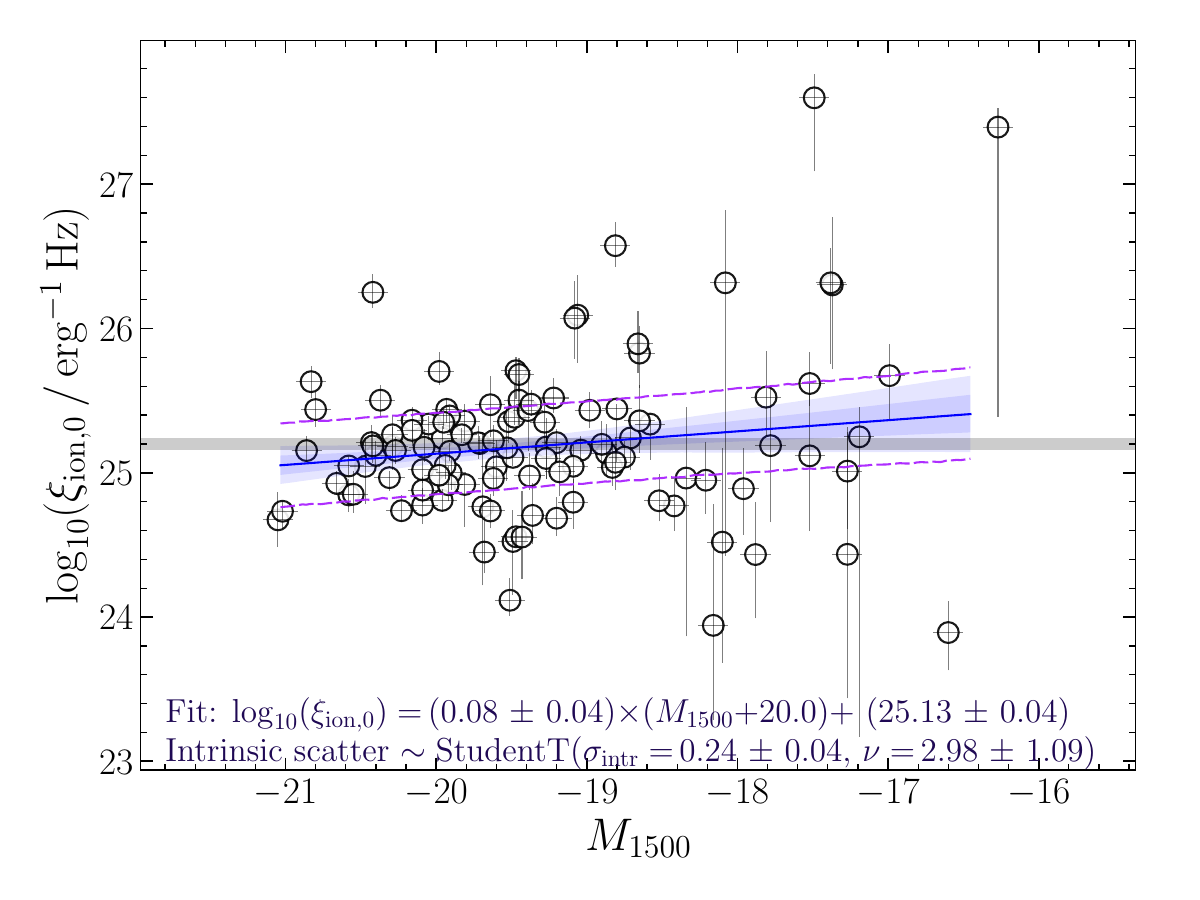}
    \includegraphics[trim={0cm 0.7cm 0cm 0.7cm},clip,width=1.\columnwidth]{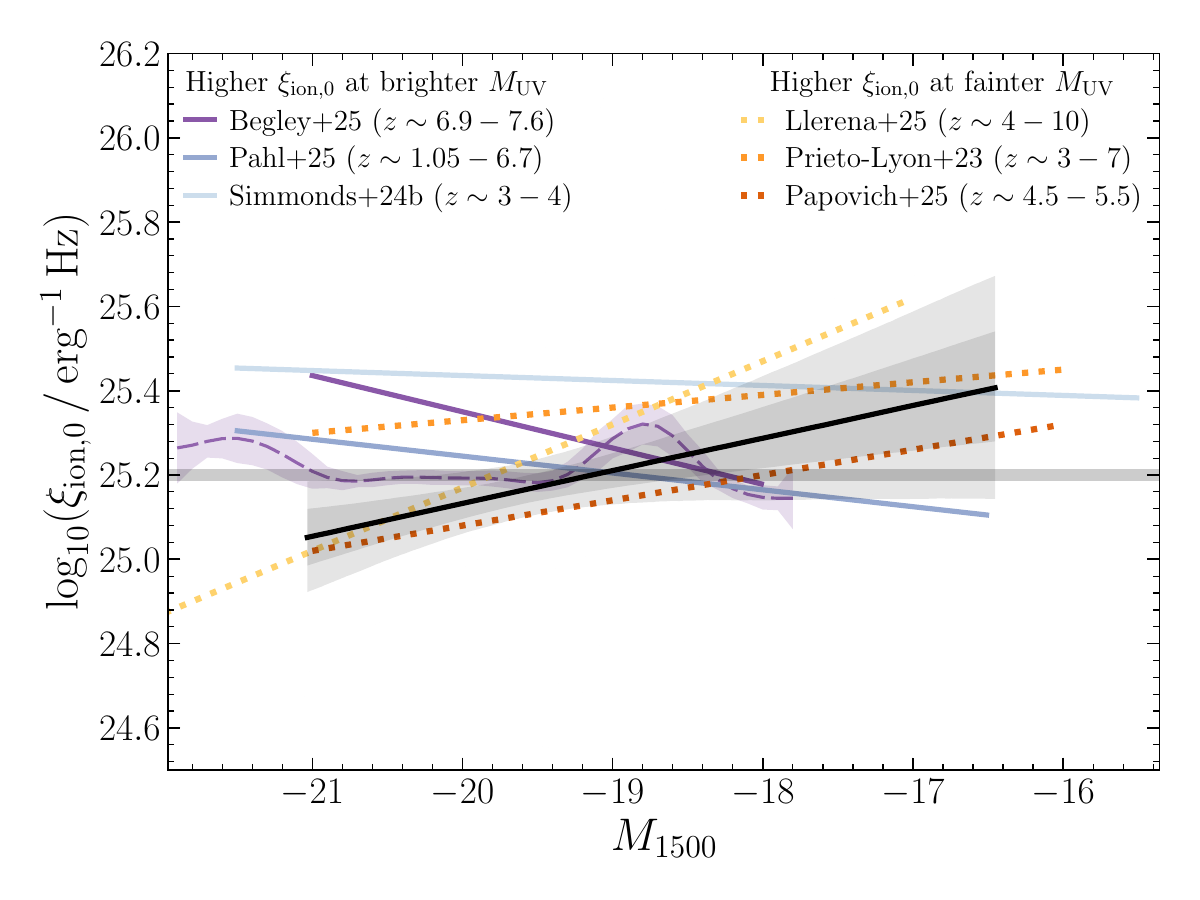}
    \includegraphics[trim={0cm 0.7cm 0cm 0.7cm},clip,width=1.\columnwidth]{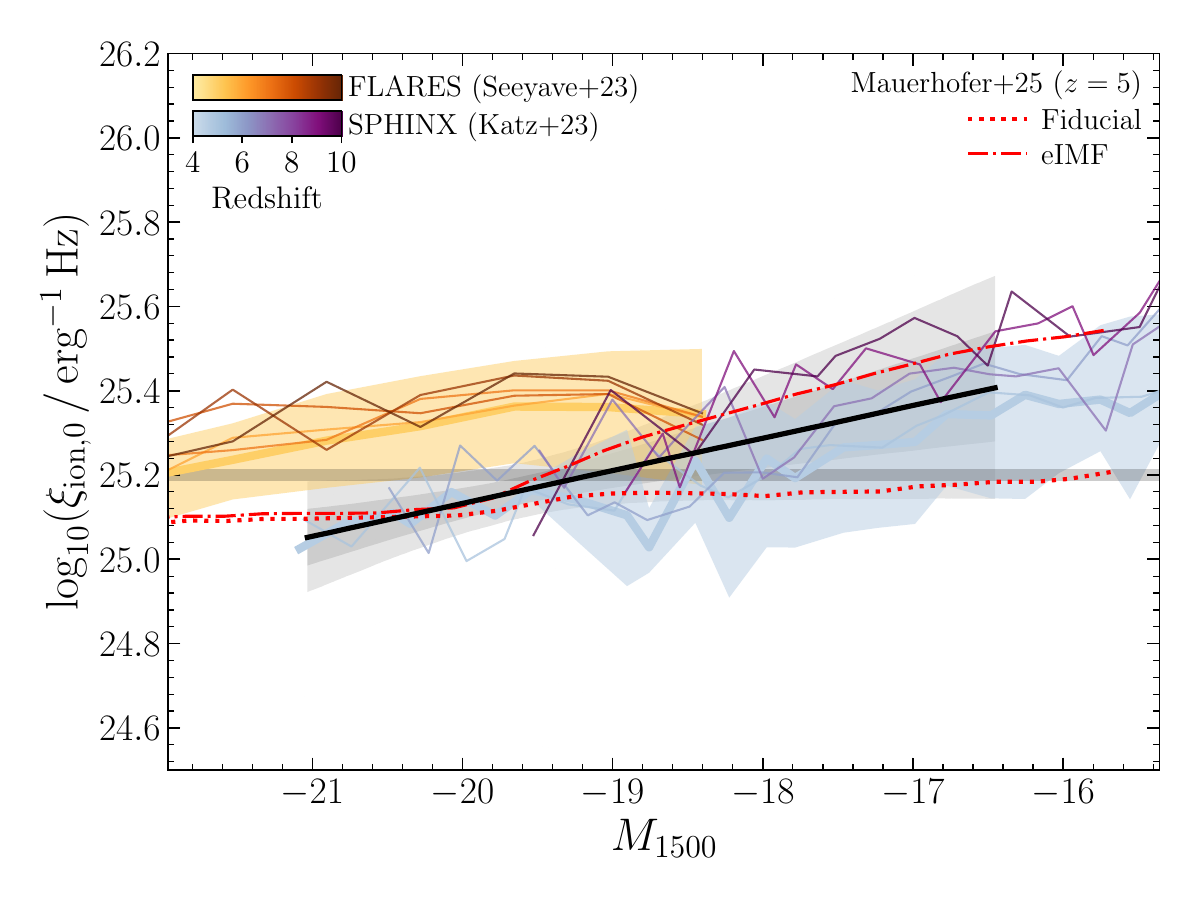}
    \caption{Results for the relation between $\xi_\mathrm{ion,0}$ and the absolute UV luminosity ($M_\mathrm{UV}$), discussed in Section \ref{subsubsec:analysis:sv_fitting:muv}. The \textbf{top} panel shows the $\log_{10}(\xi_\mathrm{ion,0}\,/\,\mathrm{Hz\,erg^{-1}})=(0.08\pm0.04)\times (M_\mathrm{UV}+20)+(25.13\pm0.04)$ relation from the Bayesian linear regression (median posterior as a blue line with $\pm1\sigma,\,\pm2\sigma$ shaded regions), with the intrinsic scatter indicated by the purple dashed lines; $\epsilon\sim \mathrm{StudentT}(\sigma_\mathrm{int}=0.24\pm0.04,\nu=2.98\pm1.09)$. The \textbf{centre panel} compares our work to observational analyses, indicating when $M_\mathrm{UV}-$faint galaxies are predicted to have elevated $\xi_\mathrm{ion,0}$ \citep[][in yellow-to-red colour dashed lines]{prieto-lyon+23,llerena+25,papovich+25} and when $M_\mathrm{UV}-$bright galaxies have higher $\xi_\mathrm{ion,0}$ \citep[][in blue-to-purple colour solid lines; also including the running median of the full $z\sim3-8$ VANDELS$+$JWST sample studied in \citealt{begley+25}]{simmonds+24b,begley+25,pahl+25a}. In parallel, we show results from leading high-resolution hydrodynamical simulations in the \textbf{bottom panel}, as a function of redshift \citep[][for FLARES and SPHINX, respectively]{seeyave+23,katz+23}, and for the latest iterations of the \textsc{Delphi} semi-analytic model \citep[e.g., see][for details]{mauerhofer+25}.
    }
    \label{fig:MUV_regression}
\end{figure}

The rest-frame UV luminosity dependence of the ionizing photon production is vital for accurate empirical modelling of the reionization epoch \citep[e.g., see][]{donnan+25}. Here we establish a $\simeq2\sigma$ relation given by: $\mathrm{log_{10}}(\xi_\mathrm{ion,0}\,/\,\mathrm{erg^{-1}\,Hz})=(0.08\pm0.04)\cdot (M_\mathrm{UV}+20) + (25.13\pm0.04)$, implying $\Delta\sim0.3\,$dex evolution in $\xi_\mathrm{ion,0}$ across a $\sim4\,$mag dynamic range ($M_\mathrm{UV}\sim-21$ to $M_\mathrm{UV}\sim-17$) (with $\epsilon\sim\mathrm{StudentT}(\sigma_\mathrm{int}=0.24\pm0.04,\nu=2.98\pm1.09)$). We show this, alongside comparisons to recent observational results in the top two panels of Fig. \ref{fig:MUV_regression}. What is fully transparent considering the landscape of recent literature on the $\xi_\mathrm{ion,0}-M_{\mathrm{UV}}$ relation (i.e., middle panel of Fig. \ref{fig:MUV_regression}), is the lack of consensus. A number of studies, like the result here, find that elevated $\xi_\mathrm{ion,0}$ are found in more $M_\mathrm{UV}-$ faint populations \citep[][]{prieto-lyon+23,llerena+25,papovich+25}. This trend would not be surprising as $M_\mathrm{UV}$-faint galaxies will typically have younger ages, lower metallicities, and lower dust content \citep[][and Stanton et al. 2025b in prep.]{reddy+18,cullen+19,cullen+20,curti+24,jain+25}, all of which are expected to drive higher $\xi_\mathrm{ion,0}$ \citep{reddy+18b,shivaei+18}. Nevertheless, we highlight that the $\xi_\mathrm{ion,0}-M_{\mathrm{UV}}$ relations are mostly still very shallow and only marginally significant (e.g., \citealt{prieto-lyon+23} with $\beta_1\simeq0.03\pm0.02$; \citealt{papovich+25} with $\beta_1\simeq0.06\pm0.03$ at $4.5\leq z\leq5.5$). This picture is made further ambiguous by the remaining confounding variables that are not taken into account when considering only $M_\mathrm{UV}$.

Caution should inherently be taken when considering inferred trends in $\xi_\mathrm{ion,0}$ with $M_\mathrm{UV}$, as there is mounting evidence that sample selection can have a nuanced but critical impact. \citet{begley+25} from a photometrically-selected ($\gtrsim90\,$per cent complete at $M_\mathrm{UV}\leq-18$) sample at $z\sim6.9-7.6$, find that brighter $M_\mathrm{UV}$ SFGs have more intense $\xi_\mathrm{ion,0}$ contrary to the literature previously discussed. Recent spectroscopic work by \citet{pahl+25a}, and a carefully selected mass-complete sample from \citet{simmonds+24b} \citep[comparing with][a clear example of the importance of the impact of completeness]{simmonds+24}, also mirror this result.
This trend can be understood from the perspective of star-formation burstiness, which increases both towards higher redshifts and lower stellar masses \citep[][]{atek+22,endsley+22,endsley+24,begley+25}. Specifically, in this scenario $M_\mathrm{UV}-$bright galaxies have undergone a recent strong up-turn in star-formation providing a boost to $\xi_\mathrm{ion,0}$ \citep[][]{endsley+25}. On the other hand, $M_\mathrm{UV}-$faint sources are in a current `lulling' phase \citep[e.g., see][but also \citealt{carvajal+25,cole+25,simmonds+25,trussler+25} for recent literature discussions on bursty star-formation]{looser+23}.
Crucially however, like in the case for relations where the fainter galaxies are seen to have higher $\xi_\mathrm{ion,0}$, the $\xi_\mathrm{ion,0}-M_\mathrm{UV}$ trends in \citet{simmonds+24b,begley+25} and \citet{pahl+25a} are mild, close to flat, and relatively insignificant ($\lesssim2\sigma$).
Here we note that when only restricting our sample to galaxies with $M_\mathrm{UV}\leq-19$, effectively making our sample more comparable to the literature discussed here (broadly removing the extreme low-redshift tail at $z<2$, and where the $M_\mathrm{UV}{}$ distribution begins to fall off towards fainter$-M_\mathrm{UV}$), the relation is fully consistent with flat at the $\lesssim1\sigma$ level ($\beta_1\simeq0.06^{+0.07}_{-0.08}$).

It is also possible that the importance (or lackthereof) of UV luminosity varies with redshift as well. The running median on the $\xi_\mathrm{ion,0}-M_\mathrm{UV}$ plane from the combined (VANDELS $+$ JWST) sample in \citet{begley+25} is entirely consistent with being flat (as shown in Fig. \ref{fig:MUV_regression}). Moreover, current evidence suggests the $\xi_\mathrm{ion,0}-M_\mathrm{UV}$ relation slope varies only mildly, if at all, with redshift \citep[][]{pahl+25a,simmonds+24b,papovich+25}.

The lower panel of Fig. \ref{fig:MUV_regression} considers results from some of the latest cosmological hydrodynamical simulations; SPHINX \citep[e.g., see][]{rosdahl+18,katz+23} and FLARES \citep[e.g., see][]{lovell+21,seeyave+23}. These two simulation suites together show an extremely mild-to-flat connection between UV luminosity and the ionizing photon production efficiency. There is also a minor modulation based on redshift, but this is due to the underlying physical property evolution. Fig. \ref{fig:MUV_regression} also includes the latest results from the \textsc{Delphi} semi-analytic model 
\citep[][]{mauerhofer+25} at $z\sim5$, including their `fiducial' model (incorporating ISM physics derived from the SPHINX high-resolution radiative-hydrodynamical simulations) and their `eIMF' model (folding in a redshift and metallicity dependent IMF). The fiducial model shown, which reliably reproduces observations of UV continuum slopes and Balmer jump strengths down to the end of the EOR, predicts an incredibly flat dependence of $\xi_\mathrm{ion,0}$ on $M_\mathrm{UV}$. In mild contrast, the \textsc{Delphi} `eIMF' model appears to agree well with the results found here, which lends support to the possibility that any $M_\mathrm{UV}$ dependence is driven by an underlying IMF evolution (although we highlight as discussed previously, our relation is not statistically significant and may be impacted by sample bias in the $M_\mathrm{UV}-$faint regime). An important aspect to consider in relation to the simulations is that they slightly under-represent the diversity seen in $\xi_\mathrm{ion,0}$ across observed populations (here; $\sim0.4\,$dex and spanning $\mathrm{log_{10}}(\xi_\mathrm{ion,0}\,/\,\mathrm{erg^{-1}\,Hz})\simeq24-27$, but see also \citealt[][]{prieto-lyon+23,simmonds+24b,begley+25,pahl+25a,laseter+25}), particularly at the low- and high-$\xi_\mathrm{ion,0}$ tails.
In contrast, for example, the SPHINX simulation has $\mathrm{log_{10}}(\xi_\mathrm{ion,0}\,/\,\mathrm{erg^{-1}\,Hz})\simeq24.8-25.8$, with $\sigma_\mathrm{sMAD}\simeq0.2\,$dex. This can be explained by the adopted SED assumptions \citep[e.g., see][their Section 3.1.12]{katz+23}, with possible impact from physics below the resolution scale of the simulations. Moreover, these values are intrinsic $\xi_\mathrm{ion,0}$ with sightline dependence and considerations of the complex stochastic SFHs not accounted for, as well as potential observational effects.

\subsubsection{Relation between $\xi_\mathrm{ion,0}$ and $E(B-V)_\mathrm{neb}$}\label{subsubsec:analysis:sv_fitting:ebv}

From Fig. \ref{fig:xiion_vs_covariates}, a clear linear and monotonic $\xi_\mathrm{ion,0}-E(B-V)_\mathrm{neb}$ relation is not expected, given the relatively shallow trend at low $E(B-V)_\mathrm{neb}$ ($<0.5$) followed by a tail of both high $\xi_\mathrm{ion,0}$ and high $E(B-V)_\mathrm{neb}$. Quantitively, things are no more transparent given the moderate Spearman rank correlation coefficient of $\rho_\mathrm{S}\simeq0.38$ ($p<10^{-4}$; and $\rho_\mathrm{partial}\simeq0.78$), that significantly drops to $\rho_\mathrm{robust}\simeq0.27$ when using a robust coefficient ($p\simeq0.01$; see Table. \ref{tab:correlations_table}). 

Fitting a linear relationship, we find median posterior parameters of $\beta_1=1.24\pm0.17$, $\beta_0=24.90\pm0.05$ and $\epsilon\sim\mathrm{StudentT}(\sigma_\mathrm{int}=0.24\pm0.04,\nu=12.3_{-7.6}^{+27.1})$. This fit however produces a relatively broad residual distribution, with (the $\sigma_\mathrm{sMAD}$ of the residual distribution; see Fig. \ref{fig:EWHa_regression_residuals} for an example) $\sigma_\mathrm{res}\simeq0.38\,$dex, compared with $\sigma_\mathrm{res}\simeq0.30\,$dex for the $\xi_\mathrm{ion,0}-W_\lambda(\mathrm{H\alpha})$ relation. Alternatively, to better fit the trend observed in Fig. \ref{fig:xiion_vs_covariates}, we fit a power-law of the form $\xi_\mathrm{ion,0}=\beta_1\cdot E(B-V)_\mathrm{neb}^{\alpha}+\beta_0$ finding: $\alpha=1.23\pm0.16$, $\beta_1=1.99\pm0.41$, and $\beta_0=25.06\pm0.05$ with $\epsilon\sim\mathrm{StudentT}(\sigma_\mathrm{int}=0.24\pm0.04,\nu=10.3_{-5.9}^{+24.2})$. The residuals from this functional fit remain broader than for $\mathrm{H\,\alpha}$ or [O\,\textsc{iii}]$\lambda5007$, with $\sigma_\mathrm{res}\simeq0.34\,$dex.

Rather than being a unique single-variate predictor of the ionizing output of a galaxy, Fig. \ref{fig:xiion_vs_covariates} qualitatively shows that $E(B-V)_\mathrm{neb}$ may be crucial in explaining strong outliers in the relatively tight connections between $\xi_\mathrm{ion,0}$ and other observables, namely $W_\lambda(\mathrm{H\alpha})$, $W_\lambda(\mathrm{O\,III})$ and $\beta_\mathrm{UV}$ (as well as the scatter in $M_\mathrm{UV}$ to a minor extent). Quantitatively, this is seen by the $\nu>10$ posterior values preferred in the scatter of the $\xi_\mathrm{ion,0}-E(B-V)_\mathrm{neb}$ relations fit, indicating the tail heaviness has been suppressed significantly (with $\simeq98.7\,$per cent of the probability within $\pm3\sigma$, compared with $\simeq99.7\,$per cent under normality). As discussed in Section \ref{subsubsec:analysis:sv_fitting:betauv}, these outliers are likely sources with obscured, but intense star-formation. These systems will have high differential nebular-to-stellar dust attenuation (and possibly more complex geometry) leading to high inferred $\xi_\mathrm{ion,0}$.

\subsubsection{Relation between $\xi_\mathrm{ion,0}$ and $z$}\label{subsubsec:analysis:sv_fitting:z}

From the EXCELS sample spanning a wide $z\sim1-8$ redshift range, we recover a very mild (non-significant) redshift evolution given as: $\mathrm{log_{10}}(\xi_\mathrm{ion,0}\,/\,\mathrm{erg^{-1}\,Hz})=(0.03\pm0.02)\cdot z + (25.03\pm0.10)$. Such a shallow redshift evolution supports recent work by \citet{simmonds+24b}, \citet{begley+25} and \citet{pahl+25a} in which the average $\xi_\mathrm{ion,0}$ only increases by $\Delta\sim0.1-0.2\,$dex between cosmic noon and the epoch reionization. The mild evolution is primarily driven by the large diversity in SFG populations unearthed by JWST (i.e., not being restricted to the brightest galaxy population subsets) which suppresses the redshift evolution that was widely anticipated by literature in the pre-JWST era \citep[][]{finkelstein+19,stefanon+22}.

Given the shallow redshift dependence, the relation intrinsic scatter significantly outweighs the underlying evolution, here given by $\epsilon\sim\mathrm{StudentT}(\sigma_{\rm int}=0.25\pm0.04,\,\nu=3.19\pm1.27)$. This is in excellent agreement with estimates of the $\xi_\mathrm{ion,0}$ population scatter in \citet{begley+25} ($\simeq0.2-0.3\,$dex). We note that we do not observe the increase in intrinsic scatter with redshift ($\sim0.2\,$dex at $z\sim3-4$ to $\sim0.3\,$dex at $z\sim7.0-7.5$; \citealt{begley+25}). This is likely due to differences in sample properties, as the $z\sim3-4$ sample in \citet{begley+25} (from VANDELS) is relatively bright ($M_\mathrm{UV}\in[-19.5,-22.0]$), whereas the range in $M_\mathrm{UV}$ for the EXCELS sample at similar redshifts is shifted $\Delta\sim1.5\,$mag fainter. In parallel, only a narrow magnitude regime ($M_\mathrm{UV}\sim-19.5\pm0.5$) is covered by the highest redshift subset of the EXCELS sample (compared to $M_\mathrm{UV}\in[-17.75,-21.0]$ at $z\sim7.0-7.5$ in \citealt{begley+25}).

\subsection{Multivariate fitting}\label{subsec:modelling:mv}

\begin{figure*}
    \centering
    \includegraphics[height=7.65cm]{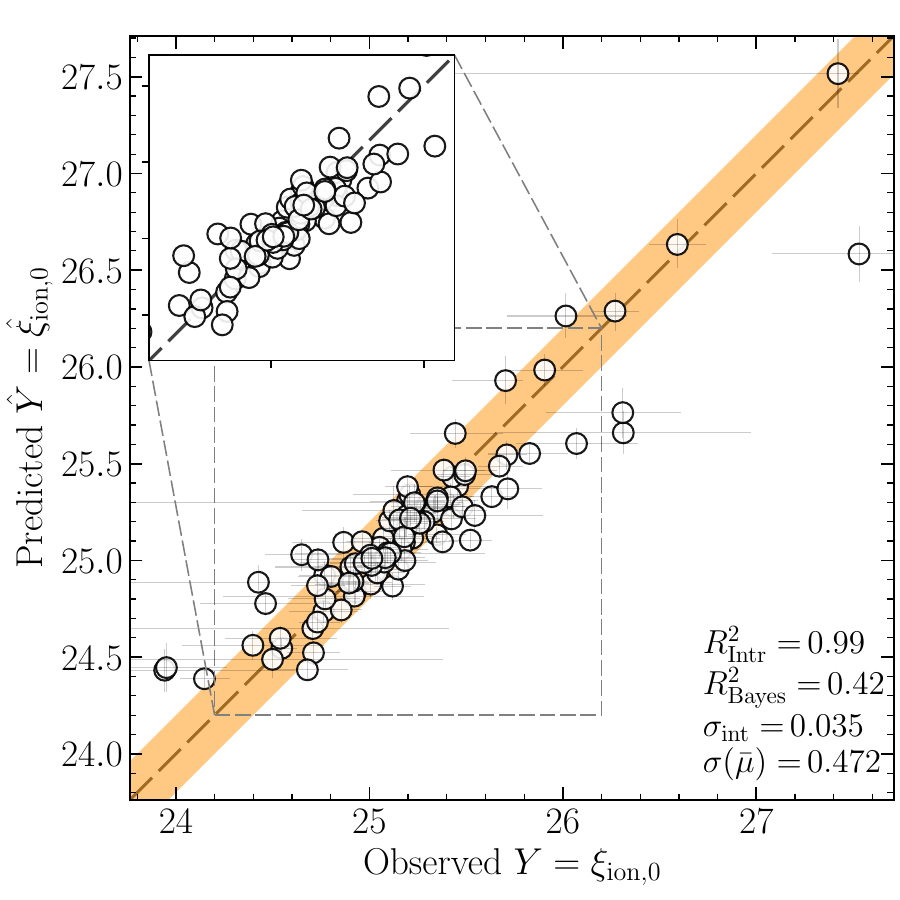}
    \includegraphics[height=7.5cm]{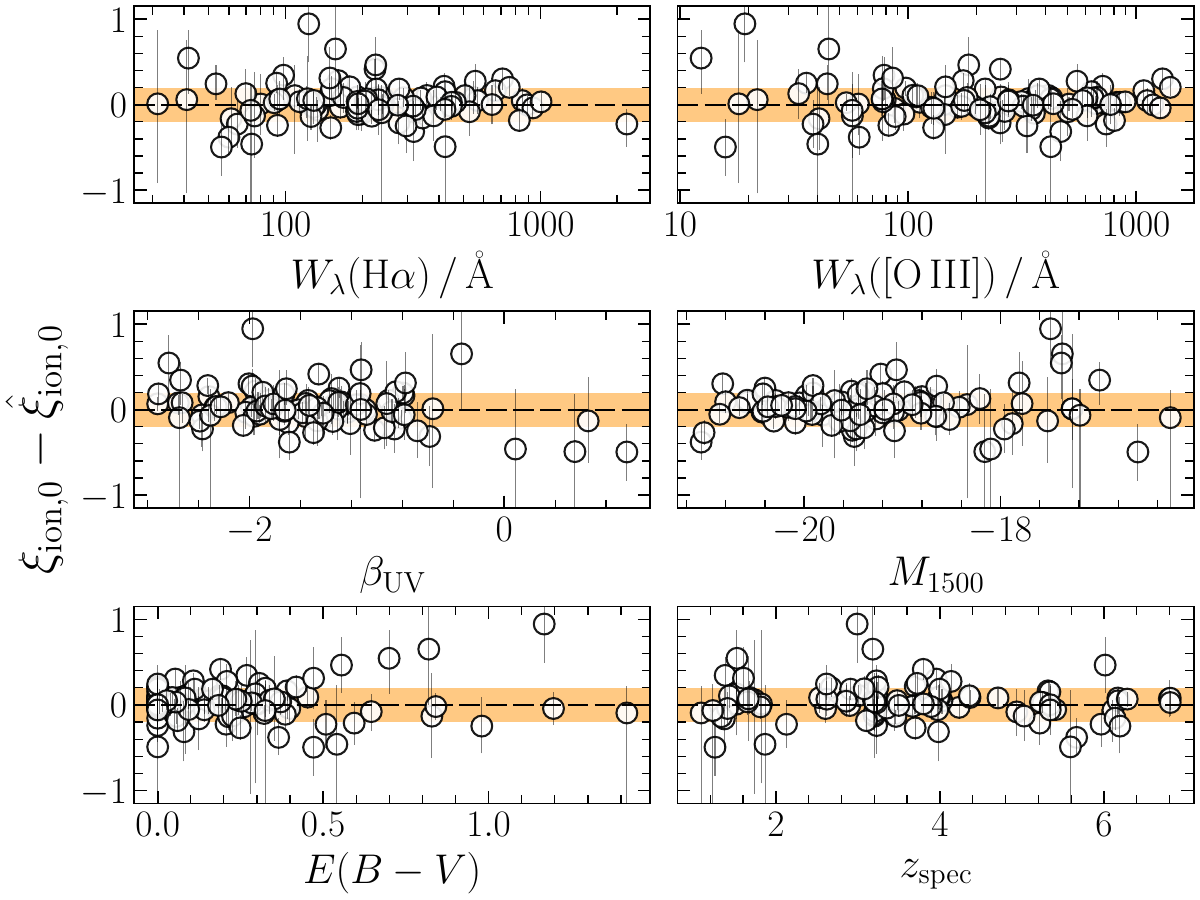}
    \caption{\textbf{Left panel:} Observed $\xi_\mathrm{ion,0}$ versus the median posterior predicted  $\xi_\mathrm{ion,0}$ from the full multivariate Bayesian linear regression to $W_\lambda(\mathrm{H\,\alpha})$,  $W_\lambda([\mathrm{O\,III}])$, $\beta_\mathrm{UV}$, $M_\mathrm{1500}$, $E(B-V)_\mathrm{neb}$ and $z_\mathrm{spec}$. The $1:1$ line is shown as black dashed diagonal, and the orange shaded region represents $\pm0.2\,$dex. The zoom inset shows the $\pm1.0\,$dex region around the canonical value required for reionziation, $\mathrm{log_{10}}(\xi_\mathrm{ion,0}\,/\,\mathrm{erg^{-1}\,Hz})=25.2$. The text within the figure shows a number of key metric and model statistics: $R^2_\mathrm{intr}=0.99$, $R^2_\mathrm{Bayes}=0.60$, $\sigma_\mathrm{int}=0.035$ and $\sigma(\bar{\mu})=0.472$ (see Section \ref{subsec:modelling:mv} for details). \textbf{Right panels:} The regression residuals ($\epsilon=\xi_\mathrm{ion,0}-\hat{\xi}_\mathrm{ion,0}$) as a function of each parameter in the model (one panel per property). The figures demonstrate that the model recovers the observed dynamic range in $\xi_\mathrm{ion,0}$, without any systematic offsets as a function of the physical properties.
    }
    \label{fig:mv_full_regression_residuals}
\end{figure*}

As shown in Section \ref{subsec:analysis:sv_fitting}, a number of observed properties trace the ionizing photon production relatively well, including $W_\lambda(\mathrm{H\,\alpha})$, $W_\lambda(\mathrm{[O\,III}])$ and $\beta_\mathrm{UV}$. For example, in the $\xi_\mathrm{ion,0}-W_\lambda(\mathrm{H\,\alpha})$ regression, we find an intrinsic scatter of $\sigma_\mathrm{int}\simeq0.2\,$dex, with the model predicting $\sigma(\mu_\theta)\equiv\sqrt{\mathrm{Var}(\mu_\theta)}\simeq0.18\,$dex galaxy-to-galaxy spread across the sample population. We can also assess the predictive power of the model using $R^2_\mathrm{intr}=\mathrm{Var}(\mu_\theta)/(\mathrm{Var}(\mu_\theta)+\sigma_\mathrm{int}^2)$, which measures the intrinsic galaxy-to-galaxy variance accounted for by the model variables (excluding measurement noise), as well as $R^2_\mathrm{Bayes}=\mathrm{Var}(\mu_\theta)/(\mathrm{Var}(\mu_\theta)+\mathrm{Var}(\epsilon_\mathrm{residuals}))$\footnote{Notably, $R^2_\mathrm{Bayes}$ accounts for the added uncertainty from the intrinsic and measurement scatter. Specifically, $\mu_\theta$ is the latent model prediction (i.e., drawing from the model posteriors), and $\epsilon_\mathrm{residuals}=\tilde{\mu_\theta}- y_\mathrm{obs}$, where $y_\mathrm{obs}$ is drawn from the $\xi_\mathrm{ion,0}$ posterior distributions (e.g., see Fig. \ref{fig:xiion_per_galaxy_distributions}) and
$\tilde{\mu_\theta}$ represents posterior-predictive draws from the model: $\tilde{\mu_\theta}=\mu_\theta+\epsilon_{\mathrm{intr}}$. The intrinsic scatter is drawn from $\epsilon^{\mathrm{intr}}\sim\mathrm{StudentT}(\sigma_\mathrm{int},\nu)$.}, which includes all sources of scatter \citep[i.e., measurement noise and intrinsic scatter; see][]{gelman+19}, and evaluates how well the \textit{full} variance is matched.

From the regression between $\xi_\mathrm{ion,0}$ and $W_\lambda($\mbox{[O\,\textsc{iii}]}), we find $R^{2}_\mathrm{intr}=0.34$ and $R^2_\mathrm{Bayes}=0.10$. 
Together these metrics indicate that $W_\lambda($\mbox{[O\,\textsc{iii}]}) show a statistically significant connection with $\xi_\mathrm{ion,0}$ and reasonable predictive power (with a $\simeq+0.3\,$dex increment in $\xi_\mathrm{ion,0}$ from a $\simeq1\,$dex increase in $W_\lambda($\mbox{[O\,\textsc{iii}]}) in addition to $\mathrm{RMAE}\simeq0.33\,$dex\footnote{We define the \textit{robust} median absolute deviation, $\mathrm{RMAE}=\langle\mid\epsilon_\mathrm{residuals}\mid\rangle$), where $\epsilon_\mathrm{residuals}$ is as above, and we take the median to mitigate against the impact of strong outliers in the metric.}). Nonetheless, $W_\lambda($\mbox{[O\,\textsc{iii}]}) (or $W_\lambda(\mathrm{H\,\alpha})$\footnote{The $\xi_\mathrm{ion,0}-W_\lambda(\mathrm{H\,\alpha})$ relation gives $R^{2}_\mathrm{intr}=0.43$ and $R^2_\mathrm{Bayes}=0.11$. According to the relation a $\simeq1\,$dex increase in $W_\lambda(\mathrm{H\,\alpha})$ produces a $\simeq0.5\,$dex increase in $\xi_\mathrm{ion,0}$, with $\mathrm{RMAE}=0.33$ (or $\mathrm{RMAE}=0.20$ from the model alone).}) as lone predictors only accounts for $\sim10\,$per cent of the total \textit{observed} variance. This is due to the fact that the intrinsic scatter ($\sigma_\mathrm{int}\simeq0.2\,$dex), combined with the measurement uncertainty across the dataset, and specifically in $\xi_\mathrm{ion,0}$ (e.g., see Fig. \ref{fig:xiion_per_galaxy_distributions}), is relatively large compared with the predicted model spread $\sigma(\mu_\theta)$. Moreover, a single property-based prediction of $\xi_\mathrm{ion,0}$ neglects the impact of other physical properties on $\xi_\mathrm{ion,0}$ (as discussed throughout this work, e.g., see Section \ref{subsec:analysis:sv_fitting} and Fig. \ref{fig:xiion_vs_covariates}).

Motivated by the necessity to improve models of the EOR, and in particular by providing more robust predictions of the ionizing photon production in the high-redshift SFG population, below we explore multivariate trends with $\xi_\mathrm{ion,0}$.  

\subsubsection{Joint modelling of $\mathrm{\xi_{ion,0}}$ and $\mathbf{X}$}\label{subsubsec:modelling:mv_full}

\begin{figure*}
    \centering
    \includegraphics[width=1.025\columnwidth]{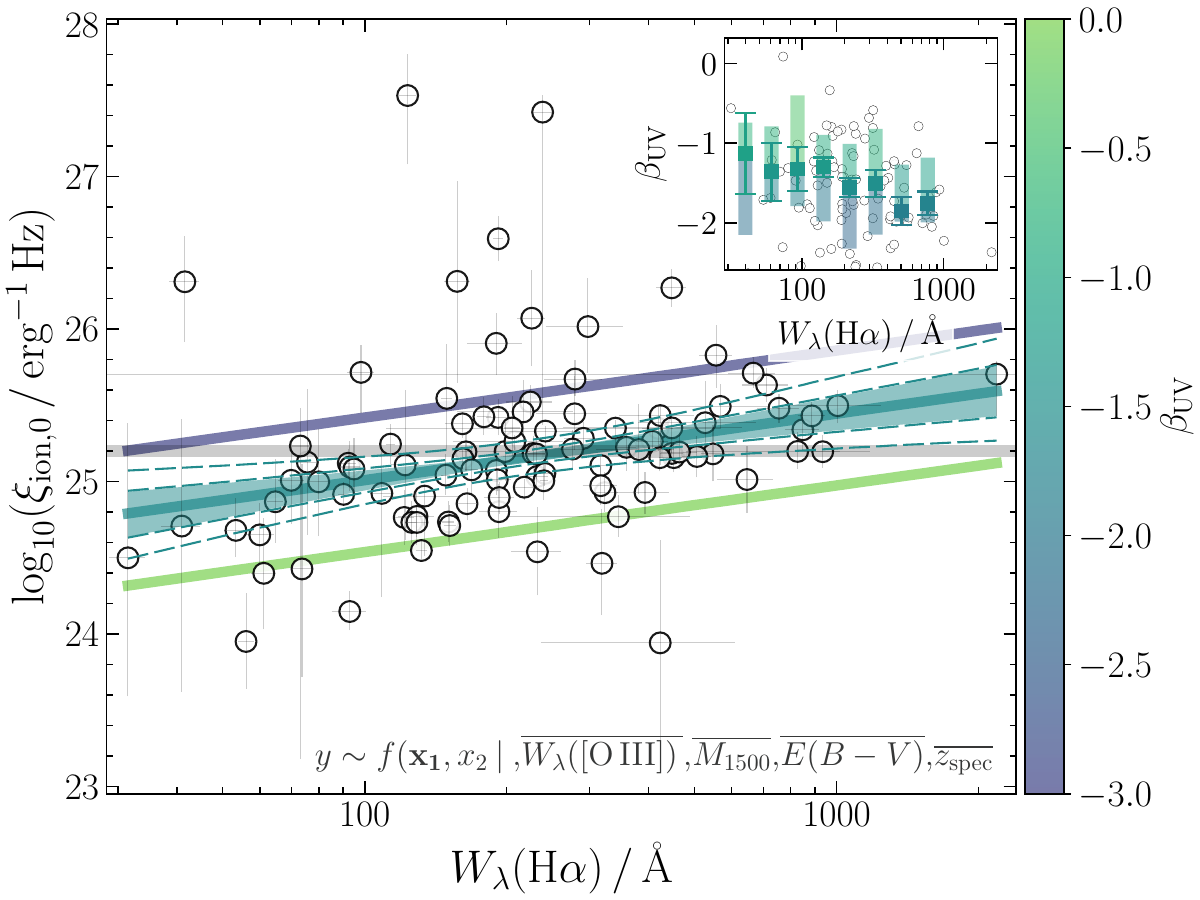}
    \includegraphics[width=1.025\columnwidth]{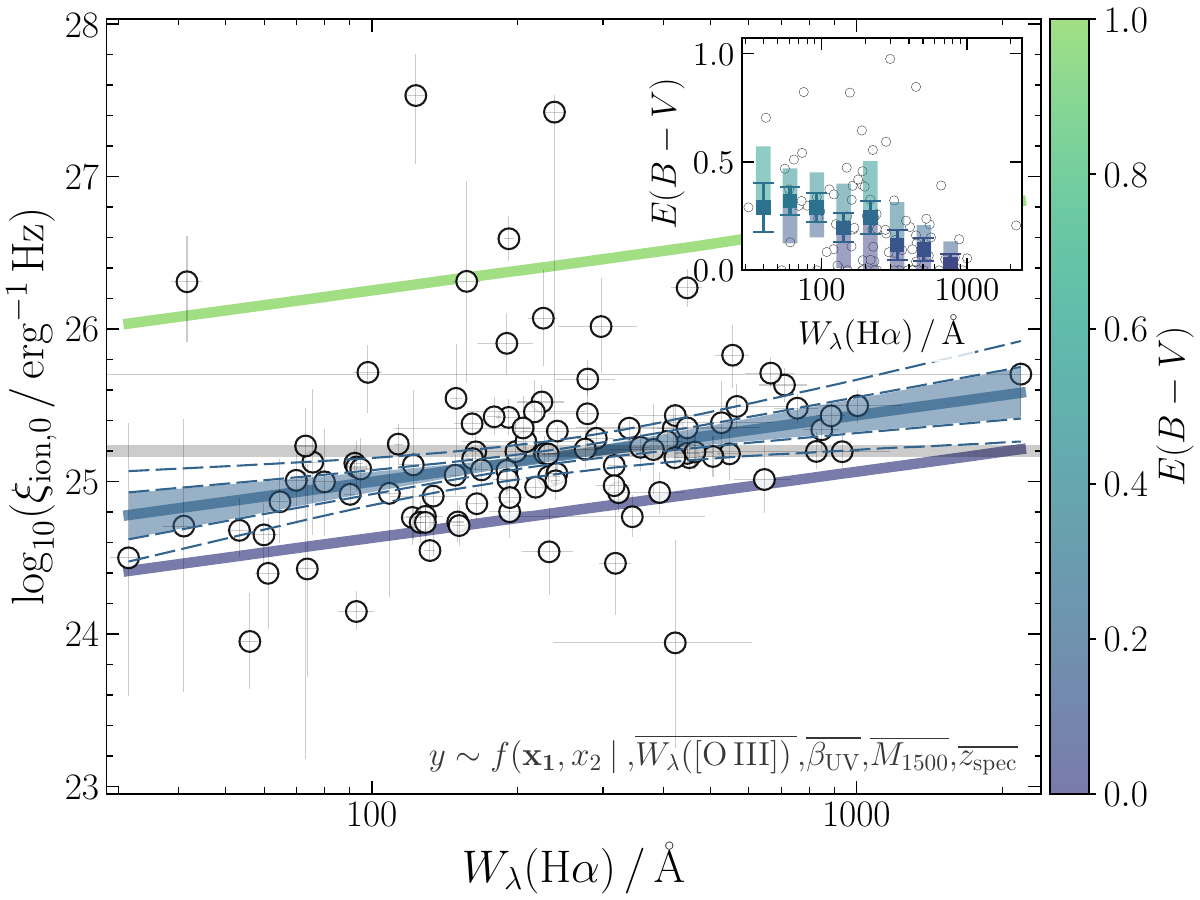}
    \caption{Relation between ionizing photon production efficiency $\xi_\mathrm{ion,0}$ and $W_\lambda(\mathrm{H\,\alpha})$ from the full multivariate linear regression (see also Fig. \ref{fig:appendix:full_mv_corner} for the parameter posterior distributions). In the \textbf{left} panel we show lines of constant $\beta_\mathrm{UV}$ with $\beta_{UV}=-3.0,\,-1.5,\,0.0$, and in the \textbf{right} panel we show lines of fixed $E(B-V)_\mathrm{neb}$, with $E(B-V)_\mathrm{neb}=0.0,\,0.25,\,1.0$. In each panel, we evaluate the remaining parameters at the sample median values ($\langle W_\lambda([\mathrm{O\,III}])\rangle\simeq250\,$\AA, $\langle M_\mathrm{1500}\rangle\simeq-19.5$, $\langle z_\mathrm{spec}\rangle\simeq3.25$, and where applicable $\langle \beta_\mathrm{UV}\rangle \simeq-1.6$ and $\langle E(B-V)_\mathrm{neb}\rangle\simeq0.18$), while marginalising over their associated gradient posterior distributions. The inset panels show the joint relation between the primary and auxiliary independent variables.  
    }
    \label{fig:mv_full_regression_relation}
\end{figure*}

Several of the measured properties explored here tie closely with the physical processes relating to the ionizing photon production efficiency; for example, the nebular emission line strengths ($W_\lambda(\mathrm{H\,\alpha})$ and $W_\lambda([\mathrm{O\,III}])$) trace the presence of young massive stars. With the goal of improving the predictive metrics for $\xi_\mathrm{ion,0}$, we apply the Bayesian linear regression framework described in Section \ref{subsec:analysis:model} to our full set of six key observables, $X_n=\{ W_\lambda(\mathrm{H\alpha}),\, W_\lambda([\mathrm{O\,III}]),\, \beta_{\mathrm{UV}},\, M_{1500},\, E(B-V),\, z_\mathrm{spec} \}$. The predictive power gained when moving from a single measurement prediction (e.g., $W_\lambda(\mathrm{H\,\alpha})$), to folding in the full set of observables is clearly illustrated in Fig. \ref{fig:mv_full_regression_residuals} and Fig. \ref{fig:mv_full_regression_relation}. Here we show a comparison between the observed and model predicted $\xi_\mathrm{ion,0}$ (with the full posterior corner plot shown in Fig. \ref{fig:appendix:full_mv_corner}), finding practically all the modelled variance can be explained by the model ($R^2_\mathrm{intr}\simeq0.99$). This indicates that together, the chosen variables thoroughly trace the underlying physical processes that drive the range in $\xi_\mathrm{ion,0}$. This is qualitatively supported by the large population diversity spanned by the model ($\sigma(\mu_\theta)\simeq0.47\,$dex), in combination with the little remaining intrinsic scatter inferred from the Bayesian model ($\sigma_\mathrm{int}\simeq0.035\pm0.02$). Crucially, we find that multivariate median posterior model has significantly more accurate predictive power, with $\mathrm{RMAE}\simeq0.16\,$dex (a factor of two lower than than the best-performing single property regression with $W_\lambda(\mathrm{H\,\alpha})$).

\subsubsection{Identifying a minimal set of predictive observables}

From the multivariate regression, we recover (conditional) relations between $\xi_\mathrm{ion,0}$ and UV luminosity and redshift of $\beta_{M_\mathrm{UV}}=0.04\pm0.03$ and $\beta_z=0.01\pm0.02$, respectively. These parameters already had weak, marginal trends with $\xi_\mathrm{ion,0}$ from their individual regressions (see Sections \ref{subsubsec:analysis:sv_fitting:muv} and \ref{subsubsec:analysis:sv_fitting:z}), and now appear consistent with no relation at the $<2\sigma$ level when fitting to the full observed data. This strengthens the argument that these weak trends (or otherwise as found in the literature; e.g., see \citealt{pahl+25a,llerena+25} and Fig. \ref{fig:MUV_regression}) are due to more fundamental evolution in the properties of star-forming galaxies.

Another important consideration is the strong collinearity between $W_\lambda(\mathrm{H\,\alpha})$ and $W_\lambda([\mathrm{O\,III}])$, as shown in Fig. \ref{fig:covariates_corner}. Such tight correlations can potentially lead to unstable Monte Carlo sampling, and inflate individual posterior slope distributions. Here we find that the uncertainties in $\beta_{W_\lambda(\mathrm{H\alpha})}$ and $\beta_{W_\lambda([\mathrm{O\,III}])}$ are a factor of $\simeq1.5-1.8$ larger than in the single predictor fits. Moreover, the individual conditional gradients become shallower ($\beta_{W_\lambda(\mathrm{H\alpha})}=0.43^{+0.17}_{-0.16}$ and $\beta_{W_\lambda([\mathrm{O\,III}])}=0.19^{+0.12}_{-0.13}$) compared to the fits in Sections \ref{subsubsec:analysis:sv_fitting:wha} and \ref{subsubsec:analysis:sv_fitting:woiii}; $\beta_{W_\lambda(\mathrm{H\alpha})}=0.49\pm0.09$ and $\beta_{W_\lambda([\mathrm{O\,III}])}=0.30\pm0.08$.

We therefore perform two complementary fits, using one of $W_\lambda(\mathrm{H\,\alpha})$ or $W_\lambda([\mathrm{O\,III}])$, combined with the remaining properties. As expected, we find steeper gradients with $\xi_\mathrm{ion,0}$ for both emission lines, and tighter posteriors compared with the full fit above: $\beta_{W_\lambda(\mathrm{H\alpha})}=0.65\pm0.06$ and $\beta_{W_\lambda([\mathrm{O\,III}])}=0.51^{+0.05}_{-0.05}$, with the remaining parameter gradients consistent to within $\pm1\sigma$.

Crucially, both these fits perform equally as well, mimicking the tight residuals relation ($R^2_\mathrm{intr}\gtrsim0.95$ and $R^2_\mathrm{Bayes}\simeq0.42$ for each) while explaining the large diversity seen in observed $\xi_\mathrm{ion,0}$ (both with $\sigma(\bar{\mu})\simeq0.47\,$dex and $\sigma_\mathrm{int}\simeq0.04\,$dex and $\simeq0.06\,$dex for $W_\lambda(\mathrm{H\,\alpha})$ or $W_\lambda([\mathrm{O\,III}])$, respectively).

Notably, these slopes are higher than in both their individual $\xi_\mathrm{ion,0}-W_\lambda(X)$ relations (e.g., see Figures \ref{fig:EWHa_regression} and \ref{fig:EWOIII_regression}). The elevated conditional slopes measure the change in $\xi_\mathrm{ion,0}$ with $W_\lambda(X)$, while holding the other properties fixed, which is particularly impactful for properties with moderate to strong covariance such as $\beta_\mathrm{UV}$ and $E(B-V)_\mathrm{neb}$. As such the steeper relations found in other literature (e.g., \citealt{tang+19} in $\xi_\mathrm{ion,0}-W_\lambda([\mathrm{O\,III}])$ or \citealt{prieto-lyon+23} in $\xi_\mathrm{ion,0}-W_\lambda(\mathrm{H\,\alpha})$), can be understood in terms of the level of diversity in the physical properties across the samples used. For example, the sample in \citet{prieto-lyon+23} \citep[see also][]{papovich+25} is much bluer on average ($\langle\beta_\mathrm{UV}\rangle=-2.1$ compared with $\langle\beta_\mathrm{UV}\rangle=-1.6$ here, see also their Fig. 3), and given the correlation between $W_\lambda(\mathrm{H\alpha})$ and $\beta_{\mathrm{UV}}$, this will drive a change in the inferred $\xi_\mathrm{ion,0}-W_\lambda(\mathrm{H\,\alpha})$ slope. Similarly, the sources in the \citet{tang+19} sample are selected as EELGs, which are an extreme subset of the SFG population, likely with more extreme ionization conditions and elevated specific star-formation rates (as well as differences in other physical properties not considered, such as metallicity).

Taking this into account, from the original set of key properties considered in this work, we conclude that the most important in predicting $\xi_\mathrm{ion,0}$ while striving to minimise the degree of systematic scatter are; either $W_\lambda(\mathrm{H\,\alpha})$ or $W_\lambda([\mathrm{O\,III}])$, the UV continuum slope $\beta_\mathrm{UV}$ and the nebular attenuation, $E(B-V)_\mathrm{neb}$.
These results support the expected physical connections between the properties and the ionizing photon output. The nebular emission lines ($W_\lambda(\mathrm{H\,\alpha})$ / $W_\lambda([\mathrm{O\,III}])$ trace the presence of recently formed massive stars (i.e., recent bursts and high sSFR, as well as responding to metallicity and ionization state), while bluer $\beta_\mathrm{UV}$ are indicative of lower metallicity, younger, and less dust attenuated systems. Lastly, $E(B-V)_\mathrm{neb}$ dictates the Balmer line attenuation correction, required to accurately infer the intrinsic $\xi_\mathrm{ion,0}$.

The nebular dust attenuation in and of itself is not a strong predictor of $\xi_\mathrm{ion,0}$ (the residual scatter is large, $\sigma_\mathrm{res}\simeq0.38\,$dex, and qualitatively seen in Fig. \ref{fig:xiion_vs_covariates}), but it is crucial to explain the outliers (and the heavy-tailed residuals). From the single variable regressions (see Section \ref{subsec:analysis:sv_fitting}) we see only weak-to-moderate behaviour at $E(B-V)\lesssim0.5$, with much larger excursions coming from the high-$E(B-V)$ tail, suggesting $E(B-V)_\mathrm{neb}$ carries the bulk of the $\simeq0.20-0.25\,$dex of intrinsic scatter that remains when it is omitted from the fitting. 
This is demonstrable by contrasting the regression results in a two-parameter fit with $\beta_\mathrm{UV}$ and $W_\lambda($\mbox{[O\,\textsc{iii}]}) against a regression additionally including $E(B-V)_\mathrm{neb}$. The former gives median posterior parameters of $\beta_\mathrm{W_\lambda([O\,III])}=0.26\pm0.08$, $\beta_{\beta_\mathrm{UV}}=-0.13\pm0.06$, $\beta_0=24.58\pm0.21$ (with $\sigma_\mathrm{int}=0.19\pm0.04$, $\nu=2.65^{+0.98}_{-0.46}$), while additionally including $E(B-V)_\mathrm{neb}$ produces: $\beta_\mathrm{W_\lambda([O\,III])}=0.55\pm0.05$, $\beta_{\beta_\mathrm{UV}}=-0.25\pm0.03$, $\beta_\mathrm{E(B-V)_\mathrm{neb}}=1.73\pm0.10$, $\beta_0=24.58\pm0.21$ (with $\sigma_\mathrm{int}=0.05\pm0.02$, $\nu=25^{+33}_{-16}$). These results clearly suggest that accounting for nebular dust attenuation is crucial for breaking degeneracies as a result of the covariance between observed properties (e.g., $W_\lambda($\mbox{O\,\textsc{iii}}) and $\beta_\mathrm{UV}$).

Moreover, the model including only $W_\lambda($\mbox{O\,\textsc{iii}}) and $\beta_\mathrm{UV}$ gives $R^2_\mathrm{intr}\simeq0.47$, $R^2_\mathrm{Bayes}\simeq0.13$, $\sigma_\mathrm{int}\simeq0.19$ and $\sigma(\bar{\mu})\simeq0.19\,$dex, whereas adding $E(B-V)_\mathrm{neb}$ gives $R^2_\mathrm{intrinsic}\simeq0.99$, $R^2_\mathrm{Bayes}\simeq0.57$, $\sigma_\mathrm{int}\simeq0.05$ and $\sigma(\bar{\mu})\simeq0.46\,$dex.  Significantly, including $E(B-V)_\mathrm{neb}$ removes the heavy-tailed residuals (the $\mathrm{StudentT}$ tail parameter rises from $\nu\simeq3$ as shown in Fig. \ref{fig:EWHa_regression_residuals} to $\nu\gg10$), indicating that the sources that appeared to be outliers are driven by dust extremes that the ($W_\lambda($\mbox{O\,\textsc{iii}}), $\beta_\mathrm{UV}$, $E(B-V)_\mathrm{neb}$) model correctly accounts for.

As a useful exercise, we assess the predictive power when using the most commonly measured properties in samples at the highest redshifts. Specifically, $M_\mathrm{UV}$, $\beta_\mathrm{UV}$ and redshift can be measured from photometric samples of SFGs selected in studies estimating the UV luminosity function for example \citep[e.g., see][]{mcleod+24,cullen+24}. Such a model (with posterior parameters; $\beta_{\beta_{\mathrm{UV}}}=-0.24\pm0.06$, $\beta_{M_\mathrm{UV}}=0.09\pm0.04$, $\beta_0(\beta_\mathrm{UV}=-2,\,M_\mathrm{UV}=-20)=25.22\pm0.04$, $\sigma_\mathrm{int}=0.23^{+0.05}_{-0.04}$, and $\nu=3.4^{+2.7}_{-1.0}$) is only able to predict $\xi_\mathrm{ion,0}$ at the $\simeq0.25\,$dex level (from the latent model parameters, or worse at $\mathrm{RMAE}\simeq0.35\,$dex when folding in intrinsic and measurement scatter).

A number of past studies have sought to infer the ionizing photon production efficiencies of SFGs from photometric samples \citep[e.g., see][]{endsley+24,simmonds+24b,begley+25}, either applying empirical relations (i.e., like those presented in Section \ref{subsubsec:analysis:sv_fitting:wha}) or from SED modelling codes. Such methods rely on the flux excess emission line signature in a photometric band, and are therefore not able to directly recover the individual line measurements that may be present \citep[e.g., the F410M band is sensitive to \mbox{[O\,\textsc{iii}]}$+\mathrm{H\,\beta}$ at $z\sim7$][]{endsley+24,begley+25}. To enable future comparisons with photometric-based samples, or to easier facilitate predictions of $\xi_\mathrm{ion,0}$ from photometric flux excess measurements, we perform our regression to the combined nebular \mbox{[O\,\textsc{iii}]}$\lambda\lambda4960,5007+\mathrm{H\,\beta}$ equivalent widths. The median posterior parameters are $\beta_1\simeq0.33^{+0.08}_{-0.09}$, $\beta_{0}=24.30^{+0.24}_{-0.21}$, $\sigma_\mathrm{int}=0.20^{+0.04}_{-0.03}$ and $\nu=2.5^{+0.8}_{-0.4}$.
Predictions based purely on the \mbox{[O\,\textsc{iii}]}$\lambda\lambda4960,5007+\mathrm{H\,\beta}$ equivalent width perform marginally better than the above model (as a function of $\beta_\mathrm{UV}$ and $M_\mathrm{UV}$), achieving $\mathrm{RMAE}\simeq0.21\,$dex from the model parameters (and $\simeq0.33$ when folding in the additional scatter). 
Similarly to other commonly used empirical relations for $\xi_\mathrm{ion,0}$ \citep[e.g., ][]{tang+19}, galaxies with high equivalent widths ($W_\lambda($\mbox{[O\,\textsc{iii}]}$\lambda\lambda4960,5007+\mathrm{H\,\beta})\gtrsim500\,$\AA) display $\mathrm{log_{10}}(\xi_\mathrm{ion,0}\,/\,\mathrm{erg^{-1}\,Hz})\gtrsim25.2$, suggesting that they are significant contributors to the ionizing photon budget \citep[][]{robertson+15}.

For a more direct comparison with the commonly used \citet{tang+19} relation, we repeat our regression isolating the ``extreme emission line'' galaxies in the EXCELS sample with $W_\lambda($\mbox{[O\,\textsc{iii}]}$\lambda\lambda4960,5007+\mathrm{H\,\beta})\gtrsim500\,$\AA. Here we find a steeper relation than found across the full EXCELS sample but still shallower than found by \citet{tang+19}, with $\beta_1=0.48\pm0.13$ (plus $\beta_0=23.88\pm0.36$, $\sigma_\mathrm{int}=0.17\pm0.03$, and $\nu\simeq20_{-14}^{+31}$). The model on the EELG subset returns $\mathrm{RMAE}\simeq0.15\,$dex  (and $\simeq0.24$ including the measurement scatter), suggesting these extreme galaxies represent a more homogeneous subset of the wider population. The steeper trend with $\xi_\mathrm{ion,0}$ is also likely driven in part by the fact these sources are typically less dust attenuated, brighter, and have bluer UV continuum slopes on average compared with the full EXCELS sample (e.g., see Fig. \ref{fig:covariates_corner}).

\subsubsection{The impact of outliers}

From inspecting Fig. \ref{fig:xiion_vs_covariates}, the panels using $E(B-V)_\mathrm{neb}$ as an auxiliary variable reveal that the majority of outliers in the relations between $\xi_\mathrm{ion,0}$ and $W_\lambda(\mathrm{H\alpha})$, $W_\lambda([\mathrm{O\,III}])$ and $\beta_\mathrm{UV}$ have elevated nebular attenuation, $E(B-V)_\mathrm{neb}\gtrsim0.5$. To test the impact of these outliers on the regression, we repeat the $\xi_\mathrm{ion,0}-W_\lambda(\mathrm{H\alpha})$ fit after removing the upper quartile of nebular attenuation from the sample $E(B-V)_\mathrm{neb}\geq0.4$. The $\xi_\mathrm{ion,0}-W_\lambda(\mathrm{H\alpha})$ slope is marginally steeper $\beta_1\simeq0.55\pm0.09$, but consistent within the uncertainties. However, removing the high $E(B-V)_\mathrm{neb}$ tail does in-fact lower the residual scatter, with $\sigma_\mathrm{int}\simeq0.18\pm0.03$ and $\nu\sim21$ indicating a return to a more typical Gaussian residuals distribution.

\subsubsection{Redshift evolution in regression gradients}\label{subsubsec:redshift_dependence_evolution}

A further aspect to consider is the possibility of an underlying evolutionary factor as a function of redshift when using nebular emission lines as a tracer of $\xi_\mathrm{ion,0}$. Such a change could be driven by the global trend  of higher ionization state, lower metallicity or increased star-formation burstiness at higher redshifts. To examine this possibility, we construct three sub-samples (of approximately equal size; $N\simeq35-40$ per bin) in redshift having $z<3$, $3\leq z<4$ and $z\geq4$. Fitting for $\xi_\mathrm{ion,0}-W_\lambda(\mathrm{H\,\alpha})$, we find no clear trend across the three redshift samples. In contrast, we find a positive (albeit not statistically significant) trend in redshift for the slope strength with the $\xi_\mathrm{ion,0}-W_\lambda([\mathrm{O\,III}])$ regressions: $\beta_{W_\lambda([\mathrm{O\,III}])}=0.14\pm0.21$ ($\beta_0=24.87\pm0.46$) at $z<3$, $\beta_{W_\lambda([\mathrm{O\,III}])}=0.27\pm0.14$ ($\beta_0=24.47\pm0.35$) at $3\leq z<4$, and $\beta_{W_\lambda([\mathrm{O\,III}])}=0.42\pm0.16$ ($\beta_0=24.14\pm0.43$) at $z\geq4$. This is physically plausible as \mbox{[O\,\textsc{iii}]} is more sensitive to the ionization state (i.e., $\mathrm{log}(U)$ and hardness) and metallicity than $\mathrm{H\,\alpha}$, which more directly tracks the ionizing photon production rate. It therefore follows, given the modest evolution in redshift with metallicity \citep[e.g.,][]{sanders+16,cullen+20,reddy+21}, that a redshift dependent trend in $\xi_\mathrm{ion,0}-W_\lambda([\mathrm{O\,III}])$ exists without an associated H$\,\alpha$ evolution. 

Lastly, we also detect a systemic decline in the intrinsic scatter (for $W_\lambda([\mathrm{O\,III}])$) with redshift; $\sigma_\mathrm{int}=0.41^{+0.14}_{-0.12}$, $0.29\pm0.06$ and $0.20^{+0.06}_{-0.05}$ in our progressively higher redshift bins. At earlier epochs, star-forming galaxies have lower metallicities and have more uniformly young stellar populations, potentially leading to a tighter mapping between \mbox{[O\,\textsc{iii}]} and ionizing photon production. Observational sample effects could also be amplifying this trend to some extent. Specifically, it is likely that we probe a narrower dynamic range in SFG properties in our higher redshift sample compared to those at the lowest redshifts in our sample (e.g., see also Fig. \ref{fig:covariates_corner}).
We stress that larger sample sizes across a large dynamic range in redshift will be required to solidify these findings, as well as robustly determine any evolutionary trends in $\xi_\mathrm{ion,0}-W_\lambda([\mathrm{O\,III}])$.

\subsubsection{The influence of metallicity}

A full investigation into the role of metallicity in the ionizing photon production efficiency of galaxies is beyond the scope of this paper. However, for a small subset ($N=56$, $\simeq35\,$per cent of our EXCELS sample), we have \mbox{[O\,\textsc{ii}]}$\lambda\lambda3727,3730$ flux measurements which can be combined with \mbox{[O\,\textsc{iii}]}$\lambda\lambda4960,5008$ to measure $\mathrm{O}_{32}$ ($=$\mbox{[O\,\textsc{iii}]}$\lambda\lambda4960,5008/$\mbox{[O\,\textsc{ii}]}$\lambda\lambda3727,3730$; after dust corrections following the method outlined in Section \ref{subsubsec:physicalproperties:dustcorrection}) as a proxy of metallicity \citep[e.g., see][]{curti+23,hirschmann+23,scholte+25,shapley+25}. 

\begin{figure}
    \centering
    \includegraphics[width=1\columnwidth]{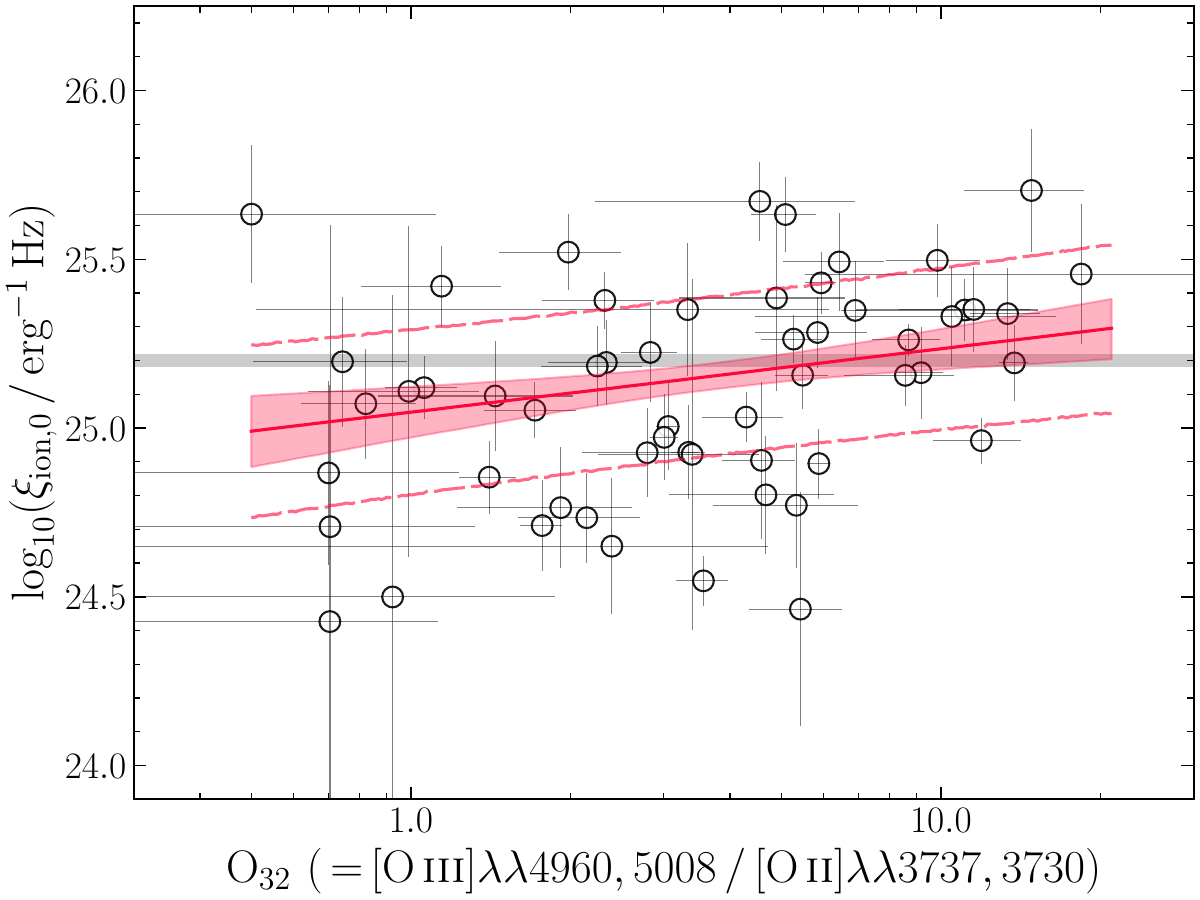}
    \caption{The ionizing-photon production efficiency \(\xi_{\mathrm{ion},0}\) as a function of \(\mathrm{O_{32}}
=[\mathrm{O\,\textsc{iii}}]\lambda\lambda4960,5008/[\mathrm{O\,\textsc{ii}}]\lambda\lambda3727,3730\),
a tracer of ionization parameter and gas-phase metallicity. From a Bayesian linear regression to the \(N=56\) galaxy subsample with \(\mathrm{O_{32}}\) measurements we obtain
\(\log_{10}(\xi_{\mathrm{ion},0}/{\rm erg^{-1}\,Hz})=(0.19\pm0.11)\,\log_{10}(\mathrm{O_{32}})+(25.05\pm0.07)\)
(the red solid line and shaded region indicating the $\pm1\sigma$ posterior uncertainty). Dashed lines show the inferred intrinsic scatter
$\sigma_{\mathrm{int}}=0.23\pm0.03\,$dex. This trend is in good qualitative agreement with recent literature \citep[e.g.][]{llerena+23,pahl+25a,papovich+25},
and is consistent with higher $\mathrm{O_{32}}$ (i.e., higher ionization parameter / lower metallicity) being associated with elevated $\xi_{\mathrm{ion},0}$.}
    \label{fig:metallicity}
\end{figure}

In Fig.~\ref{fig:metallicity}, we show $\xi_\mathrm{ion,0}$ as a function of $\mathrm{O}_{32}$, with the median posterior relation from the Bayesian linear regression: $\mathrm{log_{10}(\xi_\mathrm{ion,0}\,/\,\mathrm{erg^{-1}\,Hz})}=(0.19\pm0.11)\times\mathrm{log_{10}}(\mathrm{O_{32}})+(25.05\pm0.07)$. The gradient is only marginally significant (at the $\simeq2\sigma$ level), as expected given the relatively small sample size and the moderate correlation coefficient of $\rho_\mathrm{S}\simeq0.38$ ($p\simeq3\times10^{-3}$). This is shallower than the relation found by \citet{llerena+25} \citep[see also][]{papovich+25}, but shows excellent consistency with \citet{pahl+25a}. 

Broadly speaking, there is a clear qualitative agreement across the literature that higher $\mathrm{O}_{32}$ ratios, which trace lower metallicities and higher ionization parameters, are indicative of higher ionizing photon production efficiencies. We do note that there is moderately large intrinsic scatter $\sigma_\mathrm{int}=0.23\pm0.03$ (and $\nu\simeq27^{+33}_{-16}$) with low predictive power ($R^2_{\mathrm{intr}}\simeq0.1$, $\sigma(\mu_\theta)\simeq0.08\,$dex, $\mathrm{RMAE}\simeq0.28$), and that larger deep spectroscopic samples will be required to obtain a more robust relation between metallicity and $\xi_\mathrm{ion,0}$ while controlling for other physical properties.

\section{Conclusions}\label{sec:conclusions}

In this study we have used ultra-deep JWST/NIRSpec spectroscopy taken as part of the EXCELS survey (e.g., see Fig. \ref{fig:example_excels_spectra}) to investigate the ionizing photon production efficiencies of $N=159$ star-forming galaxies across a wide $1\lesssim z\lesssim8$ redshift range. With the aim of statistically characterising how $\xi_\mathrm{ion,0}$ connects to key physical properties; including the \mbox{[O\,\textsc{iii}]} and $\mathrm{H\,\alpha}$ nebular emission line equivalent widths, the UV luminosity ($M_\mathrm{UV}$) and UV spectral slope ($\beta_\mathrm{UV}$), as well as the dust attenuation ($E(B-V)_\mathrm{neb}$) and redshift; we adopt a Bayesian methodology to perform single and multi-variate linear regression between these and $\xi_\mathrm{ion,0}$ (as summarised in Table \ref{tab:coeff}). The key results of our analysis can be summarised as follows:
\begin{enumerate}
    \item The ionizing photon production efficiency is most tightly linked with signatures of recent massive-star formation, specifically the equivalent widths of strong optical nebular emission lines $\mathrm{H\,\alpha}$ and \mbox{[O\,\textsc{iii}]} are the best predictors. From our Bayesian linear regression fitting $\mathrm{log_{10}(\xi_\mathrm{ion,0})}=\beta_1\times\mathrm{log_{10}}(W_\lambda(\mathrm{X}))+\beta_0+\epsilon(\sigma_\mathrm{int},\nu)$, we find: $\beta_1=0.49\pm0.09$, $\beta_0=23.97\pm0.23$, $\sigma_\mathrm{int}=0.20\pm0.03$ and $\nu=2.67\pm0.75$, for $\mathrm{H\,\alpha}$ as shown in Fig. \ref{fig:EWHa_regression} and Fig. \ref{fig:EWHa_regression_residuals}. Similarly, for [\mbox{O\,\textsc{iii}}] we infer $\beta_1=0.30\pm0.08$, $\beta_0=24.44\pm0.21$, $\sigma_\mathrm{int}=0.20\pm0.03$ and $\nu=2.56\pm0.63$ (see Section \ref{subsubsec:analysis:sv_fitting:woiii} and Fig. \ref{subsubsec:analysis:sv_fitting:woiii}). 
    \item A multivariate model combining either nebular $W_\lambda$ with the UV continuum slope ($\beta_\mathrm{UV}$) and nebular attenuation level ($E(B-V)_\mathrm{neb}$) recovers essentially all the of the intrinsic galaxy-to-galaxy variation in $\xi_\mathrm{ion,0}$, showing that these three physical properties (from the full set studied; e.g., see Fig. \ref{fig:covariates_corner} and Fig. \ref{fig:xiion_vs_covariates}) capture the dominant physics setting the ionizing photon production efficiency.
    \item Trends with redshift and absolute UV luminosity ($M_\mathrm{UV}$) are shallow and marginal across the EXCELS sample, and are best interpreted as due to the underlying evolution of the primary drivers above. Broadly speaking, the literature reports a wide variety of $\xi_\mathrm{ion,0}-M_\mathrm{UV}$ relations with no clear consensus, reflecting that other physical properties (e.g., $W_\lambda(\mathrm{H\,\alpha})$ or $\beta_\mathrm{UV}$) are indeed more influential on $\xi_\mathrm{ion,0}$.
    \item Although $E(B-V)_\mathrm{neb}$ is not a strong single predictor of $\xi_\mathrm{ion,0}$, it is essential for explaining large outliers and removes the heavy tailed residuals in relations with $W_\lambda(\mathrm{H\,\alpha})$, $W_\lambda([\mathrm{O\,III}])$ or $\beta_\mathrm{UV}$ that are consistent with intensely star-forming but obscured regions producing high $\xi_\mathrm{ion,0}$.
    \item The $\xi_\mathrm{ion,0}-W_\lambda([\mathrm{O\,III}])$ relation shows a marginal steepening and a systematic decrease in intrinsic scatter at higher redshift, while no clear trend is seen for $\mathrm{H\,\alpha}$. This is physically plausible as a result of evolving nebular conditions (lower metallicity, higher / harder ionization states) impacting $W_\lambda([\mathrm{O\,III}])$ more than than $W_\lambda(\mathrm{H\,\alpha})$.
    \item Sample selection remains an important consideration when predicting $\xi_\mathrm{ion,0}$ - our diverse sample yields generally flatter $\xi_\mathrm{ion,0}-W_\lambda(X)$ relations than other literature studies with more focused samples, and differences between these samples (e.g., alternative selections, or $\xi_\mathrm{ion,0}$ measurement methods) can shift normalisations and slopes.
    \item From a practical perspective, with only a small set of observables, namely $W_\lambda(\mathrm{H\,\alpha})$ or $W_\lambda([\mathrm{O\,III}])$, $\beta_\mathrm{UV}$, and $E(B-V)_\mathrm{neb}$ to reduce the impact of dusty outliers, $\xi_\mathrm{ion,0}$ is predictable across the SFG population at the $\simeq0.15\,$dex level. 
\end{enumerate}

\section*{Acknowledgements}
R. Begley, R. J. McLure, J. S. Dunlop and D.J. McLeod acknowledge the support of the Science and Technology Facilities Council. F. Cullen and T. M. Stanton acknowledge the support from a UKRI Frontier Research Guarantee Grant [grant reference EP/X021025/1]. A. C. Carnall acknowledges support from a UKRI Frontier Research Guarantee Grant [grant reference EP/Y037065/1]. JSD acknowledges the support of the Royal Society via the award of a Royal Society Research Professorship. RSE acknowledges generous financial support from the Peter and Patricia Gruber Foundation.

This work is based in part on observations made with the
NASA/ESA/CSA James Webb Space Telescope. The data were obtained from the Mikulski Archive for Space Telescopes at the Space Telescope Science Institute, which is operated by the Association of Universities for Research in Astronomy, Inc., under NASA contract NAS 5-03127 for JWST.

This research made use of Astropy, a community-developed core Python package for Astronomy \citep{astropy13,astropy18},  NumPy \citep{numpy20} and SciPy \citep{scipy20}, Matplotlib \citep{matplotlib07}, IPython \citep{ipython07}, \textsc{PyMC} \citep[][DOI; \url{https://doi.org/10.5281/zenodo.4603970}]{pymc}, \textsc{Pingouin} \citep[][]{vallat+18} and NASA’s Astrophysics Data System Bibliographic Services.

\section*{Data Availability}
For the purpose of open access, the author has applied a Creative Commons Attribution (CC BY) licence to any Author Accepted Manuscript version arising from this submission. All raw JWST and HST data products are available via
the Mikulski Archive for Space Telescopes \url{https://mast.stsci.edu/}, and additional data underlying the main figures and tables are available from the corresponding author upon reasonable request.



\newcommand{\noopsort}[1]{}
\bibliographystyle{mnras}
\bibliography{excels_paper1} 


\newpage
\clearpage
\appendix

\section{Supplementary figures}

\subsection{Summary of Bayesian regression models}

In Table \ref{tab:coeff}, we summarise the regression coefficients $\theta=\{\beta_X,...,\alpha,\sigma_\mathrm{int},\nu\}$ from the fits throughout Section \ref{sec:results}, that may be used for predictions of $\xi_\mathrm{ion,0}$ from a variety of different measurement types (both photometric and spectroscopic).

\begin{table*}

\caption{Coefficients from the Bayesian linear regression (e.g., see Section \ref{subsec:analysis:model}), for the various models described throughout Section \ref{sec:results}.}
\begin{tabular}{l||l|c|c|c|c|c|c|c|c}
    \multicolumn{1}{|l|}{Model Parameters} & \multicolumn{8}{|l|}{Model Coefficients}\\
    \hline\hline
      $W_\lambda(\mathrm{H\,\alpha})$ & $0.49\pm0.09$ & $-$ & $-$ & $-$ & $-$ & $-$ & $0.65\pm0.06$ & $-$ \\
      $W_\lambda($[\mbox{O\,\textsc{iii}}]) & $-$ & $0.30\pm0.08$ & $-$ & $0.26\pm0.08$ & $0.55\pm0.05$ & $-$ & $-$ & $0.51\pm0.05$\\
      $\beta_\mathrm{UV}+2$ & $-$ & $-$ & $-0.22\pm0.06$ & $-0.13\pm0.06$ & $-0.25\pm0.03$ & $-0.24\pm0.06$ & $-0.33\pm0.03$ & $-0.30\pm0.04$\\
      $M_\mathrm{UV}+20$ & $-$ & $-$ & $-$ & $-$ & $-$ & $0.09\pm0.04$ & $0.05\pm0.03$ & $0.04\pm0.03$ \\
      $E(B-V)_\mathrm{neb}$ & $-$ & $-$ & $-$ & $-$ & $1.73\pm0.10$ & $-$ & $1.58\pm0.09$ &$1.74\pm0.10$ \\
      $z_\mathrm{spec}$ & $-$ & $-$ & $-$ & $-$ & $-$ & $-$ & $0.01\pm0.02$ & $0.01\pm0.02$\\
     \hline
      $\alpha$ & $23.97\pm0.23$ & $24.44\pm0.21$ & $25.25\pm0.04$ & $24.58\pm0.21$ & $24.58\pm0.21$ & $25.22\pm0.04$ & $23.39^{+0.15}_{-0.14}$ & $23.71\pm0.13$ \\
      $\sigma_\mathrm{int}$ & $0.20\pm0.03$ & $0.20\pm0.03$ & $0.23\pm0.04$ & $0.19\pm0.04$ & $0.05\pm0.02$ & $0.23^{+0.05}_{-0.04}$ & $0.04\pm0.02$ & $0.06\pm0.02$ \\
      $\nu$ & $2.67\pm0.75$ & $2.56\pm0.63$ & $3.22\pm1.46$ & $2.65^{+0.98}_{-0.46}$ & $25^{+33}_{-16}$ & $3.4^{+2.7}_{-1.0}$ & $24.2^{+32.7}_{-16.0}$ & $25^{+32}_{-16}$ \\
      \hline
      \label{tab:coeff}
\end{tabular}
\end{table*}

\subsection{Nebular dust attenuation corrections}\label{appendix:subsec:neb}

\begin{figure}
    \centering
    \includegraphics[width=1\columnwidth]{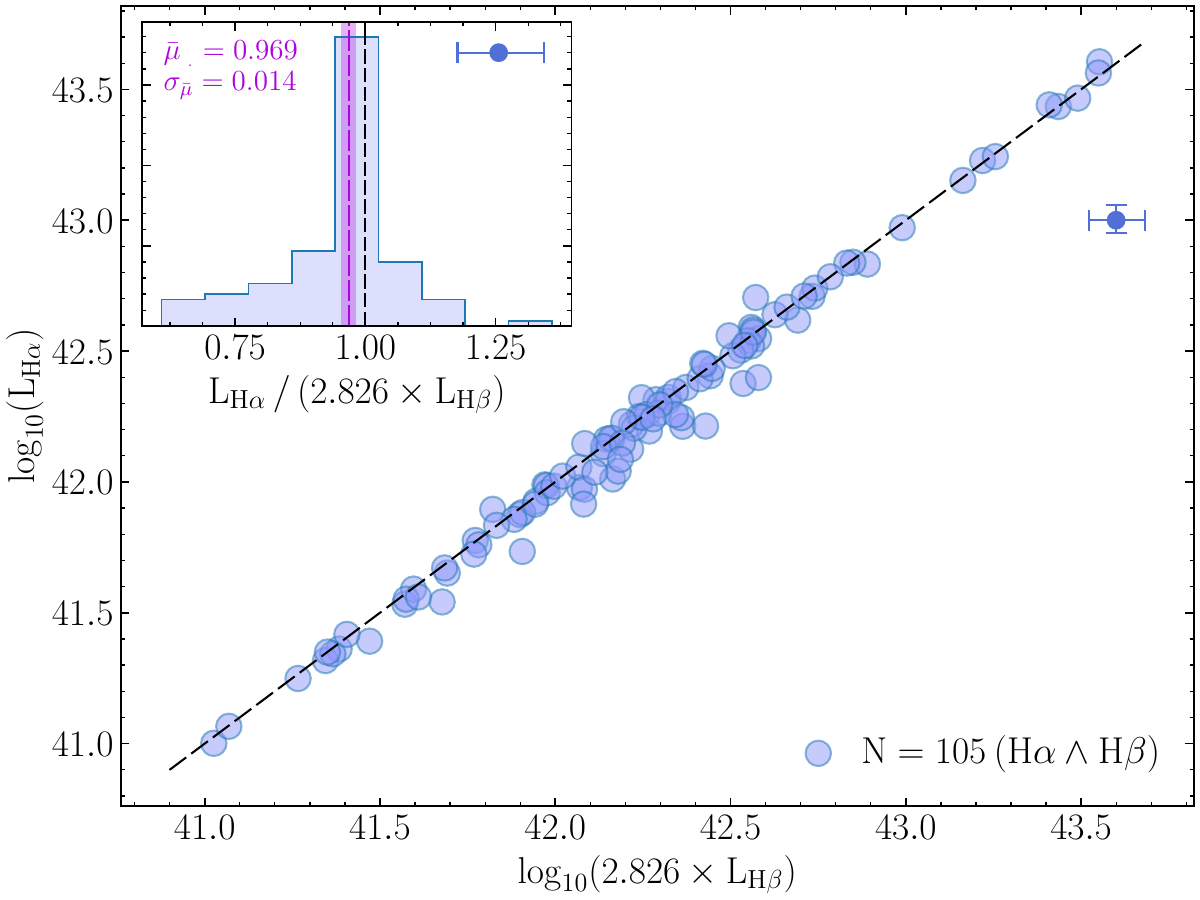}
    \caption{The $\mathrm{H\,\alpha}$ luminosity versus the $\mathrm{H\,\beta}$ luminosity, scaled assuming Case B recombination with $T_e=12,000\,\mathrm{K}$ and $n_e=100\,\mathrm{cm^{-3}}$, for the $N=105$ EXCELS galaxies in our sample with both $\mathrm{H\,\alpha}$ and $H\,\beta$ measured at $\mathrm{S/N\geq3}$. Both measurements are corrected for dust attenuation using the observed Balmer decrement, as outlined in Section \ref{subsubsec:physicalproperties:dustcorrection}. The inset shows the ratio distribution, with $\bar{\mu}=0.969$ and $\sigma_{\bar\mu}=0.014$, highlighting the excellent agreement after accounting for the differential dust attenuation under the adopted Case B assumption. This is also illustrated by the tight clustering around the (black dashed) $1:1$ line. We also indicate the typical $x-y$ error with the marker plotted in the top right.}
    \label{fig:intrinsic_ha_vs_hb}
\end{figure}

As verification for the fiducial choices of Case B recombination parameters, as well as the adopted nebular dust attenuation law (e.g., see Section \ref{subsubsec:physicalproperties:dustcorrection}), in Fig. \ref{fig:intrinsic_ha_vs_hb} we compare the intrinsic $\mathrm{H\,\alpha}$ luminosity to the scaled intrinsic $\mathrm{H\,\beta}$ luminosity. The $\mathrm{H\,\beta}$ luminosity has been scaled by a factor $=2.826$ according to case B recombination ($T_e=12,000$K and $n_e=100\,\mathrm{cm^{-3}}$). The comparison demonstrates an excellent agreement (see inset in Fig. \ref{fig:intrinsic_ha_vs_hb}) in the intrinsic Balmer line luminosities after corrections, with a median ratio of $\bar{\mu}=0.969$ and distribution standard deviation of $\sigma_{\bar{\mu}}=0.014$. Any residual difference remaining is likely due to galaxy-to-galaxy differences in the nebular dust attenuation law \citep[which are not directly measurable in our data, but see][for recent literature on the subject]{sanders+24,reddy+25}, or in a very small minority of cases, a deviation from Case B recombination conditions \citep[e.g., see][]{scarlata+24,mcclymont+25}.

\begin{figure}
    \centering
    \includegraphics[width=1.0\columnwidth]{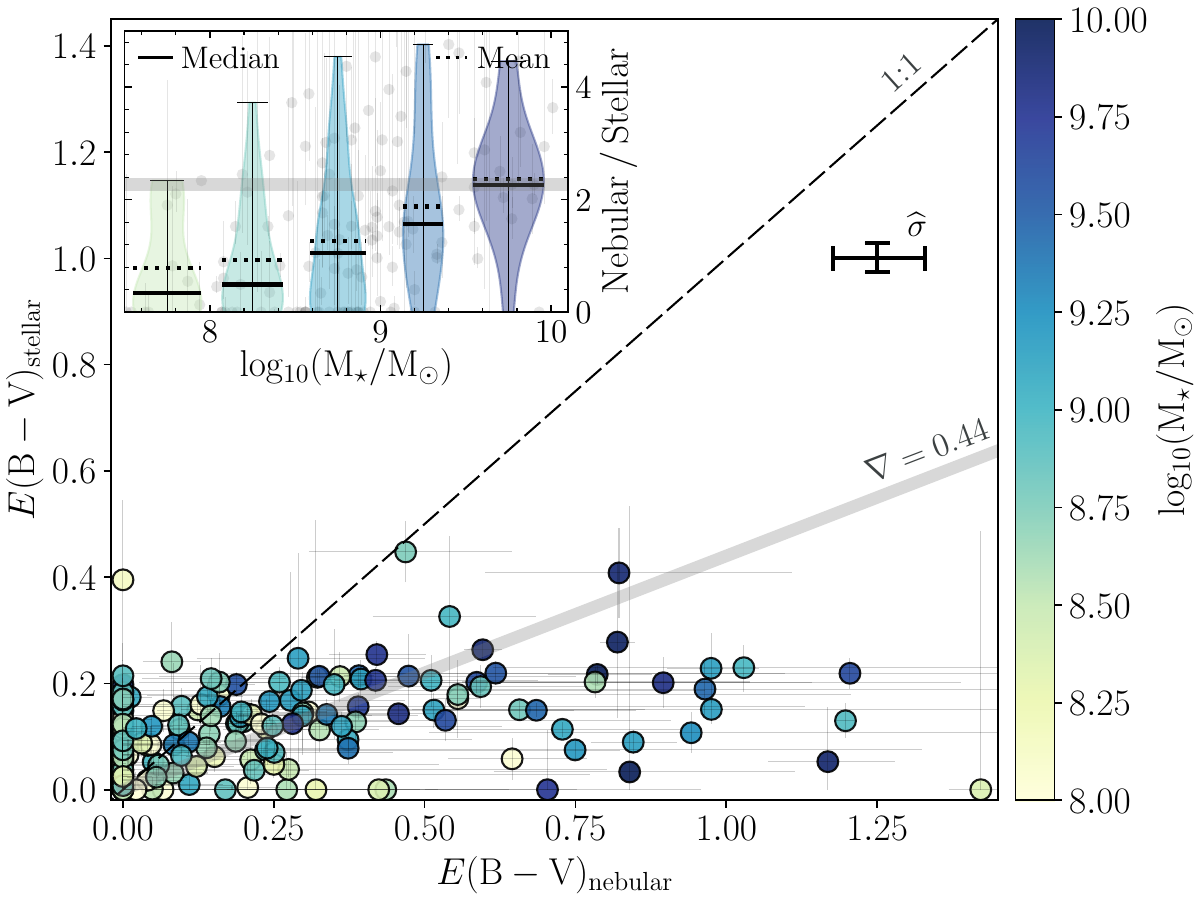}
    \caption{Stellar versus nebular dust attenuation (parameterised by colour excess, $E(\mathrm{B-V)}$, with typical error indicated in the top-right corner) for our EXCELS galaxy sample. We show the $1:1$ line (black dashed) as well as $E(B-V)_\mathrm{stellar}=0.44\times E(B-V)_\mathrm{nebular}$ relation typically adopted in the literature to account for the additional attenuation present in birth clouds \citep[e.g., see][]{calzetti+00,charlot+00}. Significant galaxy-to-galaxy scatter is seen in the EXCELS, justifying the requirement for spectroscopically measured $E(B-V)_\mathrm{neb}$ to robustly infer $\xi_\mathrm{ion,0}$. In the inset we also show the stellar-mass dependence, whereby higher mass galaxies have larger $E(B-V)_\mathrm{neb}/E(B-V)_\mathrm{stellar}$ ratios (see also Section \ref{appendix:subsec:neb} for details).}
    \label{fig:appendix:stellar_to_nebular_dust}
\end{figure}

Here we also demonstrate the importance of direct spectroscopic measurements of the nebular dust attenuation. A common assumption in the literature, based on observations of local starburst galaxies \citep[e.g.,][see also; \citealt{kriek&conroy+13} and \citealt{reddy+20}]{calzetti+00}, is that the stellar and nebular dust attenuation levels are related as $E(B-V)_\mathrm{stellar}/E(B-V)_\mathrm{nebular}=0.44$, as nebular emission originates from dustier star-forming regions, or ``birth-clouds" \citep[][]{charlot+00}. Fig. \ref{fig:appendix:stellar_to_nebular_dust} contrasts the stellar and nebular dust attenuation (parameterised via the colour excess; see Section \ref{subsubsec:physicalproperties:dustcorrection} and Sections \ref{subsubsec:physicalproperties:uvslope} and \ref{subsubsec:physicalproperties:ionizingphotonproduction} for details on $E(B-V)_\mathrm{neb}$ and $E(B-V)_\mathrm{stellar}$, respectively), revealing extremely large galaxy-to-galaxy scatter with $E(B-V)_\mathrm{stellar}/E(B-V)_\mathrm{nebular}$ ratios ranging from $\lesssim0.1$ to $\gtrsim5$. For reference, we also report a median ratio of $\langle E(B-V)_\mathrm{stellar}/E(B-V)_\mathrm{nebular}\rangle\simeq0.45$, and a $\pm1\sigma$ ($16^{\rm th}-84^{\rm th}$ percentile) range of $\simeq0.1-1.1$.

From Fig. \ref{fig:appendix:stellar_to_nebular_dust} it is clear that assuming a constant nebular-to-stellar dust attenuation ratio is a largely oversimplifying assumption, and may lead to large uncertainties in $\xi_\mathrm{ion,0}$. This is exacerbated by the underlying stellar mass dependence (with mass inferred from SED fits to emission line corrected photometry; see Stanton et al. 2025b in preparation for method details) observed in the EXCELS sample, whereby higher mass galaxies have higher $E(B-V)_\mathrm{neb}/E(B-V)_\mathrm{stellar}$ \citep[see also][for similar findings at $z\sim0$]{mingozzi+22}.

\subsection{Individual galaxy constraints on $\xi_\mathrm{ion,0}$}

\begin{figure}
    \centering
    \includegraphics[width=1\columnwidth]{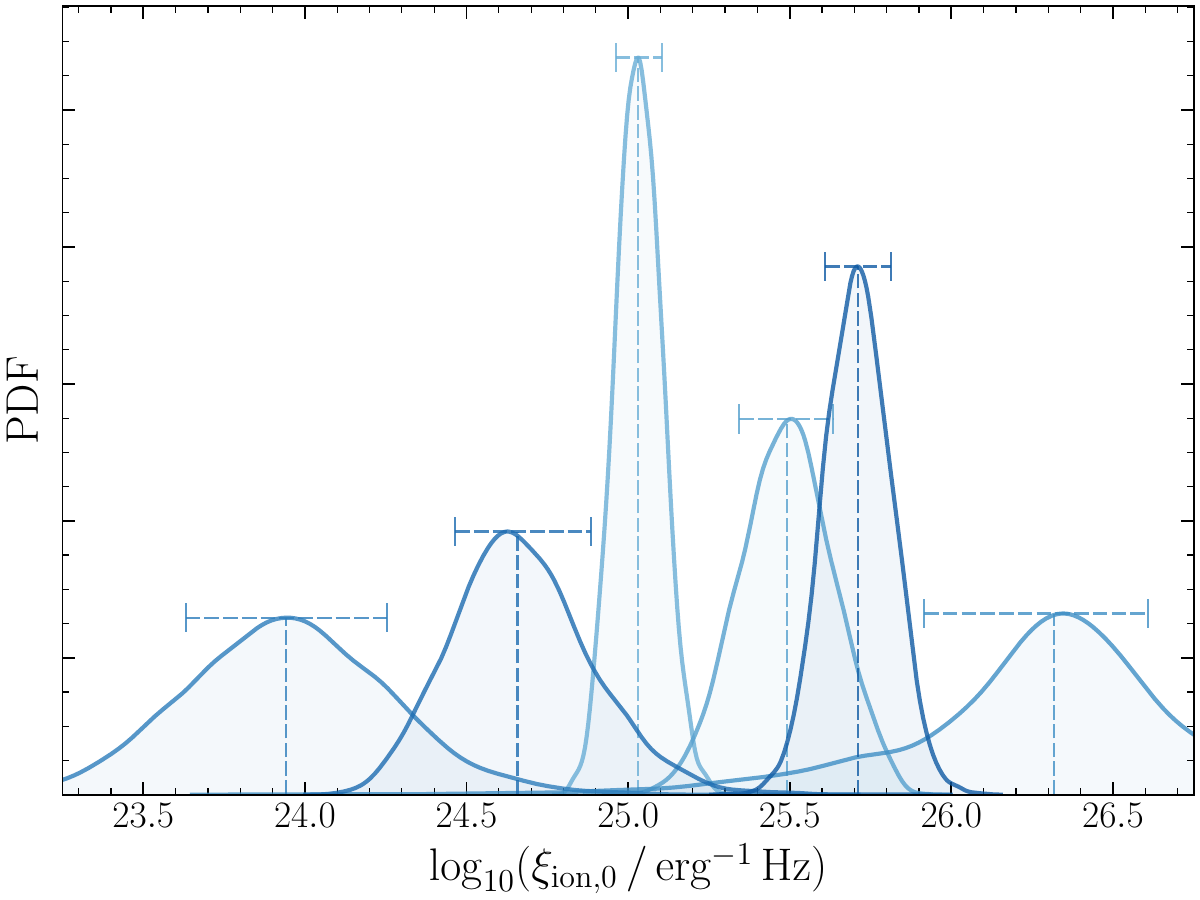}
    \caption{Example $\xi_\mathrm{ion,0}$ constraints obtained from the EXCELS spectroscopy for six sources. We show the kernel density estimates of the distributions generated from the Monte Carlo error uncertainty method used in Section \ref{subsubsec:physicalproperties:ionizingphotonproduction}, along with their median and $16^{\mathrm{th}}-84^{\mathrm{th}}$ percentiles as dashed lines. A key observation is that the galaxy-to-galaxy distributions are diverse, and as such the typical usage of a measurement and error as $x\pm \sigma_x$ (i.e., assuming Normally distributed errors, here as $\mathrm{log_{10}}(\xi_\mathrm{{ion,0}})\sim \mathcal{N}(x,\sigma^2_x)$) would generally fail to accurately represent the constraints. This then motivates the strategy taken in our Bayesian linear regression described in Section \ref{subsec:analysis:model}, whereby we marginalise over the $\xi_\mathrm{ion,0}$ distributions directly.}
    \label{fig:xiion_per_galaxy_distributions}
\end{figure}

To accurately propagate the full measurement uncertainties through to our $\xi_\mathrm{ion,0}$ constraints (e.g., see Section \ref{subsubsec:physicalproperties:ionizingphotonproduction}) and ultimately to our Bayesian linear regression (see Section \ref{subsec:analysis:model}), we have adopted a Monte Carlo approach as detailed throughout Section \ref{subsec:physicalproperties}. This method allows us to maintain more accurate constraints than offered by the standard Normally distribution measurement errors (i.e., $x\pm\sigma_x$) as shown in Fig. \ref{fig:xiion_per_galaxy_distributions}. From the figure, we can see that not only do we have a large range in $\xi_{\mathrm{ion,0}}$ across the EXCELS sample, but there is also significant diversity in the $\xi_\mathrm{ion,0}$ constraints as evident from their distributions. Failing to consider this diversity would inevitably lead to underestimations of the model parameters at best, or strong systematic biases at worst.

\subsection{Posterior distributions from multivariate regression}

\begin{figure*}
    \centering
    \includegraphics[width=\textwidth]{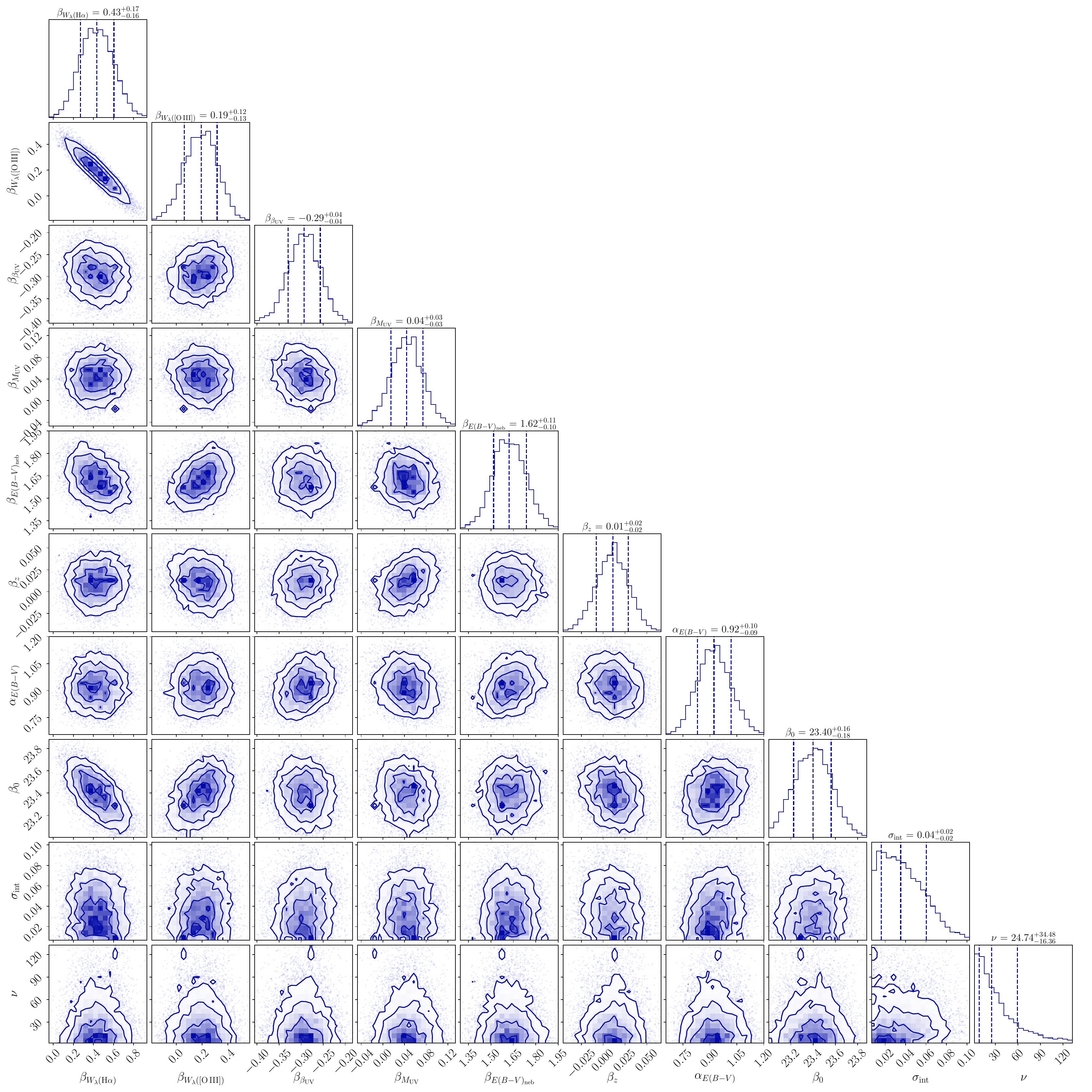}
    \caption{Corner plot for the multivariate Bayesian linear regression between $\xi_\mathrm{ion,0}$ and the six key observable properties $X_n=\{W_\lambda(\mathrm{H\alpha}),\, W_\lambda(\mathrm{[O\,III]},\, \beta_\mathrm{UV},\, M_\mathrm{UV},\, E(B-V)_\mathrm{neb},\, z_\mathrm{spec} \}$. The model has $10$ parameters, including 6 gradients ($\beta_{X_i}$), the intercept $\beta_0$, the power-law index for the assumed $g(E(B-V)_\mathrm{neb})$ functional form, and the intrinsic scatter parameters ($\sigma_\mathrm{int}$ and $\nu$).}
    \label{fig:appendix:full_mv_corner}
\end{figure*}

In Fig. \ref{fig:appendix:full_mv_corner}, we show the corner plot of the multivariate regression that includes the six principal observed properties, as discussed in Section \ref{subsubsec:modelling:mv_full}.

\newpage
\subsection{Bayesian Linear Regression Diagnostics}\label{appendix:subsec:diagnostics}

\begin{figure*}
    \centering
    \includegraphics[width=2.1\columnwidth]{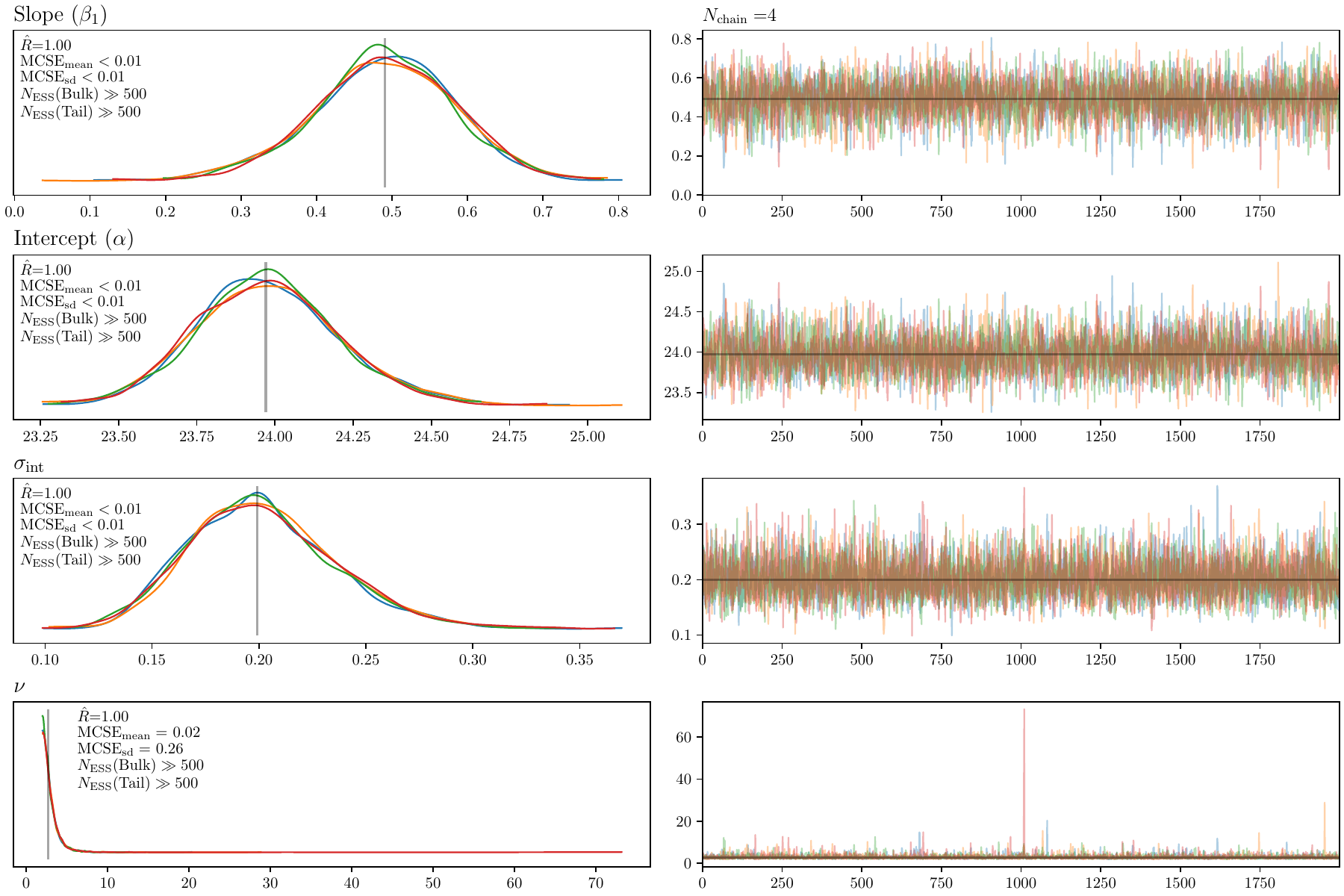}
    \caption{Diagnostic trace plot for the PyMC Bayesian linear regression methodology introduced in Section \ref{subsec:analysis:model}. Each panel on the right shows the chain traces, with the marginal posterior distributions per chain on the left, for the four key parameters (in a single-predictor model, here for $W_\lambda(\mathrm{H\,\alpha})$); $\theta=\{\beta_1,\alpha,\sigma_\mathrm{int},\nu\}$, that is the regression slope and intercept, as well as the intrinsic relation scatter and ``tail" parameter for a $\mathrm{StudentT}$ distribution. The initialisation samples ($N=1000$) are not shown, and we mark the median posterior parameter value across all chains as a grey line (vertically on the right panels, horizontally on the left panels). We also report a number of key diagnostics, as discussed in the text of Section \ref{appendix:subsec:diagnostics}.}
    \label{fig:appendix:bayes_diagnostics}
\end{figure*}

We assess the convergence of the \textsc{PyMC} sampling using diagnostic trace plots for each model fit, for example shown in Fig. \ref{fig:appendix:bayes_diagnostics} for the single-variate $\xi_\mathrm{ion,0}-W_\lambda(\mathrm{H\,\alpha})$ fit (see also summary statistics discussed in Section \ref{subsec:analysis:model}; i.e., $\hat{R}$ and effective sample size, ESS). The figure shows the four independent NUTS chain samples as well as the corresponding posterior density for each fitted parameter (in this case $\theta=\{\beta_1,\alpha,\sigma_\mathrm{int},\nu\}$). Well-behaved sampling is typically represented convergence of the different chains, in addition to the same overall shape in each panel’s posterior density, which indicate the sampler has explored the posterior reliably. If a parameter shows poor mixing or unusual excursions we treat it as a warning and re-run with the necessary adjustments (more tuning, a different parameterisation for that block of the model, or slightly stronger regularising priors) until the traces appear well mixed and stable. In the figure panels, we also report key diagnostic statistics including; $\hat{R}$, the Monte Carlo standard errors $\mathrm{MCSE_{mean}}$ and $\mathrm{MCSE_{SD}}$, and the bulk and tail effective sample sizes, $N_\mathrm{ESS}(\mathrm{Bulk})$ and  $N_\mathrm{ESS}(\mathrm{Tail})$ (see \textsc{PyMC} documentation for details). These statistics further aide interpreting the quality of the model sampling; $\hat{R}$ near unity ($\lesssim1.01$) indicates that the chains have converged to the same distribution; $\mathrm{MCSE_{mean}}$ and $\mathrm{MCSE_{SD}}$ are the Monte-Carlo uncertainties on the posterior mean and standard deviation (they should be very small compared with the posterior width); and the $N_\mathrm{ESS}$ (bulk/tail) values report the effective number of independent samples for the central region and the extremes of the posterior (larger is better, with $\gtrsim500$ typically being sufficient for accurate posteriors).

\bsp	
\label{lastpage}
\end{document}